\documentclass[twocolumn,traditabstract,longauth]{aa}
\usepackage[table,usenames,dvipsnames]{xcolor}
\usepackage{ulem}
\usepackage{enumerate}
\usepackage{ifthen}
\usepackage{perpage}
\usepackage{multirow}
\usepackage{upgreek}
\usepackage{overpic}
\usepackage{graphicx,amsmath}
\usepackage{txfonts}
\usepackage{natbib}
\bibpunct{(}{)}{;}{a}{}{,} % to follow the A&A style
\usepackage{pdfcolmk}
\usepackage{morefloats}
\usepackage{float}
\usepackage{array}
\usepackage[breaklinks, colorlinks, citecolor=blue, colorlinks=true,linkcolor=blue]{hyperref}
\usepackage{deluxetable}
\usepackage{tablefootnote}
\usepackage{pdflscape}

\makeatletter
\def\cl@chapter{\@elt {theorem}}
\makeatother
\usepackage{cleveref}

\usepackage[export]{adjustbox}

\pdfmapfile{+txfonts.map}

\def\setsymbol#1#2{\expandafter\def\csname #1\endcsname{#2}}
\def\getsymbol#1{\csname #1\endcsname}

\def\Planck{\textit{Planck}}

\newbox\tablebox    \newdimen\tablewidth
\def\leaderfil{\leaders\hbox to 5pt{\hss.\hss}\hfil}
\def\endPlancktable{\tablewidth=\columnwidth 
    $$\hss\copy\tablebox\hss$$
    \vskip-\lastskip\vskip -2pt}

\def\tablenote#1 #2\par{\begingroup \parindent=0.8em
    \abovedisplayshortskip=0pt\belowdisplayshortskip=0pt
    \noindent
    $$\hss\vbox{\hsize\tablewidth \hangindent=\parindent \hangafter=1 \noindent
    \hbox to \parindent{$^#1$\hss}\strut#2\strut\par}\hss$$
    \endgroup}
\def\doubleline{\vskip 3pt\hrule \vskip 1.5pt \hrule \vskip 5pt}

\def\L2{\ifmmode L_2\else $L_2$\fi}

\def\DeltaT{\ifmmode \Delta T\else $\Delta T$\fi}
\def\deltat{\ifmmode \Delta t\else $\Delta t$\fi}
\def\fknee{\ifmmode f_{\rm knee}\else $f_{\rm knee}$\fi}
\def\Fmax{\ifmmode F_{\rm max}\else $F_{\rm max}$\fi}
\def\solar{\ifmmode{\rm M}_{\mathord\odot}\else${\rm M}_{\mathord\odot}$\fi}
\def\Msolar{\ifmmode{\rm M}_{\mathord\odot}\else${\rm M}_{\mathord\odot}$\fi}
\def\Lsolar{\ifmmode{\rm L}_{\mathord\odot}\else${\rm L}_{\mathord\odot}$\fi}
\def\inv{\ifmmode^{-1}\else$^{-1}$\fi}
\def\mo{\ifmmode^{-1}\else$^{-1}$\fi}
\def\sup#1{\ifmmode ^{\rm #1}\else $^{\rm #1}$\fi}
\def\expo#1{\ifmmode \times 10^{#1}\else $\times 10^{#1}$\fi}
\def\,{\thinspace}
\def\lsim{\mathrel{\raise .4ex\hbox{\rlap{$<$}\lower 1.2ex\hbox{$\sim$}}}}
\def\gsim{\mathrel{\raise .4ex\hbox{\rlap{$>$}\lower 1.2ex\hbox{$\sim$}}}}
\let\lea=\lsim

\def\simprop{\mathrel{\raise .4ex\hbox{\rlap{$\propto$}\lower 1.2ex\hbox{$\sim$}}}}
\def\deg{\ifmmode^\circ\else$^\circ$\fi}
\def\pdeg{\ifmmode $\setbox0=\hbox{$^{\circ}$}\rlap{\hskip.11\wd0 .}$^{\circ}
          \else \setbox0=\hbox{$^{\circ}$}\rlap{\hskip.11\wd0 .}$^{\circ}$\fi}
\def\arcs{\ifmmode {^{\scriptstyle\prime\prime}}
          \else $^{\scriptstyle\prime\prime}$\fi}
\def\arcm{\ifmmode {^{\scriptstyle\prime}}
          \else $^{\scriptstyle\prime}$\fi}
\newdimen\sa  \newdimen\sb
\def\parcs{\sa=.07em \sb=.03em
     \ifmmode \hbox{\rlap{.}}^{\scriptstyle\prime\kern -\sb\prime}\hbox{\kern -\sa}
     \else \rlap{.}$^{\scriptstyle\prime\kern -\sb\prime}$\kern -\sa\fi}
\def\parcm{\sa=.08em \sb=.03em
     \ifmmode \hbox{\rlap{.}\kern\sa}^{\scriptstyle\prime}\hbox{\kern-\sb}
     \else \rlap{.}\kern\sa$^{\scriptstyle\prime}$\kern-\sb\fi}
\def\ra[#1 #2 #3.#4]{#1\sup{h}#2\sup{m}#3\sup{s}\llap.#4}
\def\dec[#1 #2 #3.#4]{#1\deg#2\arcm#3\arcs\llap.#4}
\def\deco[#1 #2 #3]{#1\deg#2\arcm#3\arcs}
\def\rra[#1 #2]{#1\sup{h}#2\sup{m}}

\def\dots{\relax\ifmmode \ldots\else $\ldots$\fi}
\def\WHzsr{\ifmmode $W\,Hz\mo\,sr\mo$\else W\,Hz\mo\,sr\mo\fi}
\def\mHz{\ifmmode $\,mHz$\else \,mHz\fi}
\def\GHz{\ifmmode $\,GHz$\else \,GHz\fi}
\def\mKs{\ifmmode $\,mK\,s$^{1/2}\else \,mK\,s$^{1/2}$\fi}
\def\muKs{\ifmmode \,\mu$K\,s$^{1/2}\else \,$\mu$K\,s$^{1/2}$\fi}
\def\muKRJs{\ifmmode \,\mu$K$_{\rm RJ}$\,s$^{1/2}\else \,$\mu$K$_{\rm RJ}$\,s$^{1/2}$\fi}
\def\muKHz{\ifmmode \,\mu$K\,Hz$^{-1/2}\else \,$\mu$K\,Hz$^{-1/2}$\fi}
\def\MJysr{\ifmmode \,$MJy\,sr\mo$\else \,MJy\,sr\mo\fi}
\def\MJysrmK{\ifmmode \,$MJy\,sr\mo$\,mK$_{\rm CMB}\mo\else \,MJy\,sr\mo\,mK$_{\rm CMB}\mo$\fi}
\def\microns{\ifmmode \,\mu$m$\else \,$\mu$m\fi}

\def\muK{\ifmmode \,\mu$K$\else \,$\mu$\hbox{K}\fi}
\def\microK{\ifmmode \,\mu$K$\else \,$\mu$\hbox{K}\fi}
\def\muW{\ifmmode \,\mu$W$\else \,$\mu$\hbox{W}\fi}
\def\kms{\ifmmode $\,km\,s$^{-1}\else \,km\,s$^{-1}$\fi}
\def\kmsMpc{\ifmmode $\,\kms\,Mpc\mo$\else \,\kms\,Mpc\mo\fi}
\providecommand{\sorthelp}[1]{}

\newcommand{\lcdm}{$\Lambda$CDM}
\newcommand{\vardeflect}{\langle d^2 \rangle}
\newcommand{\rmsdeflect}{\langle d^2 \rangle^{1/2}}
\newcommand{\elleq}{\ell_{\rm eq}}
\newcommand{\keq}{k_{\rm eq}}
\newcommand{\clpp}{C_{\elp}^{\phi\phi}}

\newcommand{\As}{A_{\rm s}}

\newcommand{\ns}{n_{\rm s}}

\newcommand{\neff}{N_{\rm eff}}
\newcommand{\mnu}{\sum m_\nu}
\newcommand{\sumnu}{\sum m_\nu}

\newcommand{\be}{\begin{equation}}
\newcommand{\ee}{\end{equation}}

\newcommand{\fielda}{X}
\newcommand{\fieldb}{Z}
\newcommand{\fieldc}{C}
\newcommand{\fieldd}{D}

\newcommand{\refchange}[1]{#1}

\newcommand{\bin}{{\cal B}}

\newcommand{\resp}{ {\cal R} }

\newcommand{\map}{d}
\newcommand{\ptg}{{\cal Y}}

\newcommand{\hatn}{\hat{\vec{n}}}

\newcommand{\healpix}{{\tt HEALPix}}
\newcommand{\clttfid}{C_{\elp}^{TT,\ {\rm fid}}}
\newcommand{\clppfid}{C_{\elp}^{\phi\phi,\ {\rm fid}}}
\newcommand{\cltpfid}{C_{\elp}^{T\phi,\ {\rm fid}}}

\newcommand{\muKarcmin}{\mu {\rm K}\, {\rm arcmin}}

\newcommand{\yslm}[3]{ {}_{#1}\!Y_{#2 #3} }

\newcommand{\elt}{\ell}
\newcommand{\elp}{L}
\newcommand{\ellp}{\elp}

\newcommand{\nelpbinsaggressive}{19}
\newcommand{\aggressiveLrange}{\mbox{$8\!\le\!L\!\le\!2048$}}

\newcommand{\conservativeLrange}{\mbox{$40\!\le\!L\!\le\!400$}}

\newcommand{\bmodesignif}{10\,\sigma}
\newcommand{\bmodeapproxsignif}{10\,\sigma}

\newcommand{\detectionsignif}{40\,\sigma}
\newcommand{\poldetectionsignif}{5\,\sigma}
\newcommand{\measurementpercent}{2.5\,\%}
\newcommand{\signifsmicathreexthreefiftythree}{20\,\sigma}

\newcommand{\pcntskylikelihood}{70\,\%}

\newcommand{\sigmaeightomegamconstraint}{0.591 \pm 0.021}

\newcommand{\lensingiswsignif}{3\,\sigma}
\newcommand{\lensingiswapproxsignif}{3\,\sigma}

\newcommand{\threej}[6]{\left(
    \begin{array}{ccc}
        \! #1\! & #2\!  & #3\!  \\
        \! #4\! & #5\!  & #6\!
      \end{array}
    \right)}

\def\WMAP{{WMAP}}

\newcommand{\None}{N^{(1)}}

\newcommand{\fid}{{\rm fid.}}

\newcommand{\commander}{{\tt Commander}}

\newcommand{\smica}{{\tt SMICA}}
\newcommand{\camb}{{\tt CAMB}}

\newcommand{\mksym}[1]{\ifmmode {\rm #1}\else #1\fi}

\newcommand{\lensing}{\mksym{lensing}}

\newcommand{\TT}{\mksym{TT}}

\newcommand{\planckTTonly}{\planck\ \TT}
\newcommand{\lowTEB}{\mksym{lowP}}

\newcommand{\planckTT}{\planck\ \TT+\lowTEB}

\providecommand{\Planck}{\textit{Planck}}
\providecommand{\planck}{\Planck}

\providecommand{\lea}{\la}

\providecommand{\alt}{\lea}

\providecommand{\text}[1]{\rm{#1}}

\providecommand{\muK}{\mu\rm{K}}

\providecommand{\Omk}{\Omega_K}

\providecommand{\Omb}{\Omega_{\mathrm{b}}}
\providecommand{\Omc}{\Omega_{\mathrm{c}}}
\providecommand{\Omm}{\Omega_{\mathrm{m}}}

\providecommand{\CAMB}{{\tt camb}}
\providecommand{\COSMOMC}{{\tt CosmoMC}}

\providecommand{\HALOFIT}{{\tt halofit}}

\newcommand{\vtheta}{{\boldsymbol{\theta}}}

\newcommand{\boldvec}[1]{{\mbox{\boldmath{$#1$}}}}

\MakePerPage{footnote}

\makeatletter
\renewcommand*{\@fnsymbol}[1]{\ensuremath{\ifcase#1\or \dagger\or \ddagger\or *\or
   \mathsection\or \mathparagraph\or \|\or **\or \dagger\dagger
   \or \ddagger\ddagger  \or a \or b \or c \or d \or e \or f \else g \fi }} %\@ctrerr\fi}}
\makeatother

\newcommand{\Wlens}{ {\cal W} }

\begin{document}
%This author list corresponds to \title{Author list for A17\_Lensing}
%Prepared by M. Lopez-Caniego (Marcos.Lopez.Caniego@sciops.esa.int), ESAC/ESA
%This version is from Tue Nov 17 07:55:42 2015 CET
%\subtitle{There are 229 co-authors in this list}
\author{\small
Planck Collaboration: P.~A.~R.~Ade\inst{96}
\and
N.~Aghanim\inst{65}
\and
M.~Arnaud\inst{81}
\and
M.~Ashdown\inst{77, 6}
\and
J.~Aumont\inst{65}
\and
C.~Baccigalupi\inst{95}
\and
A.~J.~Banday\inst{107, 11}
\and
R.~B.~Barreiro\inst{72}
\and
J.~G.~Bartlett\inst{1, 74}
\and
N.~Bartolo\inst{35, 73}
% ...
}
\institute{\small
APC, AstroParticule et Cosmologie, Universit\'{e} Paris Diderot, CNRS/IN2P3, CEA/lrfu, Observatoire de Paris, Sorbonne Paris Cit\'{e}, 10, rue Alice Domon et L\'{e}onie Duquet, 75205 Paris Cedex 13, France\goodbreak
\and
Aalto University Mets\"{a}hovi Radio Observatory and Dept of Radio Science and Engineering, P.O. Box 13000, FI-00076 AALTO, Finland\goodbreak
\and
African Institute for Mathematical Sciences, 6-8 Melrose Road, Muizenberg, Cape Town, South Africa\goodbreak
\and
Agenzia Spaziale Italiana Science Data Center, Via del Politecnico snc, 00133, Roma, Italy\goodbreak
\and
Aix Marseille Universit\'{e}, CNRS, LAM (Laboratoire d'Astrophysique de Marseille) UMR 7326, 13388, Marseille, France\goodbreak
\and
Astrophysics Group, Cavendish Laboratory, University of Cambridge, J J Thomson Avenue, Cambridge CB3 0HE, U.K.\goodbreak
\and
Astrophysics \& Cosmology Research Unit, School of Mathematics, Statistics \& Computer Science, University of KwaZulu-Natal, Westville Campus, Private Bag X54001, Durban 4000, South Africa\goodbreak
\and
Atacama Large Millimeter/submillimeter Array, ALMA Santiago Central Offices, Alonso de Cordova 3107, Vitacura, Casilla 763 0355, Santiago, Chile\goodbreak
\and
CGEE, SCS Qd 9, Lote C, Torre C, 4$^{\circ}$ andar, Ed. Parque Cidade Corporate, CEP 70308-200, Bras\'{i}lia, DF,Ê Brazil\goodbreak
\and
CITA, University of Toronto, 60 St. George St., Toronto, ON M5S 3H8, Canada\goodbreak
\and
CNRS, IRAP, 9 Av. colonel Roche, BP 44346, F-31028 Toulouse cedex 4, France\goodbreak
\and
CRANN, Trinity College, Dublin, Ireland\goodbreak
\and
California Institute of Technology, Pasadena, California, U.S.A.\goodbreak
\and
Centre for Theoretical Cosmology, DAMTP, University of Cambridge, Wilberforce Road, Cambridge CB3 0WA, U.K.\goodbreak
\and
Centro de Estudios de F\'{i}sica del Cosmos de Arag\'{o}n (CEFCA), Plaza San Juan, 1, planta 2, E-44001, Teruel, Spain\goodbreak
\and
Computational Cosmology Center, Lawrence Berkeley National Laboratory, Berkeley, California, U.S.A.\goodbreak
\and
Consejo Superior de Investigaciones Cient\'{\i}ficas (CSIC), Madrid, Spain\goodbreak
\and
DSM/Irfu/SPP, CEA-Saclay, F-91191 Gif-sur-Yvette Cedex, France\goodbreak
\and
DTU Space, National Space Institute, Technical University of Denmark, Elektrovej 327, DK-2800 Kgs. Lyngby, Denmark\goodbreak
\and
D\'{e}partement de Physique Th\'{e}orique, Universit\'{e} de Gen\`{e}ve, 24, Quai E. Ansermet,1211 Gen\`{e}ve 4, Switzerland\goodbreak
\and
Departamento de Astrof\'{i}sica, Universidad de La Laguna (ULL), E-38206 La Laguna, Tenerife, Spain\goodbreak
\and
Departamento de F\'{\i}sica, Universidad de Oviedo, Avda. Calvo Sotelo s/n, Oviedo, Spain\goodbreak
\and
Department of Astronomy and Astrophysics, University of Toronto, 50 Saint George Street, Toronto, Ontario, Canada\goodbreak
\and
Department of Astrophysics/IMAPP, Radboud University Nijmegen, P.O. Box 9010, 6500 GL Nijmegen, The Netherlands\goodbreak
\and
Department of Physics \& Astronomy, University of British Columbia, 6224 Agricultural Road, Vancouver, British Columbia, Canada\goodbreak
\and
Department of Physics and Astronomy, Dana and David Dornsife College of Letter, Arts and Sciences, University of Southern California, Los Angeles, CA 90089, U.S.A.\goodbreak
\and
Department of Physics and Astronomy, University College London, London WC1E 6BT, U.K.\goodbreak
\and
Department of Physics and Astronomy, University of Sussex, Brighton BN1 9QH, U.K.\goodbreak
\and
Department of Physics, Florida State University, Keen Physics Building, 77 Chieftan Way, Tallahassee, Florida, U.S.A.\goodbreak
\and
Department of Physics, Gustaf H\"{a}llstr\"{o}min katu 2a, University of Helsinki, Helsinki, Finland\goodbreak
\and
Department of Physics, Princeton University, Princeton, New Jersey, U.S.A.\goodbreak
\and
Department of Physics, University of California, Berkeley, California, U.S.A.\goodbreak
\and
Department of Physics, University of California, Santa Barbara, California, U.S.A.\goodbreak
\and
Department of Physics, University of Illinois at Urbana-Champaign, 1110 West Green Street, Urbana, Illinois, U.S.A.\goodbreak
\and
Dipartimento di Fisica e Astronomia G. Galilei, Universit\`{a} degli Studi di Padova, via Marzolo 8, 35131 Padova, Italy\goodbreak
\and
Dipartimento di Fisica e Scienze della Terra, Universit\`{a} di Ferrara, Via Saragat 1, 44122 Ferrara, Italy\goodbreak
\and
Dipartimento di Fisica, Universit\`{a} La Sapienza, P. le A. Moro 2, Roma, Italy\goodbreak
\and
Dipartimento di Fisica, Universit\`{a} degli Studi di Milano, Via Celoria, 16, Milano, Italy\goodbreak
\and
Dipartimento di Fisica, Universit\`{a} degli Studi di Trieste, via A. Valerio 2, Trieste, Italy\goodbreak
\and
Dipartimento di Matematica, Universit\`{a} di Roma Tor Vergata, Via della Ricerca Scientifica, 1, Roma, Italy\goodbreak
\and
Discovery Center, Niels Bohr Institute, Blegdamsvej 17, Copenhagen, Denmark\goodbreak
\and
Discovery Center, Niels Bohr Institute, Copenhagen University, Blegdamsvej 17, Copenhagen, Denmark\goodbreak
\and
European Southern Observatory, ESO Vitacura, Alonso de Cordova 3107, Vitacura, Casilla 19001, Santiago, Chile\goodbreak
\and
European Space Agency, ESAC, Planck Science Office, Camino bajo del Castillo, s/n, Urbanizaci\'{o}n Villafranca del Castillo, Villanueva de la Ca\~{n}ada, Madrid, Spain\goodbreak
\and
European Space Agency, ESTEC, Keplerlaan 1, 2201 AZ Noordwijk, The Netherlands\goodbreak
\and
Facolt\`{a} di Ingegneria, Universit\`{a} degli Studi e-Campus, Via Isimbardi 10, Novedrate (CO), 22060, Italy\goodbreak
\and
Gran Sasso Science Institute, INFN, viale F. Crispi 7, 67100 L'Aquila, Italy\goodbreak
\and
HGSFP and University of Heidelberg, Theoretical Physics Department, Philosophenweg 16, 69120, Heidelberg, Germany\goodbreak
\and
Helsinki Institute of Physics, Gustaf H\"{a}llstr\"{o}min katu 2, University of Helsinki, Helsinki, Finland\goodbreak
\and
INAF - Osservatorio Astronomico di Padova, Vicolo dell'Osservatorio 5, Padova, Italy\goodbreak
\and
INAF - Osservatorio Astronomico di Roma, via di Frascati 33, Monte Porzio Catone, Italy\goodbreak
\and
INAF - Osservatorio Astronomico di Trieste, Via G.B. Tiepolo 11, Trieste, Italy\goodbreak
\and
INAF/IASF Bologna, Via Gobetti 101, Bologna, Italy\goodbreak
\and
INAF/IASF Milano, Via E. Bassini 15, Milano, Italy\goodbreak
\and
INFN, Sezione di Bologna, Via Irnerio 46, I-40126, Bologna, Italy\goodbreak
\and
INFN, Sezione di Roma 1, Universit\`{a} di Roma Sapienza, Piazzale Aldo Moro 2, 00185, Roma, Italy\goodbreak
\and
INFN, Sezione di Roma 2, Universit\`{a} di Roma Tor Vergata, Via della Ricerca Scientifica, 1, Roma, Italy\goodbreak
\and
INFN/National Institute for Nuclear Physics, Via Valerio 2, I-34127 Trieste, Italy\goodbreak
\and
IPAG: Institut de Plan\'{e}tologie et d'Astrophysique de Grenoble, Universit\'{e} Grenoble Alpes, IPAG, F-38000 Grenoble, France, CNRS, IPAG, F-38000 Grenoble, France\goodbreak
\and
IUCAA, Post Bag 4, Ganeshkhind, Pune University Campus, Pune 411 007, India\goodbreak
\and
Imperial College London, Astrophysics group, Blackett Laboratory, Prince Consort Road, London, SW7 2AZ, U.K.\goodbreak
\and
Infrared Processing and Analysis Center, California Institute of Technology, Pasadena, CA 91125, U.S.A.\goodbreak
\and
Institut N\'{e}el, CNRS, Universit\'{e} Joseph Fourier Grenoble I, 25 rue des Martyrs, Grenoble, France\goodbreak
\and
Institut Universitaire de France, 103, bd Saint-Michel, 75005, Paris, France\goodbreak
\and
Institut d'Astrophysique Spatiale, CNRS (UMR8617) Universit\'{e} Paris-Sud 11, B\^{a}timent 121, Orsay, France\goodbreak
\and
Institut d'Astrophysique de Paris, CNRS (UMR7095), 98 bis Boulevard Arago, F-75014, Paris, France\goodbreak
\and
Institut f\"ur Theoretische Teilchenphysik und Kosmologie, RWTH Aachen University, D-52056 Aachen, Germany\goodbreak
\and
Institute for Space Sciences, Bucharest-Magurale, Romania\goodbreak
\and
Institute of Astronomy, University of Cambridge, Madingley Road, Cambridge CB3 0HA, U.K.\goodbreak
\and
Institute of Theoretical Astrophysics, University of Oslo, Blindern, Oslo, Norway\goodbreak
\and
Instituto de Astrof\'{\i}sica de Canarias, C/V\'{\i}a L\'{a}ctea s/n, La Laguna, Tenerife, Spain\goodbreak
\and
Instituto de F\'{\i}sica de Cantabria (CSIC-Universidad de Cantabria), Avda. de los Castros s/n, Santander, Spain\goodbreak
\and
Istituto Nazionale di Fisica Nucleare, Sezione di Padova, via Marzolo 8, I-35131 Padova, Italy\goodbreak
\and
Jet Propulsion Laboratory, California Institute of Technology, 4800 Oak Grove Drive, Pasadena, California, U.S.A.\goodbreak
\and
Jodrell Bank Centre for Astrophysics, Alan Turing Building, School of Physics and Astronomy, The University of Manchester, Oxford Road, Manchester, M13 9PL, U.K.\goodbreak
\and
Kavli Institute for Cosmological Physics, University of Chicago, Chicago, IL 60637, USA\goodbreak
\and
Kavli Institute for Cosmology Cambridge, Madingley Road, Cambridge, CB3 0HA, U.K.\goodbreak
\and
Kazan Federal University, 18 Kremlyovskaya St., Kazan, 420008, Russia\goodbreak
\and
LAL, Universit\'{e} Paris-Sud, CNRS/IN2P3, Orsay, France\goodbreak
\and
LERMA, CNRS, Observatoire de Paris, 61 Avenue de l'Observatoire, Paris, France\goodbreak
\and
Laboratoire AIM, IRFU/Service d'Astrophysique - CEA/DSM - CNRS - Universit\'{e} Paris Diderot, B\^{a}t. 709, CEA-Saclay, F-91191 Gif-sur-Yvette Cedex, France\goodbreak
\and
Laboratoire Traitement et Communication de l'Information, CNRS (UMR 5141) and T\'{e}l\'{e}com ParisTech, 46 rue Barrault F-75634 Paris Cedex 13, France\goodbreak
\and
Laboratoire de Physique Subatomique et Cosmologie, Universit\'{e} Grenoble-Alpes, CNRS/IN2P3, 53, rue des Martyrs, 38026 Grenoble Cedex, France\goodbreak
\and
Laboratoire de Physique Th\'{e}orique, Universit\'{e} Paris-Sud 11 \& CNRS, B\^{a}timent 210, 91405 Orsay, France\goodbreak
\and
Lawrence Berkeley National Laboratory, Berkeley, California, U.S.A.\goodbreak
\and
Lebedev Physical Institute of the Russian Academy of Sciences, Astro Space Centre, 84/32 Profsoyuznaya st., Moscow, GSP-7, 117997, Russia\goodbreak
\and
Max-Planck-Institut f\"{u}r Astrophysik, Karl-Schwarzschild-Str. 1, 85741 Garching, Germany\goodbreak
\and
McGill Physics, Ernest Rutherford Physics Building, McGill University, 3600 rue University, Montr\'{e}al, QC, H3A 2T8, Canada\goodbreak
\and
National University of Ireland, Department of Experimental Physics, Maynooth, Co. Kildare, Ireland\goodbreak
\and
Nicolaus Copernicus Astronomical Center, Bartycka 18, 00-716 Warsaw, Poland\goodbreak
\and
Niels Bohr Institute, Blegdamsvej 17, Copenhagen, Denmark\goodbreak
\and
Niels Bohr Institute, Copenhagen University, Blegdamsvej 17, Copenhagen, Denmark\goodbreak
\and
Nordita (Nordic Institute for Theoretical Physics), Roslagstullsbacken 23, SE-106 91 Stockholm, Sweden\goodbreak
\and
Optical Science Laboratory, University College London, Gower Street, London, U.K.\goodbreak
\and
SISSA, Astrophysics Sector, via Bonomea 265, 34136, Trieste, Italy\goodbreak
\and
School of Physics and Astronomy, Cardiff University, Queens Buildings, The Parade, Cardiff, CF24 3AA, U.K.\goodbreak
\and
School of Physics and Astronomy, University of Nottingham, Nottingham NG7 2RD, U.K.\goodbreak
\and
Sorbonne Universit\'{e}-UPMC, UMR7095, Institut d'Astrophysique de Paris, 98 bis Boulevard Arago, F-75014, Paris, France\goodbreak
\and
Space Research Institute (IKI), Russian Academy of Sciences, Profsoyuznaya Str, 84/32, Moscow, 117997, Russia\goodbreak
\and
Space Sciences Laboratory, University of California, Berkeley, California, U.S.A.\goodbreak
\and
Special Astrophysical Observatory, Russian Academy of Sciences, Nizhnij Arkhyz, Zelenchukskiy region, Karachai-Cherkessian Republic, 369167, Russia\goodbreak
\and
Stanford University, Dept of Physics, Varian Physics Bldg, 382 Via Pueblo Mall, Stanford, California, U.S.A.\goodbreak
\and
Sub-Department of Astrophysics, University of Oxford, Keble Road, Oxford OX1 3RH, U.K.\goodbreak
\and
The Oskar Klein Centre for Cosmoparticle Physics, Department of Physics,Stockholm University, AlbaNova, SE-106 91 Stockholm, Sweden\goodbreak
\and
Theory Division, PH-TH, CERN, CH-1211, Geneva 23, Switzerland\goodbreak
\and
UPMC Univ Paris 06, UMR7095, 98 bis Boulevard Arago, F-75014, Paris, France\goodbreak
\and
Universit\'{e} de Toulouse, UPS-OMP, IRAP, F-31028 Toulouse cedex 4, France\goodbreak
% ...
}

\title{\Planck\ 2015 results. XV. Gravitational lensing}

\abstract{
We present the most significant measurement of the cosmic microwave background (CMB) lensing potential to date (at a level of $\detectionsignif$),
using temperature and polarization data from the {\it Planck} 2015 full-mission release.
Using a polarization-only estimator, we detect lensing at a significance of $\poldetectionsignif$.
We cross-check the accuracy of our measurement using the wide frequency coverage
and complementarity of the temperature and polarization measurements.
Public products based on this measurement include an estimate of the lensing potential
over approximately
$\pcntskylikelihood$
of the sky, an estimate of the lensing potential power spectrum in bandpowers for the
multipole range
$\conservativeLrange$,
and an associated likelihood for cosmological parameter constraints.
We find good agreement between our measurement of the lensing potential
power spectrum and that found in the \lcdm\ model that best fits the {\it Planck} temperature and polarization power spectra.
Using the lensing likelihood alone we obtain a percent-level measurement of the parameter combination
$\sigma_8 \Omega_{\rm m}^{0.25} = \sigmaeightomegamconstraint$.
We combine our determination of the lensing potential with the $E$-mode polarization, also measured by {\it Planck}, to generate an estimate of the lensing $B$-mode.
We show that this lensing $B$-mode estimate is correlated with the $B$-modes observed directly by {\it Planck} at the expected level and with a statistical significance of $\bmodesignif$, confirming {\it Planck}'s sensitivity to this known sky signal.
We also correlate our lensing potential estimate with the large-scale
temperature anisotropies, detecting a cross-correlation at the
$\lensingiswsignif$ level, as expected because of dark energy in the concordance \lcdm\ model.
}
\renewcommand\abstractname{} %FIXME: remove.

\date{Draft compiled \today}

\titlerunning{Gravitational lensing by large-scale structures with \textit{Planck}}
\authorrunning{Planck Collaboration}
\maketitle

\section{Introduction}
\label{sec:introduction}
The cosmic microwave background (CMB) gives us a direct measurement of the early Universe when it first became
transparent to radiation just
$375\, 000$
years after the Big Bang.
It contains distinct signatures of the later Universe as well, which
were imprinted by the process of gravitational lensing.
The CMB photons that we observe today last scattered approximately 14 billion years ago,
and have travelled most of the way across the observable Universe to reach us.
During their journey, their paths were distorted by the gravitational tug of intervening matter,
a subtle effect that may be measured statistically with high angular resolution, low-noise observations of the CMB, such as those provided by {\it Planck}.
In this paper, we present lensing measurements that are based on the full-mission 2015 data release.
We produce the most powerful measurement of CMB lensing to date with a
$\measurementpercent$
constraint on the amplitude of the lensing potential power spectrum
(or alternatively, a $\detectionsignif$ detection of lensing effects).

The effect of lensing is to remap the CMB fluctuations,
so that the observed anisotropy in direction $\hatn$ is in fact the unlensed,
`primordial' anisotropy in the direction $\hatn + \nabla \phi(\hatn)$,
where $\phi(\hatn)$ is the CMB lensing potential defined by (e.g.~\citealt{Lewis:2006fu})
\be
\phi(\hatn) = -2 \int_0^{\chi_*} d\chi
\frac{ f_K( \chi_* - \chi) }{ f_K(\chi_*) f_K(\chi)}
\Psi(\chi \hatn; \eta_0 - \chi ).
\label{eqn:lensing_deflection}
\ee
Here, $\chi$ is conformal distance (with \mbox{$\chi_* \approx 14\, 000\,\mathrm{Mpc}$} denoting the distance to the CMB last-scattering surface).
The angular-diameter distance $f_K(\chi)$ depends on the curvature of the Universe, and is given by
\be
f_K(\chi)=\begin{cases}
 K^{-1/2} \sin (K^{1/2} \chi) & \text{for $K>0$ (closed)}  , \\
\chi & \text{for $K=0$ (flat)}  , \\
|K|^{-1/2} \sinh (|K|^{1/2} \chi) & \text{for $K<0$ (open)}. \\
\end{cases}
\label{f_K_def}
\ee
Finally, $\Psi( \chi \hatn; \eta )$ is the (Weyl) gravitational potential at conformal distance $\chi$ along the direction $\hatn$ at conformal time $\eta$ (the conformal time today is denoted as $\eta_0$).
The lensing potential is an integrated measure of the mass distribution back to the last-scattering surface.
The power spectrum of the lensing potential probes the matter power spectrum, which is sensitive to `late-time' parameters that modify the growth of structure such as neutrino mass \citep{Smith:2008an}.
The amplitude of lensing effects is also a sensitive probe of geometrical parameters, such as the curvature of the Universe.

The lens-induced remapping imprints distinctive statistical signatures onto the observed CMB fluctuations,
which can be mined for cosmological information in a process known as lens reconstruction \citep{Okamoto:2003zw}.
The past several years have been seen dramatic improvements in CMB lensing measurements,
moving from first detections in cross-correlation
\citep{Smith:2007rg,Hirata:2008cb} to cosmologically useful measurements of the lensing potential power spectrum
\citep{Das:2011ak,vanEngelen:2012va,planck2013-p12}.
Recently, ground-based experiments have been able to detect the effects of lensing in
polarization data as well \citep{Hanson:2013hsb,Ade:2013hjl,Ade:2013gez,vanEngelen:2014zlh}.

The results in this paper extend our earlier results in
\cite{planck2013-p12}. These were based on the
March 2013 {\it Planck}\footnote{\Planck\ (\url{http://www.esa.int/Planck}) is a project of the European Space Agency  (ESA) with instruments provided by two scientific consortia funded by ESA member states and led by Principal Investigators from France and Italy, telescope reflectors provided through a collaboration between ESA and a scientific consortium led and funded by Denmark, and additional contributions from NASA (USA).}
nominal-mission data release, which contains approximately 15 months of temperature data alone.
The additional information in the full-mission dataset, with 30 months of \planck\ HFI temperature and polarization data~\citep{planck2014-a08,planck2014-a09}, allows us to improve our reconstruction noise levels by roughly a factor of two.
Approximately half of this improvement comes from lower noise levels in temperature, and the other half from inclusion of polarization data.
The improved lensing map is included as part of the {\it Planck} 2015 public data release \citep{planck2014-a01},
as well as an estimate of the lensing potential power spectrum and associated likelihoods.
In this paper we describe the creation of these products, as well as first science results based on them. We highlight the following science results.
\begin{itemize}
\item We detect lensing $B$-modes in the {\it Planck} data at a significance of $\bmodesignif$,
using both a cross-correlation approach with the cosmic infrared background (CIB) as a tracer of the lensing potential,
as well as a CMB-only approach using the $TTEB$ trispectrum.
This provides an important confirmation that {\it Planck} is sensitive to this known source of $B$-modes on intermediate and small scales.
\item We make an improved measurement of the 3-point function
  (bispectrum) that is induced in the CMB by the correlation between
  lensing and the integrated Sachs-Wolfe (ISW) effect, the latter being sourced by late-time acceleration. The lensing-ISW bispectrum is now detected at the $\lensingiswsignif$ level.
\item Using only lensing information (along with well-motivated priors),
we constrain the parameter combination $\sigma_8 \Omega_{\rm m}^{0.25}$ to roughly $3\,\%$.
\end{itemize}

The organization of this paper is as follows. In Sect.~\ref{sec:data_and_methodology} we give a summary of our analysis pipeline for the full-mission data.
Our analysis is very similar to that presented in \cite{planck2013-p12},
with straightforward extensions to polarization data, as well as a few additional improvements,
and so we have deferred most technical details to appendices.
In Sect.~\ref{sec:results} we present our main results, including the lensing potential map, bandpower estimates of the lensing potential power spectrum, and implications for cosmological parameters.
In Sect.~\ref{sec:consistency} we present a suite of consistency and null tests to verify our bandpower estimates, and in Sect.~\ref{sec:conclusions} we conclude.
A series of appendices provide more technical details of our analysis pipeline, modeling of the CIB, and dependence of the lensing power spectrum on $\Lambda$CDM parameters.

\section{Data and methodology}
\label{sec:data_and_methodology}
Here we give a brief overview of the procedure that we use to measure the CMB lensing potential and its power spectrum from the {\Planck} maps.
We defer the detailed technical aspects of our pipeline to Appendix~\ref{appendix:pipeline}.

Our main results are based on a foreground-cleaned map of the CMB synthesized from the raw {\it Planck} 2015 full-mission frequency maps using the
\smica\ code \citep{planck2014-a11}.
This foreground-cleaned map combines all nine frequency bands from 30\,GHz to 857\,GHz with scale-dependent coefficients chosen to provide unit response to the CMB with minimal variance;
for more details see \cite{planck2014-a11}\footnote{\refchange{We use
    a slightly earlier version of \smica\ maps than those in the
    public release. They differ by pixel-smoothing and missing-pixel mask, however this has no significant effect on the results.}}.
On the small scales where lensing effects are most important (multipoles above $\elt=1000$),
most of the CMB information in the foreground-cleaned maps originates with the 143\,GHz and 217\,GHz channel data.
These channels have beams that are approximately Gaussian with full-width-at-half-maximum (FWHM) parameters of $7'$ and $5'$, respectively.
Their noise is approximately white for
$1000 \lesssim \elt \lesssim 3000$, with levels of
$30\,\muKarcmin$ ($60\,\muKarcmin$) in temperature (polarization) at 143\,GHz, and
$40\,\muKarcmin$ ($95\,\muKarcmin$) at 217\,GHz.

\begin{figure}[!ht]
\centering
\includegraphics[width=\columnwidth]{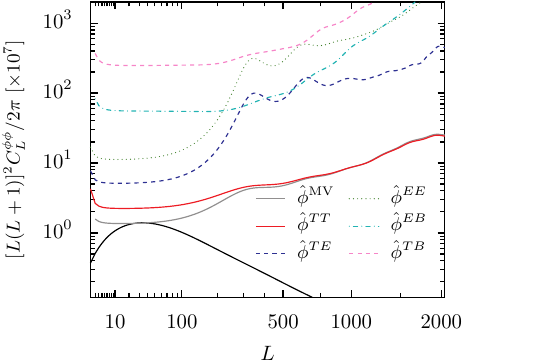}
\vspace{-0.2in}
\caption{
Lens reconstruction noise levels
$N_{\elp}^{\phi\phi}$ for the
$TT$, $TE$, $EE$, $EB$, and $TB$ estimators
applied to the \smica\ full-mission CMB map.
The noise level for their minimum-variance combination (MV) is also shown.
The fiducial \lcdm\ theory power spectrum $C_{\elp}^{\phi\phi,{\rm fid}}$ used in our Monte Carlo simulations is plotted as the black solid line.
\label{fig:clpp_nlev}
}
\end{figure}

We reconstruct the CMB lensing potential using quadratic estimators that exploit the statistical anisotropy induced by lensing, following \cite{Okamoto:2003zw}.
Neglecting the lensing of primordial $B$-modes there are five possible estimators, denoted by
$\hat{\phi}^{TT}$,
$\hat{\phi}^{TE}$,
$\hat{\phi}^{EE}$,
$\hat{\phi}^{EB}$, and
$\hat{\phi}^{TB}$, which are based on various correlations of the CMB temperature ($T$) and polarization ($E$ and $B$).
In addition, we can form a minimum-variance estimator that combines all five estimators, which we denote as $\hat{\phi}^{\rm MV}$.
In Fig.~\ref{fig:clpp_nlev} we plot the lens reconstruction noise levels for these estimators.
The most powerful estimator is $TT$, although the $TE$ and $EE$ estimators are also useful on large angular scales.

The quadratic lensing estimators take inverse-variance filtered CMB multipoles as input.
We obtain these using a filter that masks the Galaxy and point sources,
and also bandpass filters the data in harmonic space to $100\!\le\!\elt\!\le\!2048$.
For our baseline analysis, we start with the masks used in the analysis of \cite{planck2013-p12},
with a Galaxy mask that removes $30.2\,\%$ of the sky and a point-source mask that removes an additional $0.7\,\%$ of the sky.
\refchange{This removes the brightest Sunyaev-Zel'dovich
  (SZ) clusters as described in~\citet{planck2013-p12}; contamination
  of the lensing reconstruction from residual unresolved SZ is
  expected to be at the percent level for
  \planck~\citep{vanEngelen:2013rla}, even without component
  separation.}\footnote{\refchange{We mask clusters with $S/N \geq 5$ in the
    \planck\ cluster catalogue that accompanied the 2013 release
    (PSZ1;~\citealt{planck2013-p05a}). The mass limit is redshift
    dependent, with a high-redshift limit
    at 80\,\% completeness of $M_{500} \approx 6\times 10^{14}\,
    \Msolar$. If we conservatively adopt this mass limit (we
    certainly mask low-redshift clusters to lower masses), the results
    of~\citet{vanEngelen:2013rla} suggest that the SZ trispectrum at
    143\,GHz gives a positive bias of at most a few percent in the
    reconstructed lensing
    power spectrum for multipoles $L<2000$, while the lensing-SZ-SZ
    bispectrum gives a negative bias of similar size. The SZ signal is
    much smaller in the 217\,GHz temperature data, which is around the
    null of the (thermal) SZ effect.}}
This residual signal, and that from unresolved point-sources, is
further reduced by a correction that we make to the power
spectrum of the reconstructed lensing potential (see Appendix~\ref{sect:pipeline:clpp}).
Finally, we apply \smica-specific temperature and polarization masks described in
\citet{planck2014-a11}.
Combining all three sets of masks leaves a total of $67.3\,\%$ of the sky for analysis.

We estimate the power spectrum of the lensing potential $C_L^{\phi\phi}$ using the auto- and cross-spectra of the quadratic lensing estimators.\footnote{In this paper, we use multipole indices $LM$ for the lens reconstruction, reserving multipole indices $\ell m$ for the CMB fields.}
These spectra probe the 4-point function of the lensed CMB, specifically the connected (trispectrum) part of the 4-point function that is sourced by lensing.
They also contain contributions from the disconnected part of the 4-point function (which is non-zero even in the absence of lensing effects). We estimate this contribution and subtract it, as well as several other smaller bias terms, obtaining an estimate $\hat{C}_L^{\phi\phi}$ of the power spectrum.
This procedure is discussed in Appendix~\ref{sect:pipeline:clpp}.
For cosmological parameter constraints, we use a Gaussian log-likelihood in bandpowers of the estimated lensing power spectrum, given by
\be
-2 \log {\cal L}_{\phi} = \bin_i^{\elp}
( \hat{C}_{\elp}^{\phi\phi} - C_{\elp}^{\phi\phi, {\rm th}} )
\left[ \Sigma^{-1} \right]^{ij} \bin_j^{\elp'}
( \hat{C}_{\elp'}^{\phi\phi} - C_{\elp'}^{\phi\phi, {\rm th}} ).
\label{eqn:likelihood}
\ee
Here, bins are indexed by $i$ and $j$; $\bin^{L}_i$ is the bandpower binning function for the $i$th bin,
and $\Sigma$ is a covariance matrix for the bin estimates.
Paired upper/lower indices are summed over. The
$C_{\elp}^{\phi\phi, {\rm th}}$ is the theoretical expectation value of the estimated $\hat{C}_{\elp}^{\phi\phi}$ for the set of cosmological and nuisance parameters under consideration. This generally differs from the theory spectrum $C_{\elp}^{\phi\phi}$ at the same cosmological parameters due to the way that our power spectrum estimates are normalized, and corrected for additional trispectrum couplings, with a fiducial model.
Both the binning function and $C_{\elp}^{\phi\phi, {\rm th}}$ are discussed in Appendix~\ref{sect:pipeline:likelihood}.
The binning function is chosen to have unit response to a fiducial theory spectrum
$C_{\elp}^{\phi\phi,{\rm fid}}$, and so we denote
\be
\hat{A}^{\phi}_i = \bin^{L}_i \hat{C}_L^{\phi\phi}
\label{eqn:binamplitude}
\ee
as the amplitude of the power spectrum for a particular bin relative to the fiducial
expectation
(with $\hat{A}=1$ for $\hat{C}_{\elp}^{\phi\phi} = C_{\elp}^{\phi\phi, {\rm fid}}$).

To characterize the variance of our lensing potential estimates, as well as to estimate several bias terms, we use simulated {\it Planck} maps.
These are based on the Full Focal Plane 8 (FFP8) Monte Carlo simulation set described in \citet{planck2014-a14}.
As discussed there,
the {\it Planck} maps were effectively renormalized by approximately $2$--$3$\% in power in the time between the generation of FFP8 and the final \planck\ full-mission maps.
To account for this, we rescale the CMB component of the simulations by a factor of $1.0134$ before analysis.
The FFP8 simulations do not include contributions from residual foregrounds (Galactic dust, as well as unmasked point sources), and also underestimate the noise power spectra by several percent at high-$\ell$.
We account for this missing power simply by adding coloured Gaussian noise to the simulations to make their $TT$, $EE$, and $BB$ power spectra agree with the data.
This approach implicitly assumes that any non-Gaussianity of these residual components does not couple significantly to our lensing estimates.
We perform consistency tests in Sect.~\ref{sec:consistency} to check the validity of these assumptions.

Throughout this paper we use a spatially-flat fiducial cosmology with
baryon density given by
$\omega_{\rm b} = \Omega_{\rm b}h^2 = 0.0222$,
cold dark matter density
$\omega_{\rm c} =\Omega_{\rm c}h^2 = 0.1203$,
neutrino energy density
$\omega_{\nu} = \Omega_{\nu}h^2 = 0.00064$ (corresponding to two massless neutrinos and one massive with mass $0.06\,{\rm eV}$),
Hubble constant $H_0 = 100 h \,\mathrm{km}\,\mathrm{s}^{-1}\,\mathrm{Mpc}^{-1}$ with $h=0.6712$,
spectral index of the power spectrum of the primordial curvature perturbation
$n_{\rm s} = 0.96$, amplitude of the primordial power spectrum (at $k=0.05\,\mathrm{Mpc}^{-1}$) $A_{\rm s} = 2.09\times 10^{-9}$, and Thomson optical depth through reionization $\tau=0.065$.
These cosmological parameters also form the basis for the FFP8 Monte Carlo simulation set.
In addition to rescaling the FFP8 maps as already discussed, we have also adjusted the power spectra of the fiducial model by rescaling the CMB temperature and polarization spectra by a factor of $1.0134^2$, and the temperature-lensing cross-correlation $C_L^{T\phi}$ by 1.0134. We have not applied any scaling to the fiducial lensing power spectrum.
Our reconstruction methodology (in particular, the renormalization corrections\refchange{, addition of foreground power}, and realization-dependent bias corrections that we apply, discussed in Appendix~\ref{sect:pipeline:likelihood}) renders the cosmological interpretation of our lensing estimates insensitive to errors in the fiducial model power spectra and simulations.

\section{Results}
\label{sec:results}
\label{sect:results}

In this section, we provide a summary of the first science results obtained with the minimum-variance lens reconstruction from the \planck\ full-mission data.  The lensing potential map is presented in Sect.~\ref{sec:results:lensing_potential}, and this is combined in Sect.~\ref{sect:lensing_bmode_power_spectrum} with the $E$-mode polarization measured by \planck\ to obtain a map of the expected $B$-mode polarization due to lensing. We further show that this is correlated with the $B$-modes measured by \planck\ at the expected level. In Sect.~\ref{sect:results:lensingisw}, we cross-correlate the reconstructed lensing potential with the large-angle temperature anisotropies to measure the $C_{\elp}^{T\phi}$ correlation sourced by the ISW effect. Finally, the power spectrum of the lensing potential is presented in Sect.~\ref{sect:lensing_potential_power_spectrum}.
We use the associated likelihood alone, and in combination with that constructed from the \planck\ temperature and polarization power spectra~\citep{planck2014-a13}, to constrain cosmological parameters in Sect.~\ref{sec:parameters}.
	
\subsection{Lensing potential}
\label{sec:results:lensing_potential}
\begin{figure}[!ht]
	\begin{center}
	\begin{overpic}[width=\columnwidth, trim=0 70 0 0, clip=True]{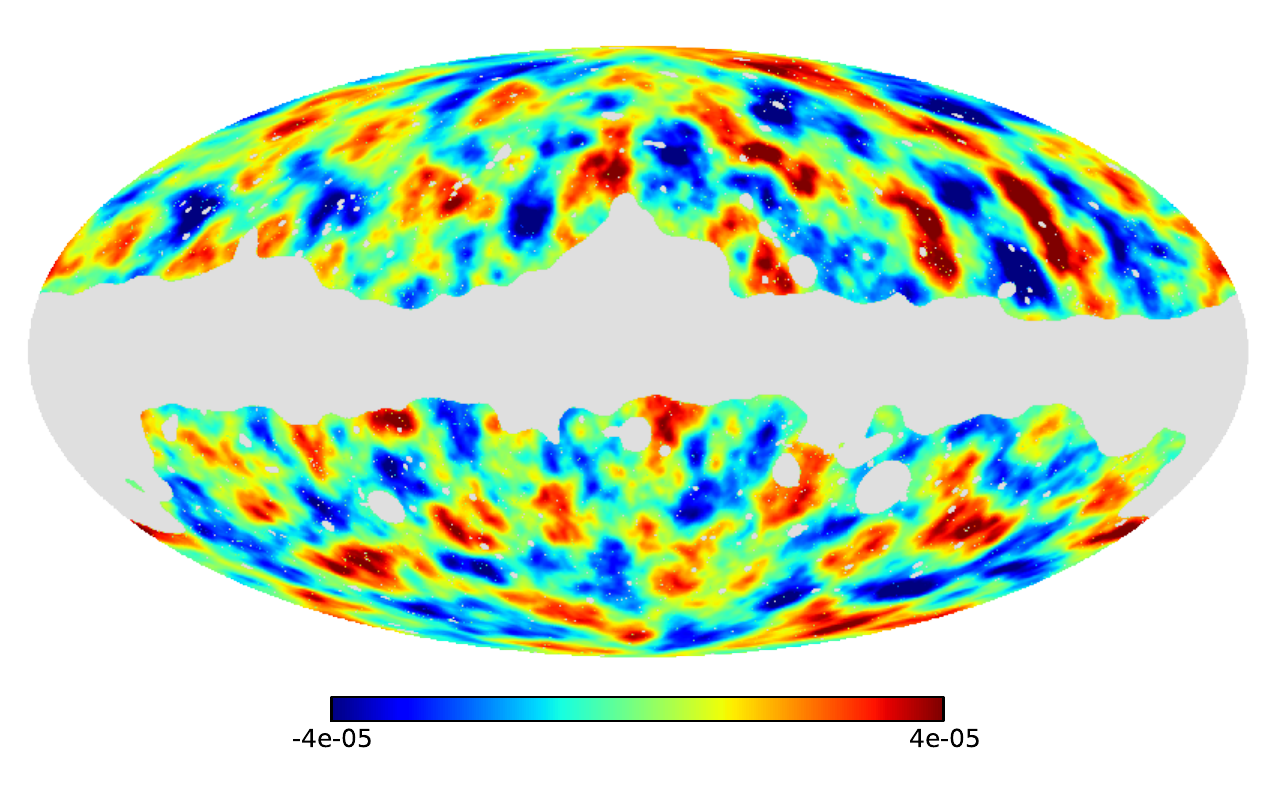}
	\put(78,1){\tiny { $\hat{\phi}^{\rm WF}$ (Data)}}
	\end{overpic}
	\vspace{-0.15in}
	\caption{
		\label{fig:phiwfdata}
		Lensing potential estimated from the \smica\ full-mission CMB maps using the MV estimator. The power spectrum of this map forms the basis of our lensing likelihood. The estimate has been Wiener filtered following Eq.~\eqref{eqn:phiwf}, and band-limited to \mbox{$8 \le L \le 2048$}.
	}
	\begin{overpic}[width=\columnwidth, trim=0 70 0 0, clip=True]{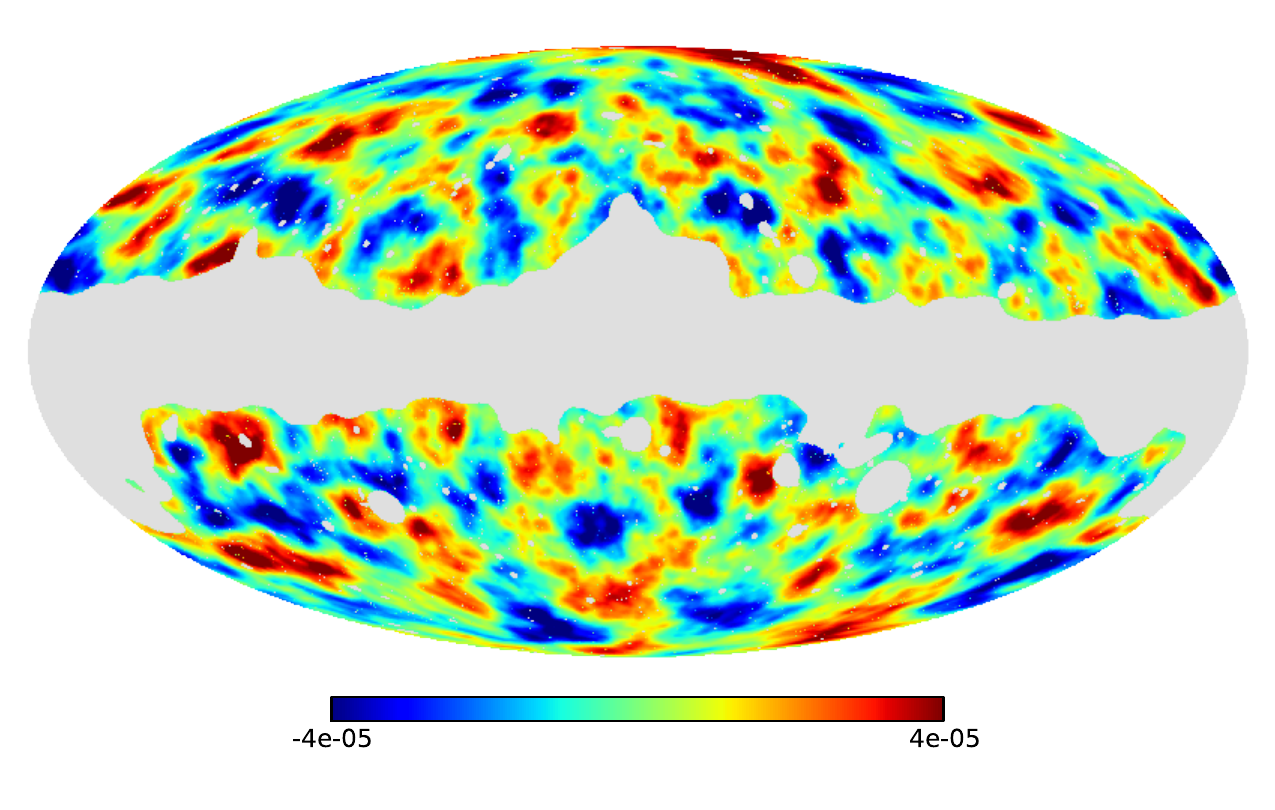}
	\put(78,1){\tiny { $\hat{\phi}^{\rm WF}$ (Sim.)}}
	\end{overpic}
	\begin{overpic}[width=\columnwidth, trim=0 70 0 0, clip=True]{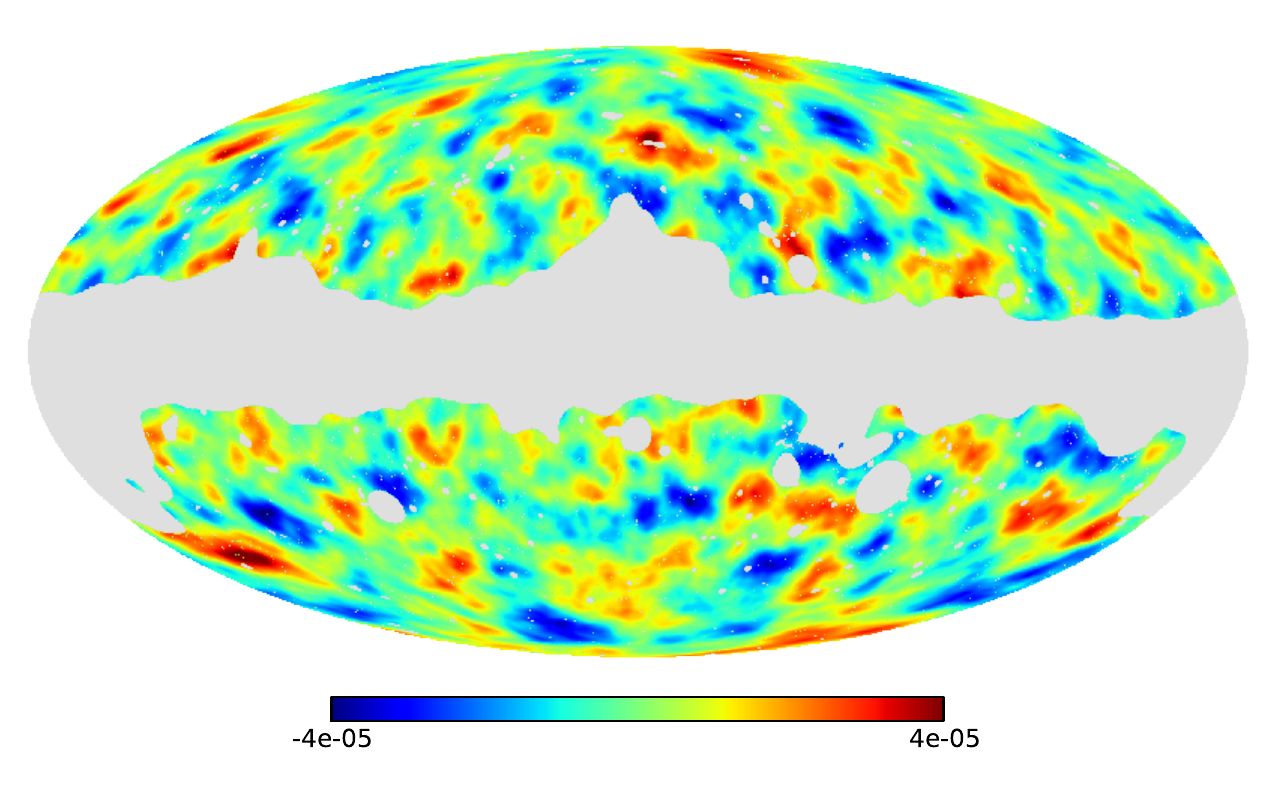}
	\put(78,1){\tiny {Input $\phi$ (Sim.)}}
	\end{overpic}
	\vspace{-0.15in}
	\caption{
		\label{fig:phiwfsim}
		Simulation of a Wiener-filtered MV lensing reconstruction (upper) and the input $\phi$ realization (lower), filtered in the same way as the MV lensing estimate.
		The reconstruction and input are clearly correlated, although the reconstruction has considerable additional power due to noise.
	}
	\end{center}
	\end{figure}

	In Fig.~\ref{fig:phiwfdata} we plot the Wiener-filtered minimum-variance lensing estimate, given by
	\be
	\hat{\phi}^{\rm WF}_{LM} = \frac{\clppfid}{\clppfid+ N_L^{\phi\phi}} \hat{\phi}^{\rm MV}_{LM},
	\label{eqn:phiwf}
	\ee
	where $\clppfid$ is the lensing potential power spectrum in our fiducial model
	and $N_L^{\phi\phi}$ is the noise power spectrum of the reconstruction.
	As we shall discuss in Sect.~\ref{sect:consistency:bias_hardened_estimators},
	the lensing potential estimate is unstable for $L < 8$,
	and so we have excluded those modes for all analyses in this paper, as well as in the MV lensing map.
		
	As a visual illustration of the signal-to-noise level in the lensing potential estimate,
	in Fig.~\ref{fig:phiwfsim} we plot a simulation of the MV reconstruction, as well as the input $\phi$ realization used.
	The reconstruction and input are clearly correlated, although the reconstruction has considerable additional power due to noise.
	As can be seen in Fig.~\ref{fig:clpp_nlev}, even the MV reconstruction only has $S/N \approx 1$ for a few modes around $L \approx 50$.

	The MV lensing estimate in Fig.~\ref{fig:phiwfdata} forms the basis for a public lensing map that we provide to the community \citep{planck2014-a01}.
	The raw lensing potential estimate has a very red power spectrum, with most of its power on large angular scales.
	This can cause leakage issues when cutting the map (for example to cross-correlate with an additional mass tracer over a small portion of the sky).
	The lensing convergence $\kappa$ defined by
	\be
	\kappa_{\elp M} = \frac{\elp(\elp+1)}{2} \phi_{LM} ,
	\ee
	has a much whiter power spectrum, particularly on large angular scales.
	The reconstruction noise on $\kappa$ is approximately white as well \citep{2012PhRvD..85d3016B}.
	For this reason, we provide a map of the estimated lensing convergence $\kappa$ rather than the lensing potential $\phi$.
	
\subsection{Lensing $B$-mode power spectrum}
	\label{sect:lensing_bmode_power_spectrum}
	The odd-parity $B$-mode component of the CMB polarization is of great importance for early-universe cosmology.
	At first order in perturbation theory it is not sourced by the scalar fluctuations that dominate the temperature and polarization anisotropies,
	and so the observation of primordial $B$-modes can be used as a uniquely powerful probe of tensor (gravitational wave) or vector perturbations in the early Universe.
	A detection of $B$-mode fluctuations on degree angular scales, where the signal from gravitational waves is expected to peak, has recently been reported at $150\, {\rm GHz}$ by the BICEP2 collaboration \citep{Ade:2014xna}. Following the joint analysis of BICEP2 and \textit{Keck Array} data (also at $150\, {\rm GHz}$) and the \planck\ polarization data, primarily at $353\, {\rm GHz}$~\citep{pb2015}, it is now understood that the $B$-mode signal detected by BICEP2 is dominated by Galactic dust emission. The joint analysis gives no statistically-significant evidence for primordial gravitational waves, and establishes a 95\,\% upper limit $r_{0.05}< 0.12$. This still represents an important milestone for $B$-mode measurements, since the direct constraint from the $B$-mode power spectrum is now as constraining as indirect, and model-dependent, constraints from the $TT$ spectrum~\citep{planck2014-a15}.

In addition to primordial sources, the effect of gravitational lensing also generates $B$-mode polarization.
	The displacement of lensing mixes $E$-mode polarization into $B$-mode as \citep{Smith:2008an}
	\be
	B^{\rm lens}_{\elt_B m_B} = (-1)^{m_B}\sum_{LM} \sum_{\elt_E m_E} \threej{\elt_E}{\elt_B}{\elp}{m_E}{-m_B}{M} W^{\phi_{EB}}_{\elt_E \elt_B L} E_{\elt_E m_E} \phi_{LM},
	\label{eqn:blens}
	\ee
	where $W^{\phi_{EB}}_{\elt_E \elt_B L}$ is a weight function and the bracketed term is a Wigner-$3j$ symbol.
	On scales $\elt_B \lea 1000$ the lensing $B$-mode power spectrum resembles that of white noise, with a level of about \mbox{$5\,\muKarcmin$}.
	This lensing power acts as a potential source of confusion for the measurement of primordial $B$-modes, which can be estimated and ultimately removed in a process of delensing.
	Given an estimate for the lensing potential $\phi$ and the $E$-mode polarization measured by {\it Planck} we can synthesize a lensed $B$-mode map for this purpose using Eq.~\eqref{eqn:blens}.
	The $5\, \muKarcmin$ level of the lensing $B$-mode power spectrum is an order of magnitude lower than the {\it Planck} 2015 noise levels,
	and so delensing does not significantly improve our $B$-mode measurements;
	however, the cross-correlation of the lensing $B$-mode template with the observed $B$-mode sky provides a useful check on the ability of {\it Planck} to measure this known source of $B$-modes.
	
\begin{figure}[!ht]
	\centering
	\includegraphics[width=\columnwidth]{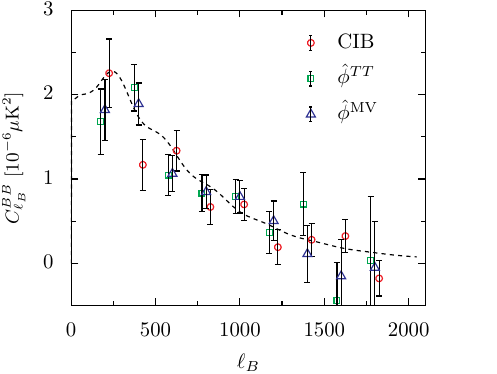}
	\vspace{-0.2in}
	\caption{
	Measurement of the lensing $B$-mode power spectrum, using cross-correlation with estimates of the lensing potential as discussed in Sect.~\ref{sect:lensing_bmode_power_spectrum}.
	The theoretical lensing $B$-mode power spectrum, for the parameters of the fiducial cosmological model of Sect.~\ref{sec:data_and_methodology}, is plotted as dashed black. Green squares and blue triangles are results using the $TT$ and MV $\phi$ reconstructions, respectively, to construct the lens-induced $B$-mode template, while red circles use the CIB (from the \planck\ $545\, {\rm GHz}$ channel) to construct a proxy for $\phi$.
	Lensing $B$-mode power is detected with the expected scale dependence and amplitude at a significance level of
	approximately $\bmodeapproxsignif$.
	\label{fig:ebxp}
	}
	\end{figure}

	We show the results of such a cross-correlation in Fig.~\ref{fig:ebxp},
	finding good agreement with the expected lensing $B$-mode power spectrum.
	In addition to our fiducial MV lensing potential estimate, which uses both temperature and polarization data,
	we have also estimated the lensing $B$-mode power spectrum using the $TT$-only lensing estimator to measure $\phi$,
	as well as the CIB fluctuations measured by the $545\, {\rm GHz}$ {\it Planck} channel.\footnote{To calculate the scaling between the CIB map and the lensing potential $\phi$, we model the CIB using the simple model of \cite{Hall:2009rv} for its redshift and frequency dependence. Further details are given in Appendix~\ref{app:cib_model}.}
	We see good agreement in all cases with the expected power;
	constraining the overall amplitude of the lensing $B$-mode power spectrum $\hat{A}^{B}$
	(relative to the predicted spectrum in our fiducial model) for a large bin from
	$8 \le \elt_B \le 2048$
	we measure amplitudes of
	\begin{align}
	\hat{A}^{B}_{8 \rightarrow 2048} &= 0.93 \pm 0.08\quad (\phi^{{\rm MV}}) , \nonumber \\
	\hat{A}^{B}_{8 \rightarrow 2048} &= 0.95 \pm 0.09\quad ({\phi^{TT}}) ,\nonumber \\
	\hat{A}^{B}_{8 \rightarrow 2048} &= 0.93 \pm 0.10\quad ({\rm CIB}) \nonumber
	\end{align}
	for the three estimates, each corresponding to a roughly
	$\bmodeapproxsignif$
	detection of lensing $B$-mode power in the {\it Planck} data.
	The shape of the cross-correlation is also in good agreement with expectation.
	Taking the bins in Fig.~\ref{fig:ebxp} as independent, forming a $\chi^2$ relative to the theory model and
	comparing to the distribution from simulations we obtain probability-to-exceed (PTE) values of
	$48\%$, $71\%$, and $78\%$
	using the MV, $TT$, and CIB lensing estimates, respectively.

\subsection{Lensing-ISW bispectrum}
\label{sect:results:lensingisw}

As photons travel towards us from the last scattering surface, they are not only deflected by gravitational lensing,
they also receive net red/blueshifts from gravitational potentials that evolve if they are crossed at late times.
This phenomenon, known as the integrated Sachs-Wolfe (ISW) effect, is believed to generate
anisotropies in the observed CMB temperature on large \mbox{($\elt \lesssim 100$)} angular scales.
It is of particular interest because the decay of gravitational potentials, which produces the ISW effect, does not occur during matter domination,
but only at redshifts $z \lesssim 2$ when dark energy becomes dynamically important.
The ISW effect can be detected statistically by cross-correlating the observed temperature
anisotropies with a tracer of structure at these redshifts.
Here we use the lensing potential, which is well-matched to the ISW effect \citep{Hu:2001fb}.
Current \planck\ results incorporating additional external tracers of large-scale structure are summarized in
\cite{planck2014-a26}.

\begin{figure}[!ht]
	\begin{center}
	\includegraphics[width=\columnwidth]{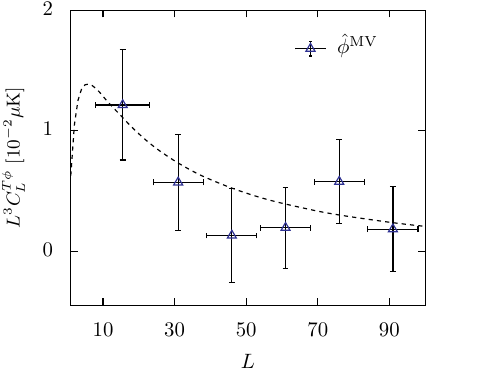}
	\vspace{-0.2in}
	\caption{
	Lensing-ISW bispectrum on large angular scales.
	The cross-spectrum between the MV lensing potential estimate and the temperature anisotropy is plotted for bins of width $\Delta \elp = 15$, covering the multipole range $\elp = 8$--$98$. The dashed line shows the predicted cross-spectrum in the fiducial model.
	The lensing-ISW bispectrum is detected at just over $\lensingiswapproxsignif$ significance.
	\label{fig:pub_cltp_mv}
	}
	\end{center}
	\end{figure}

In Fig.~\ref{fig:pub_cltp_mv} we plot the cross-correlation $C_{\elp}^{T\phi}$ between the MV lens reconstruction and the CMB temperature.
This cross-correlation probes the bispectrum, or three-point correlation function of the CMB, which is due to the correlation of the lensing and ISW effects.
The measurement is noisy, due to a combination of noise in the lens reconstruction and cosmic variance in the temperature (which ultimately limits the detection of  the lensing-ISW bispectrum to about $9\,\sigma$; \citealt{2011JCAP...03..018L}).

To determine the overall detection significance for the cross-correlation, we use the minimum-variance bispectrum estimator
\be
\hat{A}^{T\phi} = \frac{1}{N^{T\phi}}
\sum_{\elp M}
\frac{\cltpfid}{f_{\rm sky}}
\frac{ \hat{\phi}_{\elp M} }{(\clppfid + N_{\elp}^{\phi\phi}) }
\frac{ {T}^*_{\elp M} }{ \clttfid },
\ee
where $N^{T\phi}$ is a normalization determined from simulations.\footnote{
We find $N^{T\phi}$ is within
$4\%$
of the analytical expectation
\be
N^{T\phi} \approx \left[
\sum_{\elp} (2\elp+1) \left( \cltpfid \right)^2
\frac{ 1}{\clppfid + N_{\elp}^{\phi\phi} }
\frac{ 1 }{\clttfid}
\right]. \nonumber
\ee
}
For the MV lens reconstruction, using $8 \le \elp \le 100$ we measure an amplitude
\be
\hat{A}^{T\phi}_{8 \rightarrow 100} = 0.90 \pm 0.28 \quad \quad \mbox{(MV)}\, ,
\ee
which is consistent with the theoretical expectation of unity and non-zero at just over $\lensingiswsignif$.
Using the $TT$-only lensing estimate rather than the MV lensing estimate in the cross-correlation, we obtain
\be
\hat{A}^{T\phi}_{8 \rightarrow 100} = 0.68 \pm 0.32 \quad \quad \mbox{($TT$)} \, .
\ee
Using simulations, we measure an rms difference between the $TT$ and MV bispectrum amplitudes of 0.18
(roughly equal to the quadrature difference of their error bars, which is
\mbox{$\sqrt{0.32^2 -0.28^2} = 0.15$}). Therefore the difference of amplitudes,
$\Delta \hat{A}^{T\phi}_{8 \rightarrow 100} = 0.22$, is compatible with the expected scatter.

In \cite{planck2013-p12}, using the $TT$ lensing estimator on the multipole range $10 \le L \le 100$
we measured a somewhat higher value for the lensing-ISW bispectrum amplitude of
$\hat{A}^{T\phi,\ 2013}_{10 \rightarrow 100} = 0.85 \pm 0.35$.\footnote{The amplitude quoted in \cite{planck2013-p12} is actually
$\hat{A}^{T\phi,\ 2013}_{10 \rightarrow 100} = 0.78 \pm 0.32$, however this is measured with respect to
a slightly different fiducial cosmology than the one used here.
Those measurements have been renormalized to the fiducial model used for this paper with a factor of $1.09$.}
We expect the difference with respect to the 2015 $TT$ measurement to have standard deviation of approximately
\mbox{$\sqrt{0.35^2 - 0.32^2} = 0.14$},
and so the observed difference of
$0.85-0.68 = 0.17$
is reasonable.

\subsection{Lensing potential power spectrum}
	\label{sect:lensing_potential_power_spectrum}
\begin{figure*}[!ht]
	\begin{center}
	\includegraphics[width=\textwidth]{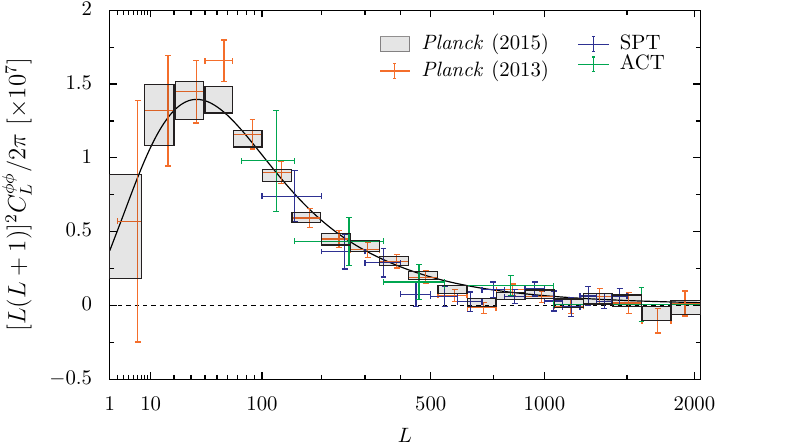}
	\vspace{-0.2in}
	\caption{
	 \planck\ 2015 full-mission MV lensing potential power spectrum measurement, as well as earlier measurements using
	the \planck\ 2013 nominal-mission temperature data \citep{planck2013-p12},
	the South Pole Telescope (SPT, \citealt{vanEngelen:2012va}),
	and the Atacama Cosmology Telescope (ACT, \citealt{Das:2013zf}).
	The fiducial $\Lambda$CDM theory power spectrum based on the parameters given in Sect.~\ref{sec:data_and_methodology} is plotted as the black solid line.
	\label{fig:clpp_mv}
	}
	\end{center}
	\end{figure*}

	In Fig.~\ref{fig:clpp_mv} we plot our estimate of the lensing potential power spectrum obtained from the MV reconstruction, as well as several earlier measurements. We see good agreement with the shape in the fiducial model, as well as earlier measurements (detailed comparisons with the 2013 spectrum are given later in this section). In Sect.~\ref{sec:consistency} we perform a suite of internal consistency and null tests to check the robustness of our lensing spectrum to different analysis and data choices.
	
	We estimate the lensing potential power spectrum in bandpowers for two sets of bins:
	a `conservative' set of eight uniformly-spaced bins with $\Delta L = 45$ in the range \mbox{$40 \le L \le 400$}; and
	an `aggressive' set of 18 bins  that are uniformly spaced in $L^{0.6}$ over the multipole range \mbox{$8 \le L \le 2048$}.
	The conservative bins cover a multipole range where the estimator signal-to-noise is greatest.
	They were used for the \planck\ 2013 lensing likelihood described in \cite{planck2013-p12}.
	The aggressive bins provide good sensitivity to the shape of the lensing power on both large and small scales, however they are more easily biased by errors in the mean-field corrections (which are large at $L<40$) and the disconnected noise bias corrections (at $L>400$).
Results for the bandpower amplitudes $\hat{A}_i^\phi$,
defined in Eq.~\eqref{eqn:binamplitude}, are given in Table~\ref{table:power_spectrum_bandpowers} for both sets of multipole bins.

% In earlier tests, the aggressive multipole range frequently failed the internal consistency tests that are presented in Sect.~\ref{sec:consistency};
%however for the final full-mission maps and FFP8 simulations we find that they pass nearly all tests at an acceptable level.
\refchange{Nearly all of the internal consistency tests that we
  present in Sect.~\ref{sec:consistency} are passed at an acceptable level.}
There is, however, mild evidence for a correlated feature in the curl-mode null-test, centred around $\elp\approx 500$. The $\elp$ range covered by this feature includes $638 \leq \elp \leq 762$, for which the \refchange{(gradient)} lensing reconstruction bandpower is $3.6\,\sigma$ low compared to the predicted power of the fiducial model. In tests of the sensitivity of parameter constraints to lensing multipole range, described later in Sect.~\ref{subsubsec:Lrange},
we find shifts in some parameters of around $1\,\sigma$ in going from the conservative to aggressive range, with negligible improvement in the parameter uncertainty. Around half of these shifts come from the outlier noted above. For this reason, we adopt the conservative multipole range as our baseline here, and in other \planck\ 2015 papers, when quoting constraints on cosmological parameters. However, where we quote constraints on amplitude parameters in this paper, we generally give these for both the aggressive and conservative binning.
	The aggressive bins are also used for all of the $C_L^{\phi\phi}$ bandpower plots in this paper.
	
	Estimating an overall lensing amplitude (following Eq.~\ref{eqn:binamplitude}) relative to our fiducial theoretical model, for a single bin over both the aggressive and conservative multipole ranges we find
    \begin{subequations}
	\begin{eqnarray}
	\qquad \qquad \qquad \qquad \hat{A}^{\phi, \,{\rm MV}}_{40 \rightarrow 400} = 0.987 \pm 0.025,  \label{eq:AhatConservative}  \\
	\qquad \qquad \qquad \qquad \hat{A}^{\phi, \,{\rm MV}}_{8 \rightarrow 2048} = 0.983 \pm 0.025.  \label{eq:AhatAggressive}
	\end{eqnarray}
\end{subequations}
	These measurements of the amplitude of the lensing power spectrum both have precision of 2.5\,\%, and are non-zero at $\detectionsignif$.
	Given the measured amplitude, it is clear that our overall lensing amplitude estimate is consistent with our fiducial $\Lambda$CDM model (which has $A =1$).
	The shape of our measurement is also in reasonable agreement with the fiducial model.
	Marginalizing over an overall amplitude parameter, for the
	$\nelpbinsaggressive$
	bins plotted in Fig.~\ref{fig:clpp_mv} we obtain a
	$\chi^2$ with respect to our fiducial model of $28$ (with $N_{\rm dof} = 18$),
	with a corresponding PTE of
	$6\,\%$.
	A large portion of this $\chi^2$ is driven by the outlier bandpower with $638\! \le \! L \! \le \! 762$.
	Removing this bandpower gives a $\chi^2$ of $16.5$ ($N_{\rm dof} = 17$), with a PTE of $49\,\%$.
	For the more conservative multipole range $\conservativeLrange$,
	using eight linear bins we obtain $\chi^2$ of $8.9$ ($N_{\rm dof} = 7$), with a PTE of $26\,\%$.
	
	Finally, we note that the lensing bandpowers measured here are in good agreement with the \planck\ 2013 results~\citep{planck2013-p12}.
	The only clear visual difference between the two measurements in Fig.~\ref{fig:clpp_mv} is for the $\elp = 40$--$65$ bin, where the 2013 bandpower is significantly higher than the current measurement.
	As previously discussed in \cite{planck2013-p12}, this bin amplitude was sensitive to the choice of foreground cleaning, and decreased when using a component-separated map for the reconstruction (as is done here), and so this difference is expected.
	To compare measurements in further detail, we calculate a $\chi^2$ for the difference between our 2013 and 2015 MV lensing bandpower estimates assuming a diagonal covariance as
	\be
	\chi^2_{2013-2015} =
	\sum_{\elp_b}
	\frac{
	\left( \left. \hat{C}_{\elp_b}^{\phi\phi} \right|_{\rm 2013} -  \left. \hat{C}_{\elp_b}^{\phi\phi} \right|_{\rm 2015} \right)^2
	}{
	{\rm Var} \left( \left. \hat{C}_{\elp_b}^{\phi\phi} \right|_{\rm 2013} \right) - {\rm Var} \left( \left. \hat{C}_{\elp_b}^{\phi\phi} \right|_{\rm 2015} \right)
	},
	\ee
	where $\elp_b$ indexes the bandpower bins. We approximate the variance of the differences of the bandpowers as the difference of the variances in forming this $\chi^2$.
	Using the conservative bins for \mbox{$40 \le L \le 400$}, which were used for the 2013 lensing likelihood, we obtain a value of
	\mbox{$\chi^2_{2013-2015} = 10.8$} ($N_{\rm dof} = 8$),
	with a corresponding PTE of $22\,\%$.
	The 2013 result is temperature only; if we compare it directly to the 2015 $TT$ lensing bandpowers, we obtain a value of
	$\chi^2 = 7.2$ ($N_{\rm dof} = 8$, ${\rm PTE}\!=\!52\%$).
	We note that some care should be taken when comparing the measured lensing amplitudes between 2013 and 2015, because the fiducial spectra against which they are measured differ significantly.
	In 2013, for example, we measured
	\mbox{$\hat{A}^{\phi, \, \text{2013}}_{40 \rightarrow 400} = 0.943 \pm 0.040$}.
	However, this was measured with respect to a fiducial $C_L^{\phi\phi}$ that is  between $6\,\%$ and $7\,\%$ higher than the one used here in this multipole range.
	Renormalizing the 2013 measurements to determine the amplitude for the 2015 fiducial cosmology, we obtain
	\mbox{$1.005 \pm 0.043$}, which can be compared directly to the measurement of
	\mbox{$\hat{A}^{\phi, \, \text{MV}}_{40 \rightarrow 400} = 0.987 \pm 0.025$} for 2015.
	It is interesting to note that there is a $2\,\%$ difference in power between the \planck\ 2013 and 2015 maps from HFI due to a change in the calibration, as well as an improved beam model \citep{planck2014-a08}. This shift does not couple to the lensing reconstruction; lensing is a geometric effect, and so an overall change in the calibration does not affect the lens reconstruction procedure, provided that a consistent power spectrum is used to analyse the maps, as is done here.
	
	\begin{table}[!tbh]
	\begingroup
	\newdimen\tblskip \tblskip=5pt
	\caption{
	Lensing potential power spectrum estimates from the MV lens reconstruction.
	The $\hat{A}^{\phi}$ values are dimensionless, with
	\mbox{$A^{\phi}=1$}
	for a measured spectrum equal to the lensing potential power spectrum of our fiducial cosmology (described in Sect.~\ref{sec:data_and_methodology}). The final column lists bandpower estimates, averaged within each bin.
	\label{table:power_spectrum_bandpowers}
	}
	\nointerlineskip
	\footnotesize
	\setbox\tablebox=\vbox{
	 \newdimen\digitwidth
	 \setbox0=\hbox{\rm 0}
	  \digitwidth=\wd0
	  \catcode`*=\active
	  \def*{\kern\digitwidth}
	  \newdimen\signwidth
	  \setbox0=\hbox{+}
	  \signwidth=\wd0
	  \catcode`!=\active
	  \def!{\kern\signwidth}
	\halign{
	\hfil$#$\hfil\tabskip=2.0em&
	  \hfil$#$\hfil\hspace{0.5cm}&
	  \hfil$#$\hfil&
	  \hfil$#$\hfil\tabskip=0pt\cr
	\noalign{\doubleline}
	\multispan4\hfil Lensing power spectrum bandpowers \hfil\cr
	\noalign{\vskip -4pt}
	\multispan4\hrulefill \cr
	\noalign{\vskip 2pt}
	L_{\rm min}&L_{\rm max}&\hat{A}^{\phi}&  10^7 [\elp(\elp+1)]^2 C_{\elp}^{\phi\phi}/2\pi  \cr
	\noalign{\vskip -4pt}
	\multispan4\hrulefill \cr
\noalign{\vskip 2pt}
\multispan4{Conservative multipole range ($\conservativeLrange$)} \hfill\cr
\noalign{\vskip 2pt}
\phantom{0}\phantom{0}40&\phantom{0}\phantom{0}84&\phantom{-}1.01\pm0.05&\phantom{-}1.28\pm0.06\cr
\phantom{0}\phantom{0}85&\phantom{0}129&\phantom{-}1.02\pm0.04&\phantom{-}1.00\pm0.04\cr
\phantom{0}130&\phantom{0}174&\phantom{-}0.93\pm0.05&\phantom{-}0.70\pm0.04\cr
\phantom{0}175&\phantom{0}219&\phantom{-}0.89\pm0.07&\phantom{-}0.52\pm0.04\cr
\phantom{0}220&\phantom{0}264&\phantom{-}0.86\pm0.10&\phantom{-}0.41\pm0.05\cr
\phantom{0}265&\phantom{0}309&\phantom{-}1.03\pm0.11&\phantom{-}0.41\pm0.05\cr
\phantom{0}310&\phantom{0}354&\phantom{-}1.24\pm0.13&\phantom{-}0.41\pm0.04\cr
\phantom{0}355&\phantom{0}400&\phantom{-}0.97\pm0.16&\phantom{-}0.27\pm0.05\cr	
\noalign{\vskip -2pt}
\multispan4\hrulefill \cr
\noalign{\vskip 2pt}
\multispan4{Aggressive multipole range ($\aggressiveLrange$)}\hfill\cr
\noalign{\vskip 2pt}
\phantom{0}\phantom{0}\phantom{0}8&\phantom{0}\phantom{0}20&\phantom{-}1.09\pm0.17&\phantom{-}1.29\pm0.21\cr
\phantom{0}\phantom{0}21&\phantom{0}\phantom{0}39&\phantom{-}1.01\pm0.09&\phantom{-}1.39\pm0.13\cr
\phantom{0}\phantom{0}40&\phantom{0}\phantom{0}65&\phantom{-}1.05\pm0.07&\phantom{-}1.40\pm0.09\cr
\phantom{0}\phantom{0}66&\phantom{0}100&\phantom{-}1.00\pm0.05&\phantom{-}1.13\pm0.06\cr
\phantom{0}101&\phantom{0}144&\phantom{-}0.99\pm0.04&\phantom{-}0.88\pm0.04\cr
\phantom{0}145&\phantom{0}198&\phantom{-}0.89\pm0.05&\phantom{-}0.60\pm0.04\cr
\phantom{0}199&\phantom{0}263&\phantom{-}0.90\pm0.08&\phantom{-}0.45\pm0.04\cr
\phantom{0}264&\phantom{0}338&\phantom{-}1.08\pm0.10&\phantom{-}0.40\pm0.04\cr
\phantom{0}339&\phantom{0}425&\phantom{-}1.08\pm0.11&\phantom{-}0.30\pm0.03\cr
\phantom{0}426&\phantom{0}525&\phantom{-}0.98\pm0.13&\phantom{-}0.20\pm0.03\cr
\phantom{0}526&\phantom{0}637&\phantom{-}0.69\pm0.17&\phantom{-}0.11\pm0.03\cr
\phantom{0}638&\phantom{0}762&\phantom{-}0.18\pm0.23&\phantom{-}0.02\pm0.03\cr
\phantom{0}763&\phantom{0}901&\phantom{-}0.73\pm0.27&\phantom{-}0.07\pm0.02\cr
\phantom{0}902&1054&\phantom{-}1.11\pm0.40&\phantom{-}0.08\pm0.03\cr
1055&1221&\phantom{-}0.27\pm0.55&\phantom{-}0.01\pm0.03\cr
1222&1404&\phantom{-}1.06\pm0.80&\phantom{-}0.05\pm0.03\cr
1405&1602&\phantom{-}0.91\pm1.18&\phantom{-}0.03\pm0.04\cr
1603&1816&-2.03\pm1.75&-0.06\pm0.05\cr
1817&2048&-0.54\pm2.16&-0.01\pm0.05\cr
	\noalign{\vskip 4pt\hrule\vskip 3pt}}}
	\endPlancktable
	\endgroup
	\end{table}
	
\subsection{Likelihoods and cosmological parameter constraints}
\label{sec:parameters}
In this section we discuss cosmological implications of the {\it Planck} lensing potential power spectrum estimate.
We use a Gaussian likelihood in the measured bandpowers, with corrections to account for errors in the fiducial CMB power spectra that are used to normalize and debias the lensing estimates.
This procedure is described in more detail in Appendix~\ref{sect:pipeline:likelihood}.

In the following subsections, we discuss the information that can be gleaned from the lensing potential power spectrum alone, or in conjunction with the {\it Planck} 2015 CMB power spectrum likelihood. Some of the results in this section summarize the parts of \cite{planck2014-a15} that are directly related to lensing.

\subsubsection{Constraints from CMB lensing alone}
\label{sec:lensonly}

	\begin{figure}[!ht]
	\centering
	\includegraphics[width=\columnwidth]{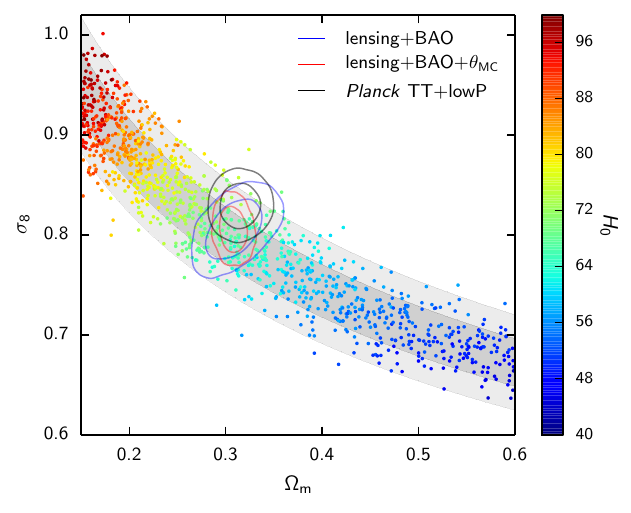}
	\vspace{-0.2in}
	\caption{
    Parameter constraints from CMB lensing alone in the \lcdm\ model (samples, colour-coded by the value of the Hubble constant) using the minimal priors described in the text; grey bands give the corresponding $1\,\sigma$ and $2\,\sigma$ constraints using the approximate fit of Eq.~\eqref{eq:LensOnlyDegen}.
    Solid coloured contours show $68\,\%$ and $95\,\%$ constraints when additional information is included: BAO data from SDSS and 6DF (\citealt{Anderson:2013zyy}; \citealt{Ross:2014qpa}; \citealt{Beutler:2011hx}; blue), and the same but fixing the CMB acoustic-scale parameter $\theta_{\rm MC}$ to a CMB power spectrum fit (red; $\theta_{\rm MC}=1.0408$). Solid black contours show the constraint from the \Planck\ CMB power spectra (\Planck\ temperature+low-$\ell$ polarization).
    	}
    \label{fig:lensonly}
	\end{figure}

The measurement of the lensing power spectrum is obtained from the 4-point function of the observed CMB anisotropies, and hence depends in general on both the lensing and CMB power spectra (see Appendix~\ref{sect:pipeline:likelihood}). However the latter are now well measured by \Planck, and there is only a weak dependence on cosmological parameters given the observed spectra  (mainly via different cosmological models changing foreground parameters, and hence the amplitude of underlying CMB power spectra). In this section we fix the CMB power spectra to a \lcdm\ \planckTT\ best fit (i.e. using $TT$ on all scales and low-$\ell$ polarization), and consider the conditional probability of different cosmological parameters given only the lensing reconstruction power spectrum.

To avoid marginalizing over very unrealistic values of poorly constrained parameters, we adopt several well-motivated priors when considering constraints from lensing alone.
 \begin{itemize}
 \item
The optical depth to reionization is fixed to $\tau=0.07$,
because lensing deflections are independent of reionization (and scattering and subsequent lensing from sources at reionization is negligible).
\item  The baryon density is given a Gaussian $1\,\sigma$  prior $\Omb h^2= 0.0223\pm 0.0009$, as measured independently from big bang nucleosynthesis models combined with quasar absorption line observations~\citep{Pettini:2012ph}.
\item The scalar spectral index is given a broad prior $\ns = 0.96 \pm 0.02$; results are only weakly sensitive to this choice, within plausible bounds.
\item A top-hat prior is used for the reduced Hubble constant, \mbox{$0.4<h<1$}. This limits the extent of the parameter degeneracy, but does not affect the results over the region of interest for joint constraints.
\end{itemize}

In addition to the priors above, we adopt the same sampling priors and methodology as \cite{planck2014-a15},\footnote{For example, we split the neutrino component into approximately two massless neutrinos and one with $\mnu=0.06\, \rm{eV}$, by default.}
using \COSMOMC\ and \CAMB\ for sampling and theoretical predictions~\citep{Lewis:2002ah,Lewis:1999bs}.
In the \lcdm\ model, as well as $\Omega_{\rm b}h^2$ and $n_{\rm s}$, we sample $\As$, $\Omega_{\rm c}h^2$, and the (approximate) acoustic-scale parameter $\theta_{\rm MC}$. Alternatively, we can think of our lensing-only results as constraining the sub-space of $\Omega_{\rm m}$, $H_0$, and $\sigma_8$. Figure~\ref{fig:lensonly} shows the corresponding constraints from CMB lensing, along with tighter constraints from combining with additional external baryon acoustic oscillation (BAO) data, compared to the constraints from the \Planck\ CMB power spectra. The contours overlap in a region of acceptable Hubble constant values, and hence are compatible. To show the multi-dimensional overlap region more clearly, the red contours show the lensing constraint when restricted to a reduced-dimensionality space with $\theta_{\rm MC}$ fixed to the value accurately measured by the CMB power spectra; the intersection of the red and black contours gives a clearer visual indication of the consistency region in the $\Omm$--$\sigma_8$ plane.

The lensing-only constraint defines a band in the $\Omm$--$\sigma_8$ plane, with the well-constrained direction corresponding approximately to the constraint
\be
\sigma_8 \Omm^{0.25} =  0.591\pm 0.021 \quad (\text{lensing only; 68\,\%}).
\label{eq:LensOnlyDegen}
\ee
This parameter combination is measured with approximately $3.5\%$ precision.

The dependence of the lensing potential power spectrum on the parameters of the \lcdm\ model is discussed in detail in Appendix~\ref{app:clpp_param_depend}; see also~\citet{Pan:2014xua}. Here, we aim to use simple physical arguments to understand the parameter degeneracies of the lensing-only constraints.
In the flat \lcdm\ model, the bulk of the lensing signal comes from high redshift ($z>0.5$) where the Universe is mostly matter-dominated (so potentials are nearly constant), and from lenses that are still nearly linear. For fixed CMB (monopole) temperature, baryon density, and $\ns$, in the \lcdm\ model the broad shape of the matter power spectrum is determined mostly by one parameter, \mbox{$k_{\rm eq}\equiv a_{\rm eq} H_{\rm eq} \propto \Omm h^2$}. The matter power spectrum also scales with the primordial amplitude $\As$; keeping $\As$ fixed, but increasing $\keq$, means that the entire spectrum shifts sideways so that lenses of the same typical potential depth $\Psi_{\rm lens}$ become smaller.
Theoretical \lcdm\ models that keep $\elleq \equiv\keq\, \chi_*$ fixed will therefore have the same number (proportional to $\keq\, \chi_*$) of lenses of each depth along the line of sight,
and distant lenses of the same depth will also maintain the same angular correlation on the sky, so that the shape of the spectrum remains roughly constant. There is therefore a shape and amplitude degeneracy where $\elleq \approx \text{constant}$, $\As \approx \text{constant}$, up to corrections from sub-dominant changes in the detailed lensing geometry, changes from late-time potential decay once dark energy becomes important, and nonlinear effects.
In terms of standard \lcdm\ parameters around the best-fit model, $\elleq \propto \Omm^{0.6} h$, with the power-law dependence on $\Omm$ only varying slowly with $\Omm$; the constraint $\elleq \propto \Omm^{0.6} h=\text{constant}$ defines the main dependence of $H_0$ on $\Omm$ seen in Fig.~\ref{fig:lensonly}.

The argument above for the parameter dependence of the lensing power spectrum ignores the effect of baryon suppression on the small-scale amplitude of the matter power spectrum (e.g.~\citealt{Eisenstein:1997ik}).
As discussed in Appendix~\ref{app:clpp_param_depend}, this introduces an explicit dependence of the lensing power spectrum on $\Omega_{\rm m}h^2$. However, since the parameter dependence of $\elleq$ is close to
$(\Omm h^2)^{1/2}$, we can still think of $\As$ and $\elleq$ as giving the
dominant dependence of $\clpp$ on parameters.

\begin{figure*}[!ht]
	\centering
	\includegraphics[width=\columnwidth]{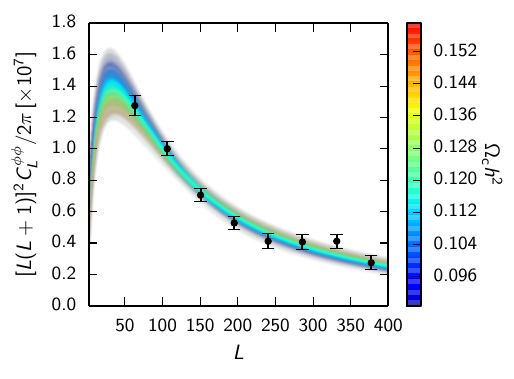}
	\includegraphics[width=\columnwidth]{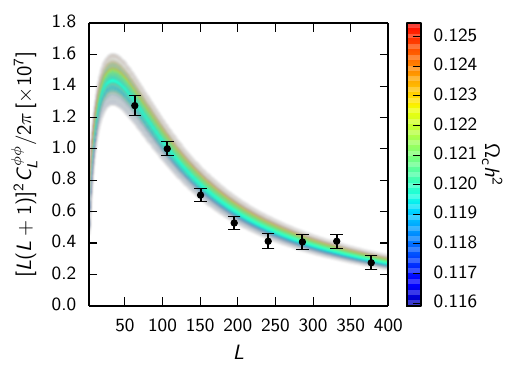}
	\vspace{-0.2in}
	\caption{
    Lensing potential power spectra in the base \lcdm\ model drawn from the posterior of the lensing-only likelihood (left) and the \planckTT\ likelihood (right), coloured by the value of the dark matter density $\Omc h^2$. For fixed CMB power spectra (i.e. the right-hand plot) the
    lensing power spectrum shape is nearly fixed, but the amplitude increases as the matter density increases across the acoustic-scale degeneracy (the accompanying reduction in $\As$ is sub-dominant). With lensing reconstruction data, as $\Omc h^2$ varies the lensing spectrum changes shape, with the amplitude around the best-measured $L\approx 150$, remaining pinned by the accuracy of the measurement there. A combination of the lensing and \planckTT\ data measures the matter density significantly better, with the higher values in the right-hand plot being excluded. Note that the colour scales on the plots are different.
}
    \label{fig:rainbows}
	\end{figure*}

In practice the shape of $\clpp$ is not measured perfectly, and there is also some degeneracy between $\As$, which directly increases the amplitude, and $\elleq$, which increases the amplitude via an increase in the number of lenses along the line of sight and indirectly by changing the depth of potential wells of a given size. On small angular scales with $L > \elleq$,
$\Psi_{\rm lens}$ becomes a strong function of scale due to the fall-off in the matter power spectrum for $k > \keq$. For a fixed comoving scale, the amplitude increases if $\keq$ becomes larger since this reduces the amount of decay of the gravitational potential during radiation domination~\citep{Pan:2014xua}, and also lessens the impact of baryon suppression. This increases the dependence of $\clpp$ on $\elleq$ beyond the linear scaling from the number of lenses, and on small scales $L^4 \clpp \propto \As \elleq^{1+n_L}$, where $n_L >0$ determines the actual $L$-dependence (inherited from the scale dependence of the small-scale matter power spectrum and the effect of baryon suppression).
For \Planck, the lensing spectrum is best measured at $L\sim 200$: since $n_L$ increases with $L$, making $\As$ larger and $ \elleq^{1+n_{200}}$ smaller to keep the power constant at $L\sim 200$ leads to an increase in power at lower $L$ and a decrease in power at higher $L$ (see Fig.~\ref{fig:rainbows}).
The lensing amplitude at the peak $L\approx50$ (and hence the rms deflection angle), is therefore slightly anti-correlated with the amount of small-scale power (as shown by the lensing only constraint in Fig.~\ref{fig:lensdirection}; see discussion below).

The actual three-parameter constraint is close to
\be
\As \left(\Omm^{0.6} h\right)^{2.3} \propto \As \elleq^{2.3} \propto \text{constant}\pm 3\,\% ,
\label{eq:threeparam1}
\ee
which, in terms of $\As$, $\elleq$, and $h$ is independent of $h$, as expected, given the parameter dependence of $C_L^{\phi\phi}$ discussed above.
Recall that the fairly high power-law dependence with $\elleq$ comes from the rapid fall-off in the matter power spectrum with scale (and from the dependence on $\Omega_{\rm m}h^2$ of baryon suppression effects).
Approximately the same scaling will also affect other measures of small-scale power, for example $\sigma_8$.
From Eq.~\eqref{sigma8Approx}, for parameter values of interest $\sigma_8$ and $\As$ are related approximately by
\be
 \sigma_8^2 \Omm^{0.6} \propto \As\elleq^{3.5}.
\label{eq:s8omegamp25:lensonly}
\ee
The parameter direction $\sigma_8 \Omm^{0.25}=\text{constant}$ is therefore close to the direction in which $\As$ (and hence $L^4\clpp$) is constant for fixed $\elleq$. The combination $\As\elleq^{3.5}$ is also not that far from $\As \elleq^{2.3}$ of Eq.~\eqref{eq:threeparam1}, which remains nearly constant as $\elleq$ and $\As$ vary within their degeneracy. The $\sigma_8 \Omm^{1/4}$ degeneracy direction shown in Fig.~\ref{fig:lensonly} and the constraint of Eq.~\eqref{eq:LensOnlyDegen} are therefore significantly tighter than the corresponding constraint in the $\As$--$\Omm$ plane, because $\sigma_8$ absorbs some of the degeneracy between matter density and primordial power at a fixed scale.

In terms of $\sigma_8$, a well-determined three-parameter combination is\footnote{This direction differs from that in Eq.~\eqref{eq:threeparam1}, since the orthogonal direction in the $\As$--$\elleq$ plane is not that much worse constrained, so the Fisher information in the three-parameter space is not fully dominated by a single direction.}
\be
\sigma_8 \Omm^{0.25} \left(\Omm h^2\right)^{-0.37} = 1.228\pm 0.028 \quad (\text{lensing only; 68\,\%}),
\label{eq:threeparam}
\ee
at $2.3\%$ precision. In terms of power, the $4.6\,\%$ precision of the square of this parameter combination is still weaker than the template amplitude measurement of Eq.~\eqref{eq:AhatConservative}. This is expected, given the approximation of a power-law fit, and that in Eq.~\eqref{eq:threeparam} we have also marginalized over variations in $\ns$ and $\Omb h^2$.
 Marginalizing out the $\Omm h^2$ dependence in Eq.~\eqref{eq:threeparam}, the constraint on $\sigma_8 \Omm^{0.25}$ of Eq.~\eqref{eq:LensOnlyDegen} is further degraded to $3.5\,\%$.

We can also quantify the amplitude of lensing by the rms deflection angle $\rmsdeflect$, which we define via
\be
\vardeflect  \equiv \sum_{L=2}^{2000} \frac{(2L+1)}{4\pi} L(L+1) \clpp.
\label{eq:deflectdef}
\ee
This is approximately an integral over all scales. It is proportional to the primordial fluctuation power $\As$, and scales roughly as a power law in $\elleq$. A simple argument based on the number of lenses along the line of sight would suggest $\vardeflect \propto \As\elleq$. This is almost correct, in particular the strong dependence of $\clpp$ on $\elleq$ integrates out when forming $\vardeflect$ (see Eq.~\ref{eq:lenspower}). However, there is a further dependence on $\Omega_{\rm m}$ and $h$ due to the effects of baryon suppression and the late-time decay of the gravitational potential, which is important at low-$L$; we find approximately
\begin{equation}
\vardeflect \propto \As \elleq (\Omega_{\rm m} h)^{1/5} ,
\label{eq:dsqlensonly}
\end{equation}
by taking derivatives of $\clpp$ computed with \camb.
The rms is measured at $2.4\,\%$ precision to be
\be
\rmsdeflect  = (2.46\pm 0.06)\,\text{arcmin} \quad (\text{lensing only; 68\,\%}).
\label{eq:rmsdeflect}
\ee

So far we have only considered a \lcdm\ model. However, for most generalizations that are close to \lcdm, the same scaling also approximately holds, and the constraint direction of $\sigma_8 \Omm^{0.25}$ is only weakly model-dependent, although the centroid can shift and also the relation between $\Omm$ and $h$. For example, for a model with massless sterile neutrinos parameterized by $\neff$, the constraint of Eq.~\eqref{eq:LensOnlyDegen} is virtually identical, but parameter samples shift towards systematically higher $\Omm h^2$ than in \lcdm\ to keep the equality scale roughly constant when $\neff > 3.046$. As a further example, in models with three massive (active) neutrinos with degenerate masses, marginalizing over the mass of massive neutrinos shifts the constraint slightly down, giving
\be
\sigma_8 \Omm^{0.25} =  0.566\pm 0.025 \quad (\text{\lcdm+$\Sigma m_\nu$, lensing only; 68\,\%}).
\ee
Note that in the neutrino case non-linear corrections from \HALOFIT\ are unreliable away from $\Omm\approx 0.3$, which can affect this result at around the $0.01$ level.

\subsubsection{Joint parameter constraints}
\label{subsec:jointparams}

The small-scale CMB power spectra are only directly sensitive to the primordial power $\As$ after damping by reionization, through the combination $\As e^{-2\tau}$. However lensing also smooths the power spectra at the several percent level, which allows the amplitude $\As$ to be constrained via the amplitude of the smoothing effect (and a small transfer of power to small scales), even when $\tau$ is not constrained precisely.\footnote{The smoothing effect of lensing in the \planck\ $TT$ power spectrum is detected at around $10\,\sigma$~\citep{planck2014-a15}. It is also detected at lower significance in the polarization power spectra.}
The lensing smoothing can be thought of as approximately a convolution of the deflection angle power spectrum with the CMB power spectrum, and hence is mostly sensitive to scales around the peak of the deflection angle power spectrum.
We can quantify the amplitude of the smoothing approximately by the mean-squared deflection angle $\vardeflect$ from Eq.~\eqref{eq:deflectdef}.

	\begin{figure}[ht]
	\centering
	\includegraphics[width=\columnwidth]{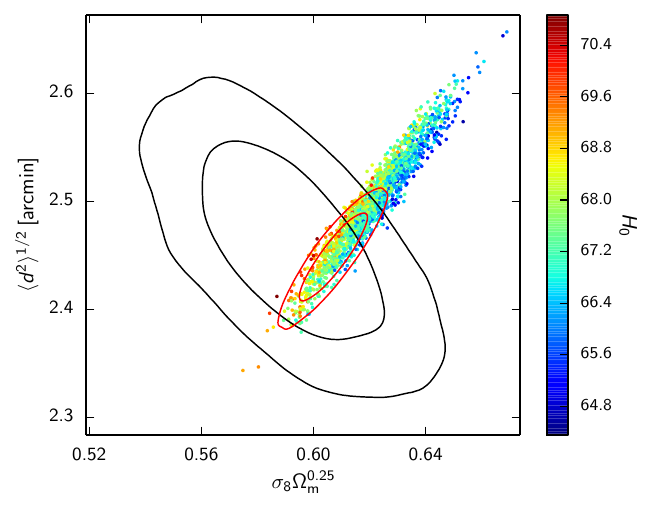}
	\vspace{-0.2in}
	\caption{
    Constraints on the rms lensing deflection angle $\rmsdeflect$ from \planckTT\ CMB power spectra in the base \lcdm\ model (samples, colour coded by the value of the Hubble constant), showing that the inferred lens deflection amplitude is strongly correlated with the parameter combination $\sigma_8 \Omm^{0.25}$. The solid black contour is the constraint from CMB lensing only, which prefers lower values of the lensing amplitude, giving the joint constraint shown by the red contours. Since the lensing reconstruction mainly constrains scales smaller than the peak of the deflection angle power spectrum,
%and determines the spectrum shape and hence matter-equality scale much less well than the primary power spectrum,
there is an additional degeneracy that makes $\rmsdeflect$ slightly anti-correlated with the measurement of lensing power on smaller scales, which is approximately proportional to (the square of) $\sigma_8 \Omm^{0.25}$}
    \label{fig:lensdirection}
	\end{figure}

The CMB power spectra determine the angular acoustic scale $\theta_{\ast}$ very accurately, which can be used to eliminate the $h$-dependence from $\vardeflect$, considered as a function $\sigma_8$, $h$, and $\Omm$. Empirically, we find that in the \lcdm\ model the CMB power spectrum posterior has $\rmsdeflect \propto  \sigma_8\Omm^{0.2}$, close to the $\sigma_8\Omm^{0.25}$ combination measured more directly by lensing reconstruction;\footnote{The empirical parameter dependence of $\rmsdeflect$ on $\Omm$ for samples from the CMB power spectrum posterior differs from what one would infer from Eqs.~\eqref{eq:s8omegamp25:lensonly} and~\eqref{eq:dsqlensonly}. These equations assume fixed $\ns$ and $\Omb h^2$, which is appropriate for our lensing-only constraints with the assumed priors, but for the CMB samples the fractional dispersion in $\ns$ and $\Omb$ is comparable to that in $\sigma_8$, $h$, and $\Omm$ and so cannot be neglected.}
see Fig.~\ref{fig:lensdirection}.
For a flat \lcdm\ model the constraint on the latter parameter is
\be
\sigma_8 \Omm^{0.25} =  0.621 \pm 0.013  \quad (\text{\planckTT; 68\,\%}).
\ee
The amplitude information here is coming from a combination of the lensing smoothing effect, and information from the CMB power spectrum amplitude combined with the constraint on $\tau$ from low-$\ell$ polarization. The lensing-only constraint of Eq.~\eqref{eq:s8omegamp25:lensonly} is slightly weaker, but somewhat lower than this (consistent at just over $1\, \sigma$); see Fig.~\ref{fig:lensonly}.
The combined constraint is tighter and restricted to slightly lower values compared to the \Planck\ power spectrum measurement:
\be
\sigma_8 \Omm^{0.25} =0.609 \pm  0.007 \quad (\text{\planckTT+\lensing; 68\,\%}).
\ee
Here, for joint constraints we no longer apply the additional priors used in the lensing-only analysis,  and allow for the full linearized CMB power spectrum dependence of the lensing renormalization in the likelihood. The overlap region in the $\Omm$--$\sigma_8$ plane also excludes some of the lower Hubble constant values allowed by the power spectrum data, giving a slight shift of the mean in the direction of the higher values preferred by some local measurements~\citep{Riess:2011yx,Humphreys:2013eja, Efstathiou:2013via}, as shown in Fig.~\ref{fig:lensdirection}. This could be an additional indication that the lower Hubble constant values from the CMB spectra are at least partly a random statistical fluctuation, in which case joint constraints should be usefully closer to the truth. The joint constraint gives
\begin{align}
H_0 &=(67.8\pm 0.9)\, {\rm km}\,{\rm s}^{-1}\,{\rm Mpc}^{-1} \nonumber \\ & \hspace{0.15\textwidth} (\text{\planckTT+\lensing; 68\,\%}),
\end{align}
about $0.5\, \sigma$ higher than \planckTT\ alone, and in good agreement with the constraint from \planckTT+BAO data.

\begin{figure}[!ht]
	\centering
	\includegraphics[width=\columnwidth]{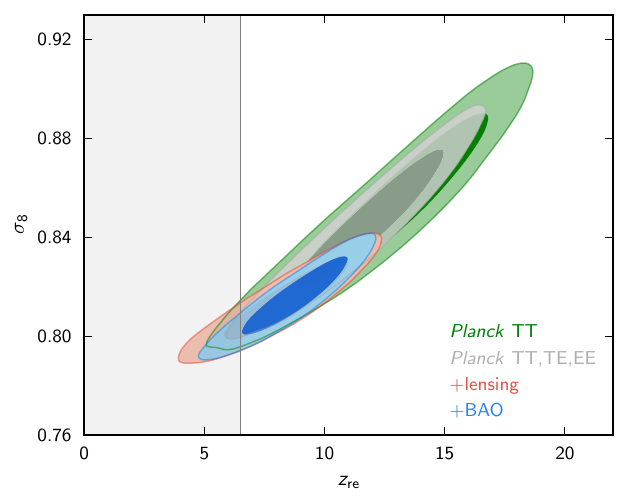}
	\vspace{-0.2in}
	\caption{
    Constraints on the reionization (mid-point) redshift $z_{\rm re}$ and $\sigma_8$ from a combination of \Planck\ CMB power spectra and the lensing reconstruction, excluding low-$\ell$ polarization information. The grey band denotes the approximate $z_{\rm re}<6.5$ region excluded by observations of the spectra of high-redshift quasars~\citep{Fan:2005es}.
    The lensing data significantly shrink the allowed region of parameter space, preferring lower values of the reionization redshift, in good agreement with the measurement from \Planck\ low-$\ell$ polarization. }
    \label{fig:sigmatau}
	\end{figure}

The measurement of the fluctuation amplitude from lensing, combined with the amplitude of the CMB power spectrum on intermediate and small scales,
can also be used to constrain the reionization optical depth independently of low-$\ell$ polarization.  The CMB power spectra themselves provide a weak constraint via the lensing smoothing, which is significantly improved by using the additional information in the lensing reconstruction.
Figure~\ref{fig:sigmatau} shows the constraint when \Planck\ CMB temperature power spectrum constraints, \textit{without low-$\ell$ polarization}, are combined with the lensing likelihood. We find
\be
\tau = 0.070\pm 0.024   \quad (\text{\planckTTonly+\lensing; 68\,\%}),
\ee
corresponding to a detection of a non-zero optical depth at more than $2\,\sigma$,
and a corresponding reionization redshift $z_{\rm re} = 9.0^{+2.5}_{-2.1}$ (68\%). These results are consistent with the baseline \Planck\ power spectrum constraints including low-$\ell$ polarization, which give $\tau =0.078\pm 0.019$  (\text{\planckTT; 68\%}). Both results indicate a consistent downward shift in mean compared to the central value from \WMAP\ ($\tau=0.089\pm 0.014$; \citealt{Hinshaw:2012aka}), in agreement with lower predicted values based on recent results for the integrated luminosity function (e.g.~\citealt{2014arXiv1410.5439F}). There is a corresponding downward shift in $\sigma_8$, with joint constraint
\be
\sigma_8 = 0.815\pm 0.009  \quad (\text{\planckTT+lensing; 68\,\%}).
\ee
For further discussion see~\cite{planck2014-a15}.

\subsubsection{Constraints on extensions to the \lcdm\ model}
	\begin{figure}[!ht]
	\centering
	\includegraphics[width=\columnwidth]{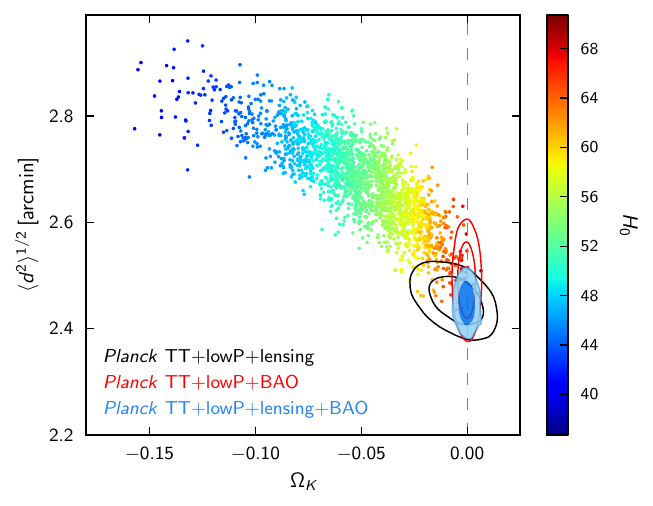}
	\vspace{-0.2in}
	\caption{ Constraints on the curvature and rms lensing deflection angle from \planckTT\ CMB power spectra (samples, colour coded by the value of the Hubble constant), showing that the closed-universe models favoured at the $2\,\sigma$ level have high lensing power that is inconsistent with the lensing reconstruction result. The combined lensing result (black contours) is much tighter, and provides additional information on top of the constraint when adding BAO data (red contours); both joint constraints, and the combined filled contours, are consistent with a flat universe.
}
    \label{fig:omegak}
	\end{figure}

The CMB power spectra only weakly constrain extended models that provide multiple ways to combine late-time parameters to give the same observed angular diameter distance to recombination. The main \Planck\ CMB power spectrum constraints on these parameters are driven largely by the impact of lensing smoothing on the acoustic peaks, which weakly breaks the geometric degeneracy. As discussed in detail in~\cite{planck2014-a15}, the power spectrum results appear to prefer larger lensing smoothing than in \lcdm\ (in this release, slightly above the $2\, \sigma$ level). It is therefore particularly useful to study joint constraints with the lensing likelihood, which constrains the lensing amplitude more directly and more tightly. Since the lensing-reconstruction power spectrum has an amplitude that peaks slightly {\it lower} than the \lcdm\ predictions from the CMB power spectra, the joint constraint eliminates a large region of parameter space with positive curvature allowed by the power spectrum results. For example, Fig.~\ref{fig:omegak} shows the constraints on the curvature parameter $\Omk$ in non-flat models; the high lensing amplitudes, as show in the figure by the rms deflection angle, are ruled out by the lensing reconstruction, and the joint constraint is consistent with a flat universe ($\Omk=0$). The improvement in error bars is dramatic, with the CMB power spectra giving
\be
\Omk = -0.052^{+0.03}_{-0.02} \quad (\text{\planckTT; 68\,\%}),
\ee
and the joint constraint
\be
\Omk  = -0.005^{+0.009}_{-0.007}   \quad (\text{\planckTT+\lensing; 68\,\%}).
\ee
This measurement of curvature from the CMB alone has sub-percent precision.
External data can also break the geometric degeneracy; the joint BAO contours are shown in Fig.~\ref{fig:omegak} and are consistent. Since the BAO results are independent of fluctuation amplitude, the lensing-reconstruction joint constraint adds significant additional information, shrinking the allowed parameter space compared to \Planck+BAO alone.

In addition to curvature, CMB results for dark energy models also suffer a geometric degeneracy, and lensing results can similarly provide additional information; for applications to a variety of dark energy models and modified gravity see~\cite{planck2014-a16}.
Lensing can also improve the limits on many other parameters, though in a less dramatic fashion if the CMB power spectra are themselves able to place a good constraint. Full results for many combinations are available online in the grid tables and parameter chains.\footnote{\url{http://www.cosmos.esa.int/web/planck/pla}}

\subsubsection{Sensitivity to $\elp$-range}
\label{subsubsec:Lrange}

The results in the previous subsections were obtained using our conservative baseline likelihood, with eight bins in the multipole range $40\le L\le 400$. We now briefly discuss how joint constraints from the \planck\ power spectra and lensing change for different cuts of the lensing data.

Using the first nine bins of the aggressive binning described in Table~\ref{table:power_spectrum_bandpowers}, covering $8 \le L\le 425$, results from \planckTT+lensing are consistent at the $0.1\,\sigma$ level with the conservative likelihood, in both \lcdm\ and \lcdm$+\mnu$ cosmologies.  Extending the range to higher $L$, the \planckTTonly+lensing constraint on the optical depth is stable, however the measured amplitude $\sigma_8\Omm^{0.25}$ shifts downwards by about $1\, \sigma$ with a negligible change in the uncertainty.
About half of this comes from including the outlying bin at $638\le L \le 762$. In the \lcdm+$\mnu$ model, this preference for lower amplitudes on smaller scales pulls joint neutrino mass constraints up by about $1\,\sigma$. The baseline result is
\be
\sumnu < 0.145 \, {\rm eV} \quad(\text{\planckTT+lensing+BAO; 68\,\%}),
\ee
but using the full $L$-range the posterior peaks slightly (though not significantly) away from zero:
\begin{align}
\sumnu &= 0.16^{+0.08}_{-0.11} \,{\rm eV} \quad(\text{\planckTT+aggressive lensing} \nonumber \\
& \hspace{0.575\columnwidth}\text{+BAO; 68\,\%}).
\end{align}
As discussed in Sect.~\ref{sec:consistency}, the multipole range $300\alt L \alt 900$ may be unreliable, specifically due the failure of the $TTTT$ curl test. A priority for future analysis should be to develop a trustworthy likelihood over the full multipole range.

\section{Consistency and null tests}
\label{sec:consistency}
With multiple frequency bands and the combination of both temperature and polarization data,
the {\it Planck} full-mission data set provides many opportunities to test internally the lens reconstruction.
In this section we present consistency and null tests for the MV lensing potential presented in Sect.~\ref{sect:results}.

For consistency tests, we perform lens reconstruction using different data/analysis choices than those used for our fiducial MV reconstruction,
obtaining an alternative measurement of the lensing potential power spectrum
$\hat{C}_L^{\phi\phi} |_{\rm test}$, which we then compare to the baseline MV reconstruction.
To compare the test and MV reconstructions more quantitatively, we use a $\chi^2$ statistic for the difference between the bandpowers, which is calculated as
\be
\chi^2_{\rm test} =
	\bin_i^{\elp}
	\left( \left. \hat{C}_{\elp}^{\phi\phi} \right|_{\rm test} -  \left. \hat{C}_{\elp}^{\phi\phi} \right|_{\rm MV} \right)
	\left[ \Sigma^{-1}_{\rm test} \right]^{ij} \bin_j^{\elp'}
	\left( \left. \hat{C}_{\elp'}^{\phi\phi} \right|_{\rm test} -  \left. \hat{C}_{\elp'}^{\phi\phi} \right|_{\rm MV} \right),
\ee
where $\Sigma_{\rm test}$ is a diagonal covariance matrix for the difference obtained from simulations of the two lens reconstructions.
We  compare this $\chi^2$ to a set of values determined from simulations, to estimate a PTE.
We do this for several possible choices of binning: a single bin
for either the conservative ($40 \le L \le 400$) likelihood range, or the low-$L$ and high-$L$ ranges ($8 \le L < 40$ and $400 < L \leq 2048$, respectively) that are additionally included in the aggressive likelihood; and the individual bins that span the conservative and high-$L$ multipole ranges.
The single-bin tests check for broad-band differences between the two reconstructions, while the multiple-bin tests provide a more sensitive check on shape differences or inaccuracies in the error bars for the two reconstructions.

\begin{figure*}[!p]
\begin{center}
\vspace{0.5in}
\includegraphics[width=\textwidth]{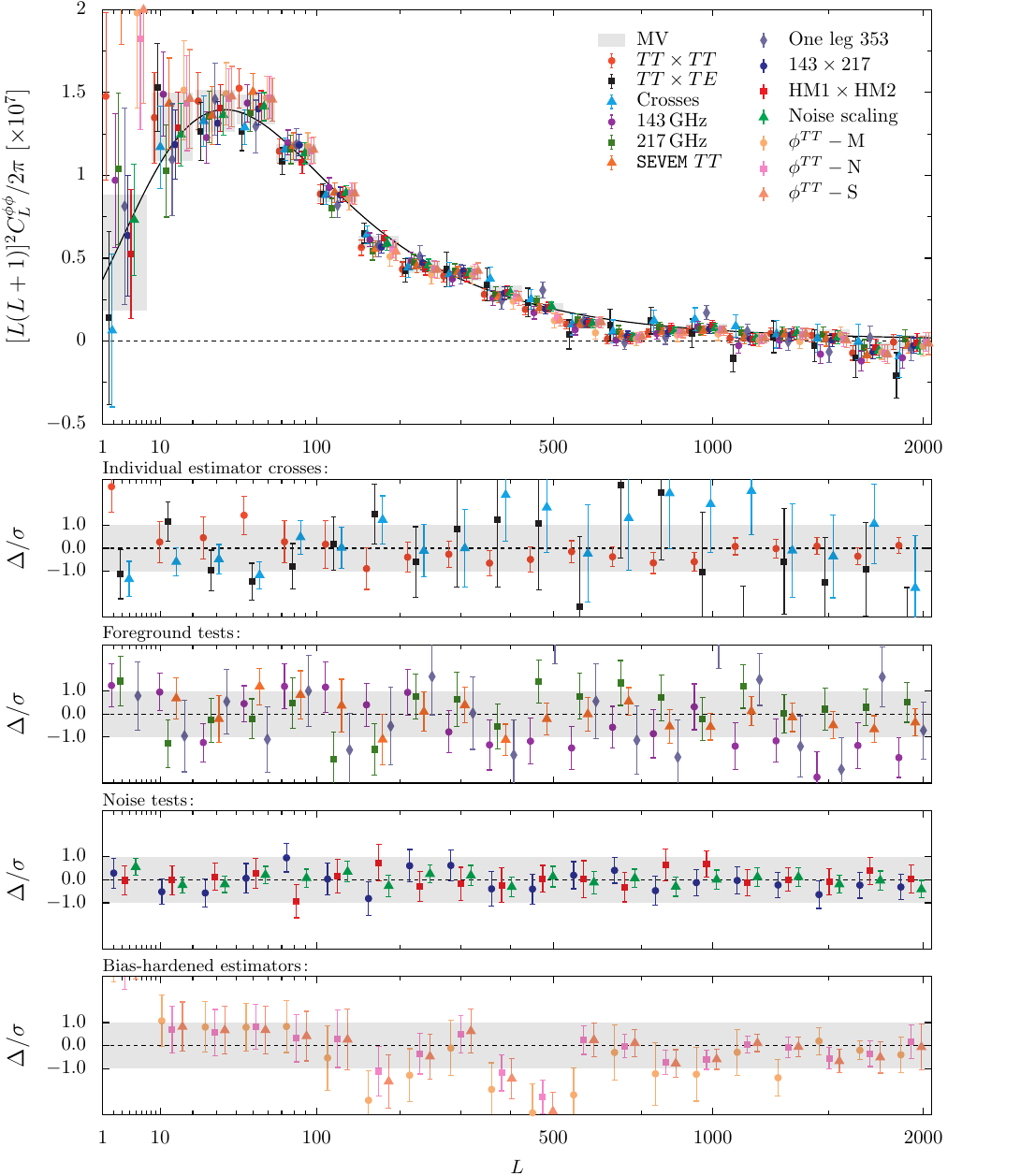}
\caption{
Summary of lensing power spectrum consistency tests.
{\it Upper panel:} comparison of the lensing power spectra for the baseline MV reconstruction and several alternative estimators with different data/analysis choices discussed in the text.
{\it Lower panels:} detailed plots of the difference between each alternative reconstruction and the MV power spectrum, in units of the MV error bars ($\sigma$).
The $\pm 1\sigma$ band is indicated in grey.
Error bars on individual data points give the standard deviation of the expected difference from the MV reconstruction for each case, estimated using Monte Carlo simulations.
\label{fig:clpp_consistency_tests}
}
\end{center}
\end{figure*}

We group the consistency tests as follows, based on the issues which they are designed to probe.
\begin{itemize}
\item `Individual estimator crosses' that test for consistency among the five quadratic estimators used to measure $\phi$.
\item `Foregrounds tests' that use single-frequency maps or different component-separation methods to test for possible foreground contamination.
\item `Noise tests' that check that our power spectrum estimates are not biased by instrumental noise, using cross-estimators designed to avoid noise biases.
\item `Analysis choices' that test our sensitivity to choices such as the unmasked sky fraction and multipole range.
\item `Bias-hardened estimators' in which we use lensing estimators that should be less sensitive to certain systematic and instrumental effects.
\end{itemize}
In Fig.~\ref{fig:clpp_consistency_tests} we plot several of the most important consistency tests, as well as their difference from the MV reconstruction.
The $\chi^2$ values associated with the full set of consistency tests are presented in Table~\ref{table:consistency_tests}.
The tests themselves are described in more detail in Sects.~\ref{sec:consistency:individual_estimator_crosses}--\ref{sect:consistency:bias_hardened_estimators}.

\begin{figure*}[!ht]
\begin{center}
\includegraphics[width=\textwidth]{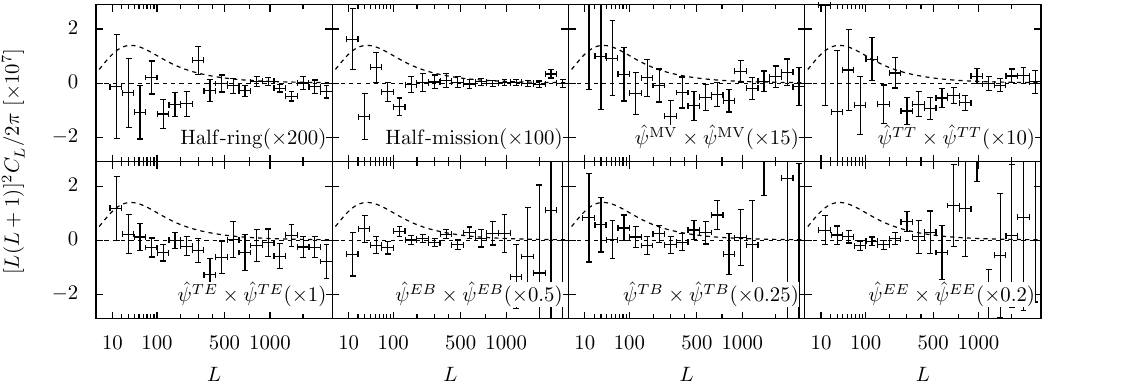}
\vspace{-0.2in}
\caption{
Power spectrum null tests discussed in Sect.~\ref{sec:consistency:null_tests}.
Curl null tests are denoted by $\psi$.
The individual null test power spectra have been scaled by the factors indicated in each panel for visual purposes. The fiducial (gradient-mode) lensing power spectrum is shown as the dashed line in each panel.
\label{fig:clxx_grid}
}
\end{center}
\end{figure*}

In addition to consistency tests, we also perform several null tests in which we apply our lens reconstruction and power spectrum estimation
procedure to maps that have been differenced to remove sky signal, or use curl-mode lensing estimators that should have zero contribution from the gradient lensing potential.
We check that the overall amplitude
(for a template given by the fiducial theory spectrum $C_{\elp}^{\phi\phi,{\rm fid}}$)
is consistent with zero, and also evaluate a $\chi^2$ computed as
\be
\chi^2_{\rm null} =
	\bin_i^{\elp}
	\left( \left. \hat{C}_{\elp}^{\phi\phi} \right|_{\rm test} \right)
	\left[ \Sigma^{-1}_{\rm null} \right]^{ij} \bin_j^{\elp'}
	\left( \left. \hat{C}_{\elp'}^{\phi\phi} \right|_{\rm test} \right),
\ee
where $\Sigma_{\rm null}$ is a diagonal covariance matrix measured from simulations of the particular null test.
A PTE for each null test is computed by comparison to simulations in the same way as for the consistency tests.
The null power spectra are plotted in Fig.~\ref{fig:clxx_grid}, and summary statistics are given in Table~\ref{table:null_tests}.
The null tests are discussed in more detail in Sect.~\ref{sec:consistency:null_tests}.

\subsection{Individual estimator crosses}
\label{sec:consistency:individual_estimator_crosses}
\begin{figure*}[!ht]
\begin{center}
\includegraphics[width=\textwidth]{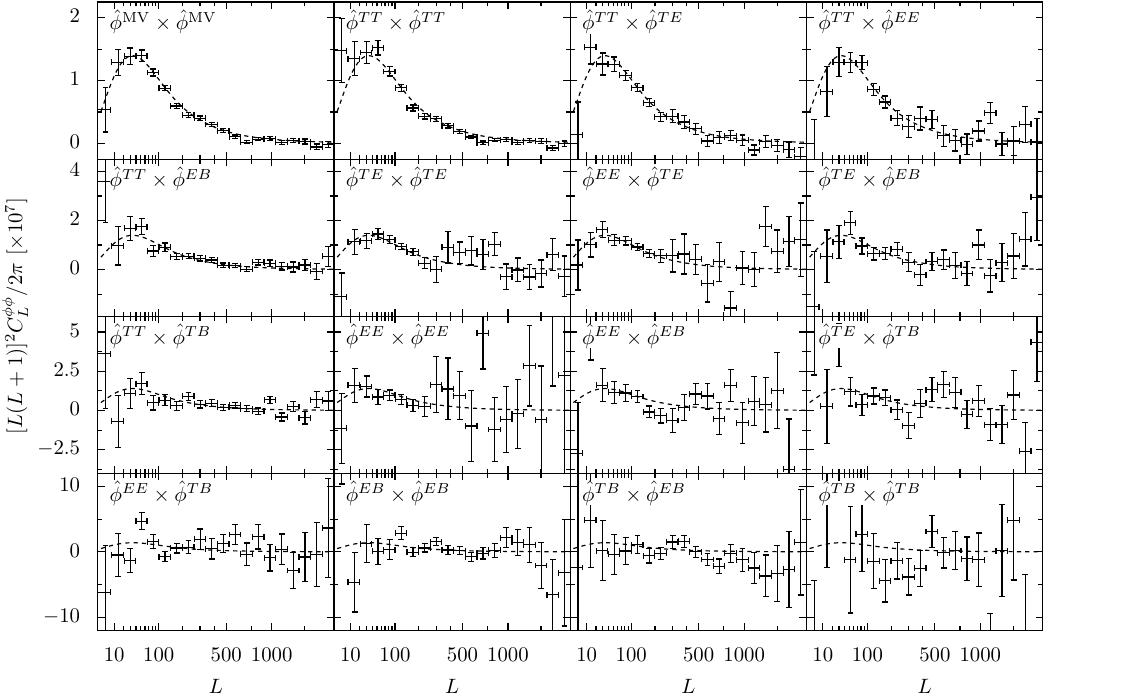}
\vspace{-0.2in}
\caption{
Grid of lensing potential power spectra for the 15 possible auto- and cross-spectra of the five quadratic lensing estimators obtained from the \smica\ foreground-cleaned maps.
The measurements are arranged in order of their total $S/N$ for a measurement of the lensing amplitude.
The MV power spectrum estimate, which combines the individual cross-spectra and is the basis for our fiducial lensing likelihood, is plotted in the top-left panel.
\label{fig:clpp_grid}
}
\end{center}
\end{figure*}

With the foreground-cleaned temperature and polarization maps from \smica\ we can form five distinct lensing estimators,
and 15 distinct auto- and cross-spectra. We plot all of these spectra individually in Fig.~\ref{fig:clpp_grid}.
The lensing amplitudes, as well as PTEs for the differences from the MV reconstruction are given in Table~\ref{table:consistency_tests}.
There is a wide variation in the sensitivity of the different estimators, although we see high
(greater than $5\,\sigma$ significance)
detections of lensing in nine of the spectra.\footnote{%
\refchange{
In the 2015 \planck\ data release, there is some evidence in the polarization
power spectra for uncorrected systematic effects associated with
temperature-to-polarization
leakage~\citep{planck2014-a15,planck2014-a13}. Although we find no
evidence for inconsistencies in the polarization-based lens reconstructions, the
statistical errors are large and may obscure small systematic effects
that are subdominant for the current analysis. Leakage from the
CMB temperature anisotropies could impact the lensing analysis in
several ways. First, the disconnected (Gaussian) part of the CMB
4-point function is modified, and this would affect both our
mean-field and \textsc{N0} corrections (see
Appendices~\ref{sect:pipeline:qe}
and~\ref{sect:pipeline:clpp}). These corrections are made using a mix
of data and simulations, such that even if the simulations do not
fully capture the leakage (note they do include leakage due to
main-beam mismatch, for example), the additive bias to the lensing
power spectrum is only second-order in any unmodelled effects. The
second way that leakage can affect the lensing power spectrum is
through changes to the lensing response functions in
Eq.~(\ref{eqn:qecov}), which propagate through to normalization
errors. This should be partially captured by a Monte-Carlo correction
that we make to the lensing power spectrum. We have not attempted to
quantify more carefully any residual effects of
temperature-to-polarization leakage, since, ultimately, the
polarization data carries relatively little weight in our final MV reconstruction.}}

As a further test, we combine several of the spectra with the highest $S/N$ that only use estimator crosses
($TT\times EE$, $TT\times TE$, $TT\times EB$, and $EE\times TE$) into a single `crosses' estimator.
This combination of cross-spectra has an order-of-magnitude smaller disconnected noise bias correction than the auto-spectrum of the fiducial MV estimator,
with
\mbox{$\Delta \hat{A}^{N0,\ \rm Crosses}_{8 \rightarrow 2048} = 0.4$},
compared to
\mbox{$\Delta \hat{A}^{N0,\ {\rm MV}}_{8 \rightarrow 2048} = 4.3$} for the baseline MV lensing reconstruction.
The results of this test are presented in Fig.~\ref{fig:clpp_consistency_tests} and Table~\ref{table:consistency_tests}.
With the crosses power spectrum estimates we are able to test that the lensing amplitude is not biased by errors in the disconnected noise bias correction
$\Delta C_L^{\hat{\phi}\hat{\phi}} |_{\textsc{N0}}$ at around the $2\%$ level.

We  also form a combined power spectrum estimate using \textit{only} polarization data (the $EE$ and $EB$ estimators).
This `Pol. only' estimator gives an amplitude estimate for the conservative multipole range of
\mbox{$\hat{A} = 0.76 \pm 0.15$}, a $\poldetectionsignif$ detection of lensing at a level consistent with the fiducial MV reconstruction.

\subsection{Foreground tests}
\label{sec:consistency:foreground_tests}
The \smica\ CMB map that is used for our baseline MV reconstruction combines multiple frequency bands as a function of angular scale to reject foregrounds and to produce a single high-fidelity estimate of the CMB.
We can also compare to a simpler approach based on individual frequency maps, which we clean by subtracting a dust template.
Our baseline results in \cite{planck2013-p12} were based on this approach using 143\,GHz and 217\,GHz data, with the 857\,GHz map
projected out during the filtering process as a dust template.
In this work we repeat this analysis. We use 857\,GHz as a dust template in temperature, and no cleaning in polarization.
The results of this test are presented in Fig.~\ref{fig:clpp_consistency_tests} and Table~\ref{table:consistency_tests} with the labels
`143\,GHz' and `217\,GHz', where it can be seen that we obtain lensing bandpowers that are consistent with the MV reconstruction for the conservative multipole range.
In the simulations used to characterize these reconstructions,
we add coloured Gaussian noise to the 143\,GHz and 217\,GHz temperature map simulations, designed to mimic the power of unsubtracted extragalactic foregrounds.
For this we use the same source model used in \cite{planck2013-p12}.
We also add Gaussian power to the polarization maps, designed to mimic the power due to Galactic dust contamination that can be seen on large angular scales.
The resulting temperature and polarization simulations have power spectra that agree with those of the data at the percent level in power.

We can also test the stability of the lensing reconstruction to the choice of the \smica\ component-separation method over other alternatives such as {\tt NILC}, {\tt SEVEM}, and {\tt Commander}.
The results of these comparisons are also shown in Table~\ref{table:consistency_tests}; all are very consistent with the \smica-based reconstruction over the conservative multipole range, although the PTEs for the overall amplitude at high-$\elp$ ($400 < L \le 2048$) are low for {\tt NILC} and {\tt Commander}.
For the {\tt SEVEM} and {\tt Commander} analyses, polarization maps were not available at the high resolution used here ($N_{\rm side}=2048$), and so we have limited the comparisons to temperature only.

As a test of Galactic contamination, we also repeat our baseline analysis (performed on a fraction $f_{\rm sky}=0.7$ of the sky) using a more aggressive ($f_{\rm sky} = 0.8$) and more conservative ($f_{\rm sky} = 0.6$) Galactic mask.
These masks are constructed by thresholding maps from the {\it Planck} high-frequency channels to achieve the desired sky fraction.
In Table~\ref{table:consistency_tests} we see that for both alternative sky fractions we estimate lensing bandpowers that are consistent with the baseline.

As a final and very powerful test of high-frequency foreground contamination, we also perform a lens reconstruction in which we use the 353\,GHz map as one of the four legs in the trispectrum used to estimate $C_{\elp}^{\phi\phi}$.
Based on the approximate frequency scaling of dust \citep{Gispert:2000np} we expect that any dust foreground bias `$b_{\rm dust}$' to the lensing power spectrum estimate that we see in this test will scale roughly as
\be
b^{353}_{\rm dust} \approx 7 b^{217}_{\rm dust} \approx 20 b^{143}_{\rm dust},
\ee
and so this should provide a sensitive test for such contamination in the \smica\ map.
The 353\,GHz map has significantly more power than CMB+noise-only simulations,
due primarily to Galactic dust, and temperature fluctuations from the CIB.
To obtain reasonable error bars from the 353\,GHz simulations used to characterize this reconstruction, as with the other reconstructions
we add an isotropic Gaussian sky signal  designed to match the simulation power spectrum to a smoothed power spectrum of the data.
When filtering the 353\,GHz map, we use the same parameters as used for the \smica\ map (except for the beam transfer function), to have the same weight given to different angular scales in the quadratic lensing estimators.
As with the 143\,GHz and 217\,GHz analyses, we project out the 857\,GHz map as a dust template from the 353\,GHz map in temperature, but make no attempt to subtract foreground contamination in polarization.
As can be seen in Fig.~\ref{fig:clpp_consistency_tests} and Table~\ref{table:consistency_tests}, we obtain good agreement between the `one leg 353\,GHz' reconstruction and the MV lensing power spectrum.
Lensing power is detected at
$\signifsmicathreexthreefiftythree$ even using 353\,GHz as one of the four estimator legs.
The stability of the lens reconstruction in the presence of the foreground power at 353\,GHz is very striking.
As a further test for the stability of the polarization-based lens reconstructions,
we can also use the `Pol. only' estimator from Sect.~\ref{sec:consistency:individual_estimator_crosses} with one leg given by the 353\,GHz  data.
This reconstruction is very noisy, however we do obtain an overall amplitude as well as bandpowers that are consistent with the MV `Pol. only' reconstruction, as shown in Table~\ref{table:consistency_tests}.

\subsection{Noise tests}
\label{sec:consistency:noisetests}
As mentioned in Sect.~\ref{sec:data_and_methodology}, the FFP8 simulations on which our results are based underestimate the noise power by several percent at high $\ell$.
We have dealt with this problem in our simulations by including isotropic Gaussian power to make up for the difference.
This neglects the fact that the missing noise component most likely has an anisotropic distribution (which could lead to errors in the mean-field subtraction),
and could be non-Gaussian as well.
To test that our treatment of the missing noise component is adequate, we have performed several tests.
Most stringently, we form quadratic lensing estimates using cross-correlation between pairs of maps with independent noise realizations.
We construct both a frequency cross $143\,{\rm GHz} \times 217\,{\rm GHz}$, as well as a cross between the first and second halves of the mission data.
These estimators have no noise mean field, and the lensing power spectrum estimated from them is insensitive to non-Gaussianity of the instrument noise.
The results of this analysis are given in Table~\ref{table:consistency_tests}, where we see results that are consistent with the MV power spectrum estimate.

As an additional test, we have recreated the analysis of \cite{planck2013-p12}, where the missing noise power was accounted for by scaling the noise component of the FFP8 simulations by a small factor.
This approach reasonably supposes that the anisotropy of the missing noise component is the same as that of the modeled component.
This test is labelled `Noise scaling' in Table~\ref{table:consistency_tests}.
Again, we see a power spectrum measurement with an amplitude that is very similar to the MV estimate
(although we note that because the only difference in this test is the construction of the simulations that are used, the expected scatter between the `Noise scaling' and baseline approaches is difficult to determine).

\subsection{Analysis choices}
\label{sec:consistency:analysis_choices}
We test the stability of our reconstruction to several analysis choices, re-running the baseline analysis implementing the following changes one at a time.
\begin{itemize}
\item Use a high-pass filter in temperature and polarization with
\mbox{$\elt_{\rm min}\!=\!1000$}
 rather than
 \mbox{$\elt_{\rm min}\!=\!100$}, as adopted in the baseline analysis.
\item Use a low-pass filter with
\mbox{$\elt_{\rm max}\!=\!1500$}
 rather than
 \mbox{$\elt_{\rm max}=2048$}, as used in the baseline analysis.
\item Apodize the lensing convergence estimate before power spectrum estimation, following the analysis in \cite{planck2013-p12}.
This change reduces the size of the Monte Carlo correction, but does not produce a significant change in the lensing bandpower estimates.
\end{itemize}
The results of these tests are given in Table~\ref{table:consistency_tests}.

\subsection{Bias-hardened estimators}
\label{sect:consistency:bias_hardened_estimators}
Throughout this paper, we have used quadratic estimators optimized to detect the anisotropic covariance between CMB modes that is induced by (fixed) lensing.
These estimators are biased by non-lensing sources of such mode coupling,
including effects such as galaxy and point-source masking, inhomogeneous instrumental noise, and beam asymmetries.
As discussed in Sect.~\ref{sec:data_and_methodology},
we estimate these biases using Monte Carlo simulations and subtract them as a `mean-field' contribution from the estimated lensing potential (as in Eq.~\ref{eqn:phihat}).
A complementary method, which is less reliant on the quality of the Monte Carlo simulations used to perform this debiasing, is the bias-hardening procedure advocated by \cite{Namikawa:2012pe}.
In this approach, one constructs a new quadratic lensing estimator that is orthogonalized to worrisome sources of bias.
The construction of these estimators is reviewed in Appendix~\ref{sect:pipeline:bias_hardened}.
In the analysis here we focus on temperature-only bias-hardened estimators to investigate the effects of bias-hardening,
because mean-field corrections are much smaller in polarization than in temperature.
We denote these estimators as $\hat{\phi}^{TT}-X$, where $X$ denotes the bias(es) that are orthogonalized against.
Here, we consider the biases due to masking ($M$), inhomogeneous noise ($N$), and the shot noise due to unresolved point sources ($S$).
As discussed in \cite{Osborne:2013nna}, the source-hardened estimator $\hat{\phi}^{TT}-S$ also reduces contamination from effects such as the cross-correlation between the unresolved sources and the lensing potential.

\begin{figure}[!ht]
\label{fig:lowl_mean_fields}
\centering
\includegraphics[width=\columnwidth, trim=0 28 0 0, clip=True]{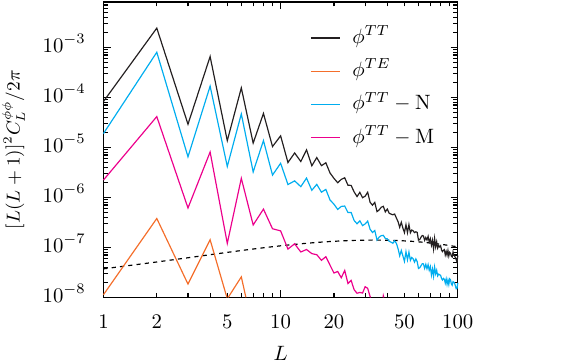}
\includegraphics[width=\columnwidth]{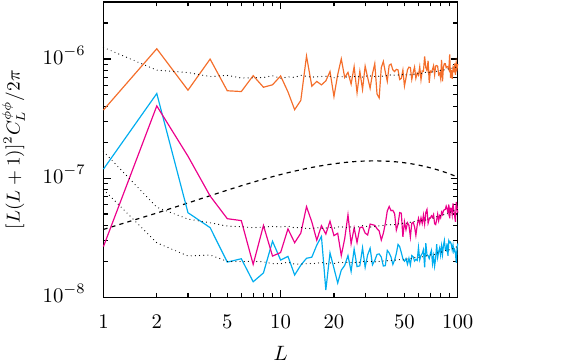}
\caption{
\textit{Upper}:
power spectrum of the low-$L$ mean-field (measured from Monte Carlo simulations) for the $TT$ lensing estimator as well as several bias-hardened estimators.
The mean-field power spectrum of the $TE$ lensing estimator is plotted for comparison, and the fiducial lensing potential power spectrum is shown in dashed black.
To avoid visual clutter we have not shown the source-hardened estimator $TT-S$; its spectrum is midway between the noise-hardened ($TT-N$) and mask-hardened ($TT-M$) cases.
\textit{Lower}:
power spectra of the differences between the $TT$ lensing estimate and the bias-hardened reconstructions or the $TE$ reconstruction (with the same colour scheme as in the top panel).
The expected power spectra of the differences, estimated from simulations, are plotted as black dotted lines.
}
\end{figure}

Mean-field corrections are largest at low-$L$ (the power spectrum of the mean-field correction is larger than the fiducial lensing potential power spectrum for $\elp \la 100$),
and so we begin by looking closely at these scales.
In Fig.~\ref{fig:lowl_mean_fields} we plot the power spectra of the mean-fields for each of the bias-hardened estimators.
In the lower panel we also plot the difference between each estimator and the non-bias-hardened $TT$ estimator.
For $L\ge 8$, we see agreement between all of the bias-hardened estimators and the non-bias-hardened reconstruction, regardless of the much smaller mean-field corrections for the former.
We have also used the bias-hardened estimators to measure the lensing potential power spectrum on smaller scales.
In Fig.~\ref{fig:clpp_consistency_tests} and Table~\ref{table:consistency_tests}
we report lensing bandpower estimates using the bias-hardened estimators, which we find to be consistent with the baseline MV reconstruction for $L \leq 400$.

\subsection{Null tests}
\label{sec:consistency:null_tests}
In addition to the consistency tests described above, we have also performed several direct null tests,
either using estimators or maps for which the lensing signal is expected to be zero.
These are plotted in Fig.~\ref{fig:clxx_grid}, and significance statistics for each are given in Table~\ref{table:null_tests}.
We discuss them in more detail below.

\paragraph{Half-ring difference:} The {\it Planck} observation strategy consists of scanning the sky in rings with an opening angle of approximately $85^\circ$.
Each scan ring is observed for between $39$ and $65$ minutes, and then the satellite is repointed.
This leads to a natural `half-ring' null test in which we run our standard lensing estimator on the
half-difference of maps constructed from the
first and second half of each scanning ring, which should consist entirely of noise.
The PTE for the spectrum of the lensing reconstruction estimated from these half-ring difference maps is rather low for the conservative multipole range (see Table~\ref{table:null_tests}). Since the maps should be only noise by construction, to obtain an acceptable $\chi^2$ for the power spectra of reconstructions based on such maps requires very good modeling of the instrument noise. As can be seen from the factor of 200 scaling applied in Fig.~\ref{fig:clxx_grid}, the impact on the reconstructed $\hat{C}_L^{\phi\phi}$ of minor errors in the noise modeling that likely drive the low PTEs for the null test should be negligible.
\paragraph{Half-mission difference:} Noise that is correlated between rings (for example, due to glitches and destriping) can be nulled in the half-ring difference.
Another differencing approach, which avoids this problem, is to divide the entire mission into two halves that are then differenced.
This half-mission difference requires a slightly larger analysis mask, to account for the fact that the sky coverage is imperfect in both the first and second halves of the mission and some pixels are missed. The PTEs for the half-mission test are reasonable.
%
% hack to make paragraph fit on this page
\\[-1.5em]
\paragraph{Curl-mode estimators:} In general, deflections due to lensing can be decomposed into gradient ($\phi$) and curl ($\psi$) modes.
The curl modes are expected to be consistent with zero for lensing by scalar perturbations at {\it Planck} reconstruction noise levels.
We construct curl-mode power spectrum estimators, analogous to those used for the gradient lensing modes, and measure their power spectra.
Full-sky weight functions for these estimators are presented in \cite{Namikawa:2011cs}.
Note that although they do not couple directly to the lensing potential, the curl power spectra do receive `N1'-type contributions from the lensing trispectrum,
which we estimate analytically and subtract as with the $C_{\elp}^{\phi\phi}$ power spectrum estimates.
We do not make `PS' or `MC' corrections for the curl-mode power spectra.
In Table~\ref{table:null_tests} we report estimated amplitudes for a signal with the shape of the fiducial lensing power spectrum $C_{\elp}^{\phi\phi,\ {\rm fid}}$ in the curl estimates.
These amplitudes are all consistent with zero at better than $2\,\sigma$, with the exception of the $TT$ reconstruction over the high-$L$ range ($400 < L \leq 2048$).
Indeed, there is mild evidence for a broad feature in the $TT$
curl-mode reconstruction centred around $L=500$ (see
Fig.~\ref{fig:clxx_grid}), which drives the low PTE of the $TT$
curl-mode power spectrum over the high-$L$ multipole
range. \refchange{Considering $TT$ alone, the average curl power is
  low by about $2.9\,\sigma$ in the high-$L$ range, which is suggestive but not clearly
  unacceptable given the number of tests performed. Combined with the
  polarization reconstructions, which on their own appear consistent,
  the MV estimate remains low but only at $2\,\sigma$ over the
  high-$L$ range.} Since the origin of this feature in the curl-mode
spectrum is currently not understood, we have chosen to adopt the
conservative multipole range for the baseline \planck\ 2015 lensing
likelihood. This multipole range has little overlap with that covered
by the curl-mode feature \refchange{and was determined in advance of
  the curl result as it} retains the majority of the $S/N$ in the reconstruction.

\clearpage
\newlength{\classpagewidth} \setlength{\classpagewidth}{\pdfpagewidth}
\eject \pdfpagewidth=10in
\thispagestyle{empty}
\begin{landscape}
\newcounter{ttonlyb}
\newcommand{\ttonlyb}{\dagger\dagger}
\newcounter{polonlyb}
\newcommand{\polonlyb}{\dagger}
\newcounter{simonlyb}
\newcommand{\simonlyb}{\dagger\dagger\dagger}
\begin{table}[hp]
\begingroup
\newdimen\tblskip \tblskip=5pt
\caption{
Amplitude fits for a suite of alternative reconstructions of the lensing power spectrum and comparison to the baseline MV lensing power spectrum estimate.
The measured overall lensing amplitude $\hat{A}$ is given for the conservative ($\conservativeLrange$) multipole range, as well as the lower signal-to-noise bins on either side.
The difference between the test amplitudes and the corresponding MV result (with amplitudes of
$\hat{A}^{\phi, \,{\rm MV}}_{8 \rightarrow 40} = 1.023 \pm 0.085$,
$\hat{A}^{\phi, \,{\rm MV}}_{40 \rightarrow 400} = 0.987 \pm 0.025$, and
$\hat{A}^{\phi, \,{\rm MV}}_{400 \rightarrow 2048} = 0.788 \pm 0.097$ for the three multipole ranges)
is given as $\Delta_{\rm MV}$.
The quoted uncertainty on the difference is obtained using Monte Carlo simulations and therefore accounts for any correlations between the MV and test estimates.
The PTE values are estimated by comparison of the $\chi^2$ values to the distribution from simulations.
PTEs are given for the square of the overall amplitude difference,
as well as for the $\chi^2$ of the difference between the bandpowers of the binned spectra.
\label{table:consistency_tests}
}
\nointerlineskip
\vskip 1mm
\footnotesize
\setbox\tablebox=\vbox{
 \newdimen\digitwidth
 \setbox0=\hbox{\rm 0}
  \digitwidth=\wd0
  \catcode`*=\active
  \def*{\kern\digitwidth}
  \newdimen\signwidth
  \setbox0=\hbox{+}
  \signwidth=\wd0
  \catcode`!=\active
  \def!{\kern\signwidth}
\halign{\hbox to 3.75cm{$#$\hfil}\tabskip=0.5em&
  \hfil$#$\hfil\tabskip=1.0em&
  \hfil$#$\hfil&
  \hfil$#$\hfil&
  \hfil$#$\hfil\tabskip=1.0em&
  \hfil$#$\hfil&
  \hfil$#$\hfil&
  \hfil$#$\hfil&
  \hfil$#$\hfil\tabskip=1.0em&
  \hfil$#$\hfil&
  \hfil$#$\hfil&
  \hfil$#$\hfil\tabskip=0pt\cr
  \noalign{\vskip -8pt}
\noalign{\doubleline}
\omit&\multispan{11}\hfil Consistency test summary values \hfil\cr
\noalign{\vskip -4pt}
\omit&\multispan{11}\hrulefill \cr
\noalign{\vskip 2pt}
\omit&\multispan3 \hfill Low-$\elp$ \hfill & \multispan4 \hfill Conservative \hfill & \multispan4 \hfill High-$\elp$ \hfill \cr
\noalign{\vskip 2pt}
\omit&\multispan3 \hfill 8 $\le \elp < 40$ \hfill & \multispan4 \hfill $40 \le \elp \le 400$ \hfill & \multispan4 \hfill $400 < \elp \le 2048$ \hfill \cr
\noalign{\vskip 2pt}
\omit&\hat{A}\pm\sigma_{A}&\Delta_{\rm MV} \pm \sigma_{\Delta}&$PTE$_{\Delta_{\rm MV}=0}&\hat{A}\pm\sigma_{A}&\Delta_{\rm MV} \pm \sigma_{\Delta}&$PTE$_{\Delta_{\rm MV}=0}&$PTE$_{\rm bins}&\hat{A}\pm\sigma_{A}&\Delta_{\rm MV} \pm \sigma_{\Delta}&$PTE$_{\Delta_{\rm MV}=0}&$PTE$_{\rm bins}\cr
\noalign{\vskip -4pt}
&\multispan3\hrulefill &\multispan4\hrulefill &\multispan4\hrulefill \cr
\noalign{\vskip 2pt}
\text{\textit{Individual Estimator Crosses}}\hfill\cr
\noalign{\vskip 2pt}
TT\times TT&\phantom{-}1.044\pm0.111&-0.035\pm0.075&0.65&\phantom{-}0.990\pm0.032&-0.003\pm0.023&0.90&0.47&\phantom{-}0.724\pm0.096&\phantom{-}0.058\pm0.042&0.17&0.81\cr
 TT \times TE&\phantom{-}1.000\pm0.108&\phantom{-}0.009\pm0.070&0.91&\phantom{-}0.969\pm0.039&\phantom{-}0.018\pm0.028&0.54&0.12&\phantom{-}0.628\pm0.273&\phantom{-}0.155\pm0.243&0.52&0.52\cr
 TT \times EE&\phantom{-}0.872\pm0.149&\phantom{-}0.137\pm0.126&0.29&\phantom{-}1.012\pm0.055&-0.025\pm0.051&0.65&0.89&\phantom{-}1.616\pm0.448&-0.833\pm0.446&0.06&0.18\cr
 TT \times EB&\phantom{-}1.105\pm0.308&-0.096\pm0.295&0.76&\phantom{-}0.988\pm0.085&-0.001\pm0.081&0.99&0.83&\phantom{-}1.118\pm0.270&-0.335\pm0.269&0.22&0.48\cr
 TE\times TE&\phantom{-}0.890\pm0.194&\phantom{-}0.118\pm0.172&0.47&\phantom{-}1.057\pm0.079&-0.070\pm0.075&0.35&0.94&\phantom{-}3.607\pm1.472&-2.824\pm1.461&0.05&0.41\cr
 EE\times TE&\phantom{-}1.067\pm0.209&-0.058\pm0.194&0.77&\phantom{-}0.995\pm0.085&-0.008\pm0.080&0.93&0.77&-0.441\pm1.961&\phantom{-}1.224\pm1.964&0.54&0.18\cr
 TE\times EB&\phantom{-}0.604\pm0.424&\phantom{-}0.405\pm0.415&0.31&\phantom{-}1.016\pm0.157&-0.029\pm0.156&0.88&0.86&\phantom{-}0.786\pm1.235&-0.003\pm1.223&1.00&0.57\cr
 TT \times TB&\phantom{-}0.496\pm0.631&\phantom{-}0.512\pm0.630&0.42&\phantom{-}0.890\pm0.180&\phantom{-}0.097\pm0.178&0.59&0.76&\phantom{-}1.413\pm0.575&-0.630\pm0.571&0.26&0.06\cr
 EE\times EE&\phantom{-}1.132\pm0.407&-0.124\pm0.397&0.72&\phantom{-}0.748\pm0.189&\phantom{-}0.239\pm0.186&0.20&0.50&\phantom{-}4.503\pm5.723&-3.720\pm5.717&0.49&0.56\cr
 EE\times EB&\phantom{-}1.530\pm0.664&-0.522\pm0.659&0.44&\phantom{-}0.643\pm0.247&\phantom{-}0.344\pm0.246&0.15&0.20&\phantom{-}4.131\pm2.432&-3.348\pm2.429&0.19&0.29\cr
 TE\times TB&\phantom{-}2.426\pm0.837&-1.417\pm0.833&0.09&\phantom{-}0.722\pm0.297&\phantom{-}0.266\pm0.294&0.37&0.75&\phantom{-}5.242\pm2.431&-4.459\pm2.425&0.07&0.20\cr
 EE\times TB&-0.981\pm1.212&\phantom{-}1.990\pm1.209&0.10&\phantom{-}1.288\pm0.463&-0.301\pm0.463&0.52&0.01&\phantom{-}3.154\pm4.492&-2.371\pm4.482&0.61&0.81\cr
 EB\times EB&\phantom{-}0.588\pm1.819&\phantom{-}0.421\pm1.817&0.82&\phantom{-}1.401\pm0.522&-0.414\pm0.521&0.42&0.39&-0.422\pm2.274&\phantom{-}1.205\pm2.263&0.60&0.80\cr
TB\times EB&\phantom{-}0.887\pm2.954&\phantom{-}0.122\pm2.947&0.95&\phantom{-}0.618\pm0.798&\phantom{-}0.369\pm0.798&0.63&0.34&-2.591\pm2.955&\phantom{-}3.374\pm2.945&0.26&0.43\cr
TB\times TB& \phantom{\,}13.790\pm8.353&\!\!\! -12.781\pm8.351&0.12&-2.809\pm2.189&\phantom{-}3.796\pm2.189&0.09&0.08&-0.532\pm8.268&\phantom{-}1.315\pm8.260&0.86&0.20\cr
$Crosses$&\phantom{-}0.958\pm0.097&\phantom{-}0.051\pm0.053&0.33&\phantom{-}0.989\pm0.033&-0.002\pm0.022&0.91&0.65&\phantom{-}1.028\pm0.193&-0.245\pm0.174&0.15&0.84\cr
$Pol.\phantom{-}only\hyperref[polonlyb]{\tablenotemark{\polonlyb}}$&\phantom{-}1.252\pm0.350&-0.243\pm0.339&0.46&\phantom{-}0.761\pm0.145&\phantom{-}0.226\pm0.142&0.11&0.39&\phantom{-}1.787\pm1.598&-1.004\pm1.586&0.53&0.84\cr
\noalign{\vskip 2pt}
\text{\textit{Foreground tests}}\hfill\cr
\noalign{\vskip 2pt}
$143\,GHz$&\phantom{-}0.958\pm0.096&\phantom{-}0.029\pm0.069&0.64&\phantom{-}1.024\pm0.034&-0.034\pm0.024&0.16&0.22&\phantom{-}0.583\pm0.122&\phantom{-}0.198\pm0.093&0.04&0.02\cr
$217\,GHz$&\phantom{-}0.941\pm0.092&\phantom{-}0.046\pm0.077&0.57&\phantom{-}0.963\pm0.034&\phantom{-}0.027\pm0.024&0.30&0.71&\phantom{-}0.929\pm0.109&-0.148\pm0.088&0.08&0.80\cr
${\tt NILC}$&\phantom{-}0.977\pm0.073&\phantom{-}0.010\pm0.027&0.75&\phantom{-}0.988\pm0.026&\phantom{-}0.002\pm0.011&0.83&0.94&\phantom{-}0.877\pm0.082&-0.096\pm0.031&0.01&0.09\cr
${\tt SEVEM}\phantom{-}$TT$\hyperref[ttonlyb]{\tablenotemark{\ttonlyb}}$&\phantom{-}1.015\pm0.108&\phantom{-}0.016\pm0.053&0.78&\phantom{-}1.000\pm0.032&-0.011\pm0.017&0.53&0.63&\phantom{-}0.758\pm0.094&-0.039\pm0.047&0.40&0.73\cr
$\commander \phantom{-}$TT$\hyperref[ttonlyb]{\tablenotemark{\ttonlyb}}$&\phantom{-}0.899\pm0.124&\phantom{-}0.132\pm0.092&0.15&\phantom{-}1.018\pm0.038&-0.029\pm0.028&0.30&0.42&\phantom{-}0.950\pm0.113&-0.230\pm0.099&0.02&0.21\cr
$\mbox{$f_{\rm{sky}}=0.8$}$&\phantom{-}1.009\pm0.064&-0.022\pm0.028&0.36&\phantom{-}0.987\pm0.025&\phantom{-}0.003\pm0.009&0.73&0.14&\phantom{-}0.761\pm0.086&\phantom{-}0.019\pm0.031&0.57&0.22\cr
$\mbox{$f_{\rm{sky}}=0.6$}$&\phantom{-}1.000\pm0.078&-0.014\pm0.034&0.71&\phantom{-}0.991\pm0.026&-0.000\pm0.010&0.97&0.90&\phantom{-}0.777\pm0.091&\phantom{-}0.004\pm0.037&0.92&0.68\cr
$(\smica)$^3$353\,GHz$&\phantom{-}1.000\pm0.133&-0.013\pm0.114&0.93&\phantom{-}0.970\pm0.048&\phantom{-}0.021\pm0.041&0.64&0.76&\phantom{-}0.946\pm0.134&-0.165\pm0.128&0.21&0.05\cr
$(\smica)$^3$353\,GHz\phantom{-}pol.\phantom{-}only\hyperref[polonlyb]{\tablenotemark{\polonlyb}}$&\phantom{-}0.277\pm0.996&\phantom{-}1.016\pm1.076&0.33&\phantom{-}0.264\pm0.451&\phantom{-}0.531\pm0.495&0.27&0.80&\phantom{-}7.045\pm5.895&-5.056\pm6.440&0.40&0.31\cr
\noalign{\vskip 2pt}
\text{\textit{Noise tests}}\hfill\cr
\noalign{\vskip 2pt}
143\times 217\,{\rm GHz}&\phantom{-}0.949\pm0.077&\phantom{-}0.038\pm0.050&0.44&\phantom{-}0.999\pm0.027&-0.008\pm0.015&0.61&0.97&\phantom{-}0.755\pm0.095&\phantom{-}0.025\pm0.056&0.66&0.94\cr
\text{HM1}\times\text{HM2}&\phantom{-}1.028\pm0.081&-0.041\pm0.050&0.39&\phantom{-}0.981\pm0.029&\phantom{-}0.009\pm0.016&0.63&0.67&\phantom{-}0.827\pm0.090&-0.047\pm0.058&0.40&0.88\cr
\text{Noise scaling}\hyperref[simonlyb]{\tablenotemark{\simonlyb}}&\phantom{-}0.989\pm0.070&-0.002\pm0.030&0.93&\phantom{-}0.994\pm0.026&-0.003\pm0.011&0.73&1.00&\phantom{-}0.779\pm0.084&\phantom{-}0.001\pm0.037&0.94&1.00\cr
\noalign{\vskip 2pt}
\text{\textit{Analysis choices}}\hfill\cr
\noalign{\vskip 2pt}
$\mbox{$\ell_{\rm{min}}=1000$}$&\phantom{-}1.136\pm0.104&-0.149\pm0.081&0.06&\phantom{-}0.990\pm0.036&\phantom{-}0.000\pm0.029&1.00&0.08&\phantom{-}0.718\pm0.153&\phantom{-}0.063\pm0.128&0.57&0.44\cr
$\mbox{$\ell_{\rm{max}}=1500$}$&\phantom{-}1.013\pm0.086&-0.026\pm0.060&0.67&\phantom{-}0.985\pm0.035&\phantom{-}0.006\pm0.022&0.76&0.19&\phantom{-}0.625\pm0.181&\phantom{-}0.156\pm0.160&0.41&0.31\cr
$Apodization$&\phantom{-}1.009\pm0.088&-0.001\pm0.023&0.98&\phantom{-}0.985\pm0.026&\phantom{-}0.002\pm0.008&0.85&0.58&\phantom{-}0.753\pm0.097&\phantom{-}0.030\pm0.042&0.50&0.91\cr
\noalign{\vskip 2pt}
\text{\textit{Bias-hardened estimators}}\hfill\cr
\noalign{\vskip 2pt}
$$\hat{\phi}^{TT}-{\rm M}$\hyperref[ttonlyb]{\tablenotemark{\ttonlyb}}$&\phantom{-}1.117\pm0.120&-0.073\pm0.054&0.17&\phantom{-}0.968\pm0.039&\phantom{-}0.022\pm0.021&0.30&0.33&\phantom{-}0.451\pm0.136&\phantom{-}0.273\pm0.098&0.01&0.11\cr
$$\hat{\phi}^{TT}-{\rm N}$\hyperref[ttonlyb]{\tablenotemark{\ttonlyb}}$&\phantom{-}1.068\pm0.116&-0.024\pm0.036&0.52&\phantom{-}0.986\pm0.036&\phantom{-}0.004\pm0.014&0.79&0.79&\phantom{-}0.651\pm0.103&\phantom{-}0.074\pm0.035&0.04&0.09\cr
$$\hat{\phi}^{TT}-{\rm S}$\hyperref[ttonlyb]{\tablenotemark{\ttonlyb}}$&\phantom{-}1.082\pm0.118&-0.038\pm0.045&0.43&\phantom{-}0.982\pm0.038&\phantom{-}0.008\pm0.017&0.64&0.62&\phantom{-}0.617\pm0.107&\phantom{-}0.107\pm0.045&0.01&0.06\cr
\noalign{\vskip -6pt}
}
 \tablenotetext{\polonlyb}{{\refstepcounter{polonlyb}{}\label{polonlyb}{Comparison is against the {\tt SMICA} polarization-only reconstruction, rather than the MV reconstruction. }}}
 \tablenotetext{\ttonlyb}{{\refstepcounter{ttonlyb}{}\label{ttonlyb}{Comparison is against the {\tt SMICA} TT reconstruction, rather than the MV reconstruction.}}}
 \tablenotetext{\simonlyb}{{\refstepcounter{simonlyb}{}\label{simonlyb}{This test only involves a change in the simulations used to perform the debiasing corrections and estimate error bars. We have quoted the scatter between the two sets of simulations, although this is an upper limit on the difference that we should actually see for the data.}}}
}
\endPlancktable
\endgroup
\end{table}
\end{landscape}
\eject \pdfpagewidth=\classpagewidth
\clearpage

\clearpage
\thispagestyle{empty}
\begin{landscape}
\begin{table}[hp]
\begingroup
\newdimen\tblskip \tblskip=5pt
\caption{
Null tests for a set of lens reconstructions based on maps that should contain no CMB signal, or estimators that should be orthogonal to the lensing signal.
The measured overall lensing amplitude $\hat{A}$ is given for the conservative ($\conservativeLrange$) multipole range, as well as lower and higher multipoles. All should be consistent with zero.
The PTE values are estimated by comparison of the $\chi^2$ values to the distributions from simulations.
PTEs are given for the difference of the overall amplitude from zero,
as well as for the $\chi^2$ of the difference from zero of the binned spectra.
\label{table:null_tests}
}
\nointerlineskip
\vskip 1mm
\footnotesize
\setbox\tablebox=\vbox{
 \newdimen\digitwidth
 \setbox0=\hbox{\rm 0}
  \digitwidth=\wd0
  \catcode`*=\active
  \def*{\kern\digitwidth}
  \newdimen\signwidth
  \setbox0=\hbox{+}
  \signwidth=\wd0
  \catcode`!=\active
  \def!{\kern\signwidth}
\halign{\hbox to 3.75cm{$#$}\tabskip=0.5em&
  \hfil$#$\hfil\tabskip=1.0em&
  \hfil$#$\hfil&
  \hfil$#$\hfil&
  \hfil$#$\hfil\tabskip=1.0em&
  \hfil$#$\hfil&
  \hfil$#$\hfil&
  \hfil$#$\hfil&
  \hfil$#$\hfil\tabskip=1.0em&
  \hfil$#$\hfil&
  \hfil$#$\hfil&
  \hfil$#$\hfil\tabskip=0pt\cr
  \noalign{\vskip 0pt}
\noalign{\doubleline}
\omit&\multispan{11}\hfil Null test summary values \hfil\cr
\noalign{\vskip -4pt}
\omit&\multispan{11}\hrulefill \cr
\noalign{\vskip 2pt}
\omit&\multispan3 \hfill Low-$\elp$ \hfill & \multispan4 \hfill Conservative \hfill & \multispan4 \hfill High-$\elp$ \hfill \cr
\noalign{\vskip 2pt}
\omit&\multispan3 \hfill 8 $\le \elp < 40$ \hfill & \multispan4 \hfill $40 \le \elp \le 400$ \hfill & \multispan4 \hfill $400 < \elp \le 2048$ \hfill \cr
\noalign{\vskip 2pt}
\omit&\multispan2 \hfill $\hat{A}\pm\sigma_{A}$ \hfill&$PTE$_{\hat{A}=0}&\multispan2 \hfill $\hat{A}\pm\sigma_{A}$ \hfill&$PTE$_{\hat{A}=0}&$PTE$_{\rm bins}&\multispan2 \hfill $\hat{A}\pm\sigma_{A}$ \hfill&$PTE$_{\hat{A}=0}&$PTE$_{\rm bins}\cr
\noalign{\vskip -4pt}
&\multispan3\hrulefill &\multispan4\hrulefill &\multispan4\hrulefill \cr
$Half-ring\phantom{-}difference$&\multispan2\phantom{-}$(-0.116\pm0.371)\times10^{-2}$\hfill&0.78&\multispan2\phantom{-}$(-0.331\pm0.154)\times10^{-2}$\hfill&0.03&0.04&\multispan2\phantom{-}$(-0.214\pm0.434)\times10^{-2}$\hfill&0.54&0.18\cr
$Half-mission\phantom{-}difference$&\multispan2\phantom{-}$(-0.282\pm0.565)\times10^{-2}$\hfill&0.63&\multispan2\phantom{-}$(-0.215\pm0.177)\times10^{-2}$\hfill&0.21&0.39&\multispan2\phantom{-}$(\phantom{-}0.075\pm0.545)\times10^{-2}$\hfill&0.93&0.83\cr
$$\psi^{\rm MV}$$&\multispan2\phantom{-}$(\phantom{-}0.658\pm0.843)\times10^{-1}$\hfill&0.43&\multispan2\phantom{-}$(-0.099\pm0.290)\times10^{-1}$\hfill&0.74&0.46&\multispan2\phantom{-}$(-1.891\pm0.967)\times10^{-1}$\hfill&0.05&0.41\cr
$$\psi^{TT}$$&\multispan2\phantom{-}$(-0.241\pm1.413)\times10^{-1}$\hfill&0.87&\multispan2\phantom{-}$(-0.415\pm0.428)\times10^{-1}$\hfill&0.34&0.27&\multispan2\phantom{-}$(-3.034\pm1.047)\times10^{-1}$\hfill&0.00&0.03\cr
$$\psi^{EE}$$&\multispan2\phantom{-}$(\phantom{-}1.138\pm1.075)\times10^{0}$\hfill&0.29&\multispan2\phantom{-}$(-0.266\pm0.463)\times10^{0}$\hfill&0.59&0.66&\multispan2\phantom{-}$(\phantom{-}7.748\pm13.27)\times10^{0}$\hfill&0.53&0.56\cr
$$\psi^{TE}$$&\multispan2\phantom{-}$(\phantom{-}0.214\pm0.463)\times10^{0}$\hfill&0.64&\multispan2\phantom{-}$(-0.249\pm0.185)\times10^{0}$\hfill&0.18&0.25&\multispan2\phantom{-}$(-3.183\pm1.938)\times10^{0}$\hfill&0.11&0.75\cr
$$\psi^{EB}$$&\multispan2\phantom{-}$(\phantom{-}0.484\pm0.655)\times10^{0}$\hfill&0.46&\multispan2\phantom{-}$(\phantom{-}0.057\pm0.213)\times10^{0}$\hfill&0.79&0.41&\multispan2\phantom{-}$(\phantom{-}1.014\pm1.222)\times10^{0}$\hfill&0.43&0.97\cr
$$\psi^{TB}$$&\multispan2\phantom{-}$(\phantom{-}1.352\pm2.563)\times10^{0}$\hfill&0.60&\multispan2\phantom{-}$(\phantom{-}0.378\pm0.860)\times10^{0}$\hfill&0.63&0.81&\multispan2\phantom{-}$(\phantom{-}6.597\pm5.144)\times10^{0}$\hfill&0.21&0.24\cr
\noalign{\vskip 4pt\hrule\vskip 3pt}
}
}
\endPlancktable
\endgroup
\end{table}
\end{landscape}
\clearpage

\section{Conclusions}
\label{sec:conclusions}
We have presented a first analysis of gravitational lensing in the \Planck\ 2015 full-mission data set. Using temperature and polarization data, we make the most powerful measurement of CMB lensing to date,
with an overall significance of around $40\,\sigma$.
This paper is accompanied by the public release of the lensing potential map
(as well as simulations that can be used to characterize the map, e.g. for cross-correlation analyses with external data)
and a conservative lensing bandpower likelihood.

We have performed a number of tests to verify the internal consistency of the lens reconstruction,
in particular we have tested the compatibility between individual temperature and polarization estimators,
sensitivity to various analysis choices,
and possible foreground contamination.

Given {\it Planck}'s unique full-sky coverage,
the full-mission maps are unlikely to be displaced as the most powerful data set for
cosmological lensing analysis for the next several years
(until high signal-to-noise $E$- and $B$-mode polarization measurements can be made on sufficiently large sky fractions).
There is some room, however, for improvement on the analysis presented here.
The inverse-variance filtering we have used does not account for variations in the instrument noise level across the sky,
which could be incorporated to reduce the reconstruction noise (particularly in polarization, where instrumental noise dominates the signal).
We have high-pass filtered all results to $\elp \ge 8$, due to concerns about the lensing mean-field correction.
A more aggressive analysis could potentially remove this cut.
The conservative $L \le 400$ cut that we make in our baseline likelihood, imposed because of mild evidence of a broad feature in curl-mode tests beyond this range, might also be relaxed in a future analysis, although only with minor improvements in the statistical power of the likelihood.

The microwave background is a unique source for lensing studies, as the most distant and well-understood source plane that can be observed.
The measurement presented here represents an integrated measurement of the total matter distribution
in the entire observable Universe, and another powerful test of the veracity of the concordance $\Lambda$CDM cosmological model. 

\begin{acknowledgements}
The Planck Collaboration acknowledges the support of: ESA; CNES and CNRS/INSU-IN2P3-INP (France); ASI, CNR, and INAF (Italy); NASA and DoE (USA); STFC and UKSA (UK); CSIC, MINECO, JA, and RES (Spain); Tekes, AoF, and CSC (Finland); DLR and MPG (Germany); CSA (Canada); DTU Space (Denmark); SER/SSO (Switzerland); RCN (Norway); SFI (Ireland); FCT/MCTES (Portugal); ERC and PRACE (EU). A description of the Planck Collaboration and a list of its members, indicating which technical or scientific activities they have been involved in, can be found at \href{http://www.cosmos.esa.int/web/planck/planck-collaboration}{http://www.cosmos.esa.int/web/planck/planck-collaboration}.
 Some of the results in this paper have been derived using the HEALPix package. %\citep{gorski2005}.
 We acknowledge support from the Science and Technology Facilities Council [grant number  ST/L000652/1].
The research leading to these results has received funding from the European Research Council under the European Union's Seventh Framework Programme (FP/2007-2013) / ERC Grant Agreement No. [616170].
Part of this work was undertaken on the STFC DiRAC HPC Facilities at the University of Cambridge funded by UK BIS National E-infrastructure capital grants.

 \end{acknowledgements}

\bibliographystyle{aat}%hacked version of aa.bst to include titles.
\bibliography{planck_lensing_2015,Planck_bib,parameters,cosmomc}

\def\eprinttmppp@#1arXiv:@{#1}
\providecommand{\arxivlink[1]}{\href{http://arxiv.org/abs/#1}{arXiv:#1}}
\def\eprinttmp@#1arXiv:#2 [#3]#4@{\ifthenelse{\equal{#3}{x}}{\ifthenelse{
\equal{#1}{}}{\arxivlink{\eprinttmppp@#2@}}{\arxivlink{#1}}}{\arxivlink{#2}
  [#3]}}
\providecommand{\eprintlink}[1]{\eprinttmp@#1arXiv: [x]@}
\providecommand{\eprint}[1]{\eprintlink{#1}}
\providecommand{\adsurl}[1]{\href{#1}{ADS}}
\begin{thebibliography}{60}
\expandafter\ifx\csname natexlab\endcsname\relax\def\natexlab#1{#1}\fi

\bibitem[{Ade {et~al.}(2014)}]{Ade:2014xna}
Ade, P. {et~al.}, {Detection of $B$-Mode Polarization at Degree Angular Scales
  by BICEP2}. 2014, Phys. Rev. Lett., 112, 241101, \eprint{1403.3985}

\bibitem[{Anderson {et~al.}(2014)}]{Anderson:2013zyy}
Anderson, L. {et~al.}, {The clustering of galaxies in the SDSS-III Baryon
  Oscillation Spectroscopic Survey: Baryon Acoustic Oscillations in the Data
  Release 10 and 11 galaxy samples}. 2014, \mnras, 441, 24, \eprint{1312.4877}

\bibitem[{Beutler {et~al.}(2011)Beutler, Blake, Colless, Jones, Staveley-Smith,
  {et~al.}}]{Beutler:2011hx}
Beutler, F., Blake, C., Colless, M., {et~al.}, {The 6dF Galaxy Survey: Baryon
  Acoustic Oscillations and the Local Hubble Constant}. 2011,
  Mon.Not.Roy.Astron.Soc., 416, 3017, \eprint{1106.3366}

\bibitem[{{BICEP2/Keck Array and Planck Collaborations}(2015)}]{pb2015}
{BICEP2/Keck Array and Planck Collaborations}, {Joint Analysis of BICEP2/Keck
  Array and Planck Data}. 2015, \prl, 114, 101301, \eprint{1502.00612}

\bibitem[{{Bucher} {et~al.}(2012){Bucher}, {Carvalho}, {Moodley}, \&
  {Remazeilles}}]{2012PhRvD..85d3016B}
{Bucher}, M., {Carvalho}, C.~S., {Moodley}, K., \& {Remazeilles}, M., {CMB
  lensing reconstruction in real space}. 2012, Phys. Rev., D85, 043016,
  \eprint{1004.3285}

\bibitem[{Bunn {et~al.}(1994)Bunn, Fisher, Hoffman, Lahav, Silk,
  {et~al.}}]{Bunn:1994xn}
Bunn, E.~F., Fisher, K.~B., Hoffman, Y., {et~al.}, {Wiener filtering of the
  COBE Differential Microwave Radiometer data}. 1994, ApJ, 432, L75,
  \eprint{astro-ph/9404007}

\bibitem[{Das {et~al.}(2014)Das, Louis, Nolta, Addison, Battistelli,
  {et~al.}}]{Das:2013zf}
Das, S., Louis, T., Nolta, M.~R., {et~al.}, {The Atacama Cosmology Telescope:
  temperature and gravitational lensing power spectrum measurements from three
  seasons of data}. 2014, JCAP, 1404, 014, \eprint{1301.1037}

\bibitem[{Das {et~al.}(2011)Das, Sherwin, Aguirre, Appel, Bond,
  {et~al.}}]{Das:2011ak}
Das, S., Sherwin, B.~D., Aguirre, P., {et~al.}, {Detection of the Power
  Spectrum of Cosmic Microwave Background Lensing by the Atacama Cosmology
  Telescope}. 2011, Phys. Rev. Lett., 107, 021301, \eprint{1103.2124}

\bibitem[{Dvorkin \& Smith(2009)}]{Dvorkin:2008tf}
Dvorkin, C. \& Smith, K.~M., {Reconstructing Patchy Reionization from the
  Cosmic Microwave Background}. 2009, Phys. Rev., D79, 043003,
  \eprint{0812.1566}

\bibitem[{Efstathiou(2014)}]{Efstathiou:2013via}
Efstathiou, G., {H0 Revisited}. 2014, MNRAS, 440, 1138, \eprint{1311.3461}

\bibitem[{Eisenstein \& Hu(1998)}]{Eisenstein:1997ik}
Eisenstein, D.~J. \& Hu, W., {Baryonic features in the matter transfer
  function}. 1998, ApJ, 496, 605, \eprint{astro-ph/9709112}

\bibitem[{Fan {et~al.}(2006)Fan, Strauss, Becker, White, Gunn,
  {et~al.}}]{Fan:2005es}
Fan, X.-H., Strauss, M.~A., Becker, R.~H., {et~al.}, {Constraining the
  evolution of the ionizing background and the epoch of reionization with z~6
  quasars. 2. a sample of 19 quasars}. 2006, AJ, 132, 117,
  \eprint{astro-ph/0512082}

\bibitem[{{Finkelstein} {et~al.}(2015){Finkelstein}, {Ryan}, {Papovich},
  {Dickinson}, {Song}, {Somerville}, {Ferguson}, {Salmon}, {Giavalisco},
  {Koekemoer}, {Ashby}, {Behroozi}, {Castellano}, {Dunlop}, {Faber}, {Fazio},
  {Fontana}, {Grogin}, {Hathi}, {Jaacks}, {Kocevski}, {Livermore}, {McLure},
  {Merlin}, {Mobasher}, {Newman}, {Rafelski}, {Tilvi}, \&
  {Willner}}]{2014arXiv1410.5439F}
{Finkelstein}, S.~L., {Ryan}, Jr., R.~E., {Papovich}, C., {et~al.}, {The
  Evolution of the Galaxy Rest-frame Ultraviolet Luminosity Function over the
  First Two Billion Years}. 2015, \apj, 810, 71, \eprint{1410.5439}

\bibitem[{Gispert {et~al.}(2000)Gispert, Lagache, \& Puget}]{Gispert:2000np}
Gispert, R., Lagache, G., \& Puget, J., {Implications of the cosmic infrared
  background for light production and the star formation history in the
  universe}. 2000, Astron.Astrophys., 360, 1, \eprint{astro-ph/0005554}

\bibitem[{{G{\'o}rski} {et~al.}(2005){G{\'o}rski}, {Hivon}, {Banday},
  {Wandelt}, {Hansen}, {Reinecke}, \& {Bartelmann}}]{gorski2005}
{G{\'o}rski}, K.~M., {Hivon}, E., {Banday}, A.~J., {et~al.}, {HEALPix: A
  Framework for High-Resolution Discretization and Fast Analysis of Data
  Distributed on the Sphere}. 2005, \apj, 622, 759, \eprint{astro-ph/0409513}

\bibitem[{Hall {et~al.}(2010)Hall, Knox, Reichardt, Ade, Aird,
  {et~al.}}]{Hall:2009rv}
Hall, N., Knox, L., Reichardt, C., {et~al.}, {Angular Power Spectra of the
  Millimeter Wavelength Background Light from Dusty Star-forming Galaxies with
  the South Pole Telescope}. 2010, ApJ, 718, 632, \eprint{0912.4315}

\bibitem[{{Hanson} {et~al.}(2011){Hanson}, {Challinor}, {Efstathiou}, \&
  {Bielewicz}}]{2011PhRvD..83d3005H}
{Hanson}, D., {Challinor}, A., {Efstathiou}, G., \& {Bielewicz}, P., {CMB
  temperature lensing power reconstruction}. 2011, Phys. Rev., D83, 043005,
  \eprint{1008.4403}

\bibitem[{Hanson {et~al.}(2013)}]{Hanson:2013hsb}
Hanson, D. {et~al.}, {Detection of B-mode Polarization in the Cosmic Microwave
  Background with Data from the South Pole Telescope}. 2013, Phys. Rev. Lett.,
  111, 141301, \eprint{1307.5830}

\bibitem[{Hinshaw {et~al.}(2013)}]{Hinshaw:2012aka}
Hinshaw, G. {et~al.}, {Nine-Year Wilkinson Microwave Anisotropy Probe (WMAP)
  Observations: Cosmological Parameter Results}. 2013, ApJS, 208, 19,
  \eprint{1212.5226}

\bibitem[{Hirata {et~al.}(2008)Hirata, Ho, Padmanabhan, Seljak, \&
  Bahcall}]{Hirata:2008cb}
Hirata, C.~M., Ho, S., Padmanabhan, N., Seljak, U., \& Bahcall, N.~A.,
  {Correlation of CMB with large-scale structure: II. Weak lensing}. 2008,
  Phys. Rev., D78, 043520, \eprint{0801.0644}

\bibitem[{Hu(2001)}]{Hu:2001fa}
Hu, W., {Angular trispectrum of the CMB}. 2001, Phys. Rev., D64, 083005,
  \eprint{astro-ph/0105117}

\bibitem[{Hu(2002)}]{Hu:2001fb}
Hu, W., {Dark synergy: Gravitational lensing and the CMB}. 2002, Phys. Rev.,
  D65, 023003, \eprint{astro-ph/0108090}

\bibitem[{Hu \& Okamoto(2002)}]{Hu:2001kj}
Hu, W. \& Okamoto, T., {Mass reconstruction with {CMB} polarization}. 2002,
  ApJ, 574, 566, \eprint{astro-ph/0111606}

\bibitem[{Humphreys {et~al.}(2013)Humphreys, Reid, Moran, Greenhill, \&
  Argon}]{Humphreys:2013eja}
Humphreys, E., Reid, M.~J., Moran, J.~M., Greenhill, L.~J., \& Argon, A.~L.,
  {Toward a New Geometric Distance to the Active Galaxy NGC 4258. III. Final
  Results and the Hubble Constant}. 2013, ApJ, 775, 13, \eprint{1307.6031}

\bibitem[{Kesden {et~al.}(2003)Kesden, Cooray, \& Kamionkowski}]{Kesden:2003cc}
Kesden, M.~H., Cooray, A., \& Kamionkowski, M., {Lensing reconstruction with
  CMB temperature and polarization}. 2003, Phys. Rev., D67, 123507,
  \eprint{astro-ph/0302536}

\bibitem[{Lewis \& Bridle(2002)}]{Lewis:2002ah}
Lewis, A. \& Bridle, S., {Cosmological parameters from CMB and other data: a
  Monte- Carlo approach}. 2002, Phys. Rev., D66, 103511,
  \eprint{astro-ph/0205436}

\bibitem[{Lewis \& Challinor(2006)}]{Lewis:2006fu}
Lewis, A. \& Challinor, A., {Weak gravitational lensing of the {CMB}}. 2006,
  Phys. Rept., 429, 1, \eprint{astro-ph/0601594}

\bibitem[{{Lewis} {et~al.}(2011){Lewis}, {Challinor}, \&
  {Hanson}}]{2011JCAP...03..018L}
{Lewis}, A., {Challinor}, A., \& {Hanson}, D., {The shape of the CMB lensing
  bispectrum}. 2011, \jcap, 3, 18, \eprint{1101.2234}

\bibitem[{Lewis {et~al.}(2000)Lewis, Challinor, \& Lasenby}]{Lewis:1999bs}
Lewis, A., Challinor, A., \& Lasenby, A., Efficient Computation of {CMB}
  anisotropies in closed {FRW} models. 2000, Astrophys. J., 538, 473,
  \eprint{astro-ph/9911177}

\bibitem[{Limber(1954)}]{Limber1954}
Limber, D.~N., {Analysis of counts of the extragalactic nebulae in terms of a
  fluctuating density field II}. 1954, ApJ, 119:655-681

\bibitem[{Namikawa {et~al.}(2013)Namikawa, Hanson, \&
  Takahashi}]{Namikawa:2012pe}
Namikawa, T., Hanson, D., \& Takahashi, R., {Bias-Hardened CMB Lensing}. 2013,
  MNRAS, 431, 609, \eprint{1209.0091}

\bibitem[{Namikawa {et~al.}(2012)Namikawa, Yamauchi, \&
  Taruya}]{Namikawa:2011cs}
Namikawa, T., Yamauchi, D., \& Taruya, A., {Full-sky lensing reconstruction of
  gradient and curl modes from CMB maps}. 2012, \jcap, 1201, 007,
  \eprint{1110.1718}

\bibitem[{Okamoto \& Hu(2003)}]{Okamoto:2003zw}
Okamoto, T. \& Hu, W., {CMB lensing reconstruction on the full sky}. 2003,
  Phys. Rev., D67, 083002, \eprint{astro-ph/0301031}

\bibitem[{Osborne {et~al.}(2014)Osborne, Hanson, \& Dor{\'e}}]{Osborne:2013nna}
Osborne, S.~J., Hanson, D., \& Dor{\'e}, O., {Extragalactic Foreground
  Contamination in Temperature-based CMB Lens Reconstruction}. 2014, \jcap,
  1403, 024, \eprint{1310.7547}

\bibitem[{Pan {et~al.}(2014)Pan, Knox, \& White}]{Pan:2014xua}
Pan, Z., Knox, L., \& White, M., {Dependence of the Cosmic Microwave Background
  Lensing Power Spectrum on the Matter Density}. 2014, MNRAS, 445, 2941,
  \eprint{1406.5459}

\bibitem[{Pettini \& Cooke(2012)}]{Pettini:2012ph}
Pettini, M. \& Cooke, R., {A new, precise measurement of the primordial
  abundance of Deuterium}. 2012, MNRAS, 425, 2477, \eprint{1205.3785}

\bibitem[{{\sorthelp{Planck Collaboration 2014Q}}{Planck Collaboration
  XVII}(2014)}]{planck2013-p12}
{\sorthelp{Planck Collaboration 2014Q}}{Planck Collaboration XVII},
  {\textit{Planck} 2013 results. XVII. Gravitational lensing by large-scale
  structure}. 2014, \aap, 571, A17, \eprint{1303.5077}

\bibitem[{{\sorthelp{Planck Collaboration 2014R}}{Planck Collaboration
  XVIII}(2014)}]{planck2013-p13}
{\sorthelp{Planck Collaboration 2014R}}{Planck Collaboration XVIII},
  {\textit{Planck} 2013 results. XVIII. The gravitational lensing-infrared
  background correlation}. 2014, \aap, 571, A18, \eprint{1303.5078}

\bibitem[{{\sorthelp{Planck Collaboration 2014ZD}}{Planck Collaboration
  XXIX}(2014)}]{planck2013-p05a}
{\sorthelp{Planck Collaboration 2014ZD}}{Planck Collaboration XXIX},
  {\textit{Planck} 2013 results. XXIX. The Planck catalogue of
  Sunyaev-Zeldovich sources}. 2014, \aap, 571, A29, \eprint{1303.5089}

\bibitem[{{\sorthelp{Planck Collaboration 2015A}}{Planck Collaboration
  I}(2016)}]{planck2014-a01}
{\sorthelp{Planck Collaboration 2015A}}{Planck Collaboration I},
  {\textit{Planck} 2015 results. I. Overview of products and results}. 2016,
  \aap, 594, A1, \eprint{1502.01582}

\bibitem[{{\sorthelp{Planck Collaboration 2015G}}{Planck Collaboration
  VII}(2016)}]{planck2014-a08}
{\sorthelp{Planck Collaboration 2015G}}{Planck Collaboration VII},
  {\textit{Planck} 2015 results. VII. High Frequency Instrument data
  processing: Time-ordered information and beam processing}. 2016, \aap, 594,
  A7, \eprint{1502.01586}

\bibitem[{{\sorthelp{Planck Collaboration 2015H}}{Planck Collaboration
  VIII}(2016)}]{planck2014-a09}
{\sorthelp{Planck Collaboration 2015H}}{Planck Collaboration VIII},
  {\textit{Planck} 2015 results. VIII. High Frequency Instrument data
  processing: Calibration and maps}. 2016, \aap, 594, A8, \eprint{1502.01587}

\bibitem[{{\sorthelp{Planck Collaboration 2015I}}{Planck Collaboration
  IX}(2016)}]{planck2014-a11}
{\sorthelp{Planck Collaboration 2015I}}{Planck Collaboration IX},
  {\textit{Planck} 2015 results. IX. Diffuse component separation: CMB maps}.
  2016, \aap, 594, A9, \eprint{1502.05956}

\bibitem[{{\sorthelp{Planck Collaboration 2015K}}{Planck Collaboration
  XI}(2016)}]{planck2014-a13}
{\sorthelp{Planck Collaboration 2015K}}{Planck Collaboration XI},
  {\textit{Planck} 2015 results. XI. CMB power spectra, likelihoods, and
  robustness of parameters}. 2016, \aap, 594, A11, \eprint{1507.02704}

\bibitem[{{\sorthelp{Planck Collaboration 2015L}}{Planck Collaboration
  XII}(2016)}]{planck2014-a14}
{\sorthelp{Planck Collaboration 2015L}}{Planck Collaboration XII},
  {\textit{Planck} 2015 results. XII. Full Focal Plane simulations}. 2016,
  \aap, 594, A12, \eprint{1509.06348}

\bibitem[{{\sorthelp{Planck Collaboration 2015M}}{Planck Collaboration
  XIII}(2016)}]{planck2014-a15}
{\sorthelp{Planck Collaboration 2015M}}{Planck Collaboration XIII},
  {\textit{Planck} 2015 results. XIII. Cosmological parameters}. 2016, \aap,
  594, A13, \eprint{1502.01589}

\bibitem[{{\sorthelp{Planck Collaboration 2015N}}{Planck Collaboration
  XIV}(2016)}]{planck2014-a16}
{\sorthelp{Planck Collaboration 2015N}}{Planck Collaboration XIV},
  {\textit{Planck} 2015 results. XIV. Dark energy and modified gravity}. 2016,
  \aap, 594, A14, \eprint{1502.01590}

\bibitem[{{\sorthelp{Planck Collaboration 2015U}}{Planck Collaboration
  XXI}(2016)}]{planck2014-a26}
{\sorthelp{Planck Collaboration 2015U}}{Planck Collaboration XXI},
  {\textit{Planck} 2015 results. XXI. The integrated Sachs-Wolfe effect}. 2016,
  \aap, 594, A21, \eprint{1502.01595}

\bibitem[{{POLARBEAR Collaboration}(2014{\natexlab{a}})}]{Ade:2013hjl}
{POLARBEAR Collaboration}, {Evidence for Gravitational Lensing of the Cosmic
  Microwave Background Polarization from Cross-correlation with the Cosmic
  Infrared Background}. 2014{\natexlab{a}}, Phys.Rev.Lett., 112, 131302,
  \eprint{1312.6645}

\bibitem[{{POLARBEAR Collaboration}(2014{\natexlab{b}})}]{Ade:2013gez}
{POLARBEAR Collaboration}, {Measurement of the Cosmic Microwave Background
  Polarization Lensing Power Spectrum with the POLARBEAR experiment}.
  2014{\natexlab{b}}, Phys.Rev.Lett., 113, 021301, \eprint{1312.6646}

\bibitem[{Regan {et~al.}(2010)Regan, Shellard, \& Fergusson}]{Regan:2010cn}
Regan, D., Shellard, E., \& Fergusson, J., {General CMB and Primordial
  Trispectrum Estimation}. 2010, Phys. Rev., D82, 023520, \eprint{1004.2915}

\bibitem[{Riess {et~al.}(2011)Riess, Macri, Casertano, Lampeitl, Ferguson,
  {et~al.}}]{Riess:2011yx}
Riess, A.~G., Macri, L., Casertano, S., {et~al.}, {A 3% Solution: Determination
  of the Hubble Constant with the Hubble Space Telescope and Wide Field Camera
  3}. 2011, Astrophys.J., 730, 119, \eprint{1103.2976}

\bibitem[{Ross {et~al.}(2015)Ross, Samushia, Howlett, Percival, Burden, \&
  Manera}]{Ross:2014qpa}
Ross, A.~J., Samushia, L., Howlett, C., {et~al.}, {The clustering of the SDSS
  DR7 main Galaxy sample – I. A 4 per cent distance measure at $z = 0.15$}.
  2015, Mon. Not. Roy. Astron. Soc., 449, 835, \eprint{1409.3242}

\bibitem[{Schmittfull {et~al.}(2013)Schmittfull, Challinor, Hanson, \&
  Lewis}]{Schmittfull:2013uea}
Schmittfull, M.~M., Challinor, A., Hanson, D., \& Lewis, A., {Joint analysis of
  CMB temperature and lensing-reconstruction power spectra}. 2013, Phys. Rev.,
  D88, 063012, \eprint{1308.0286}

\bibitem[{Smith {et~al.}(2009)Smith, Cooray, Das, Dore, Hanson,
  {et~al.}}]{Smith:2008an}
Smith, K.~M., Cooray, A., Das, S., {et~al.}, {CMBPol Mission Concept Study:
  Gravitational Lensing}. 2009, AIP Conf. Proc., 1141, 121, \eprint{0811.3916}

\bibitem[{Smith {et~al.}(2007)Smith, Zahn, Dore, \& Nolta}]{Smith:2007rg}
Smith, K.~M., Zahn, O., Dore, O., \& Nolta, M.~R., {Detection of Gravitational
  Lensing in the Cosmic Microwave Background}. 2007, Phys. Rev., D76, 043510,
  \eprint{0705.3980}

\bibitem[{Song {et~al.}(2003)Song, Cooray, Knox, \& Zaldarriaga}]{Song:2002sg}
Song, Y.-S., Cooray, A., Knox, L., \& Zaldarriaga, M., {The Far-infrared
  background correlation with CMB lensing}. 2003, ApJ, 590, 664,
  \eprint{astro-ph/0209001}

\bibitem[{van Engelen {et~al.}(2014)van Engelen, Bhattacharya, Sehgal, Holder,
  Zahn, \& Nagai}]{vanEngelen:2013rla}
van Engelen, A., Bhattacharya, S., Sehgal, N., {et~al.}, {CMB Lensing Power
  Spectrum Biases from Galaxies and Clusters using High-angular Resolution
  Temperature Maps}. 2014, Astrophys. J., 786, 13, \eprint{1310.7023}

\bibitem[{van Engelen {et~al.}(2012)van Engelen, Keisler, Zahn, Aird, Benson,
  {et~al.}}]{vanEngelen:2012va}
van Engelen, A., Keisler, R., Zahn, O., {et~al.}, {A measurement of
  gravitational lensing of the microwave background using South Pole Telescope
  data}. 2012, ApJ, 756, 142, \eprint{1202.0546}

\bibitem[{van Engelen {et~al.}(2015)}]{vanEngelen:2014zlh}
van Engelen, A. {et~al.}, {The Atacama Cosmology Telescope: Lensing of CMB
  Temperature and Polarization Derived from Cosmic Infrared Background
  Cross-Correlation}. 2015, Astrophys. J., 808, 7, \eprint{1412.0626}

\end{thebibliography}

\appendix
\section{Pipeline}
\label{appendix:pipeline}
	
	In this appendix we provide a more detailed technical overview of the pipeline that we use to extract lensing information from the {\it Planck} maps.
	Starting from input CMB sky maps the pipeline has the following three distinct analysis stages.
	\begin{enumerate}[(1)]
	\item \textit{Filtering}. In this step, we perform a linear operation on the maps, which performs three tasks.
		\begin{enumerate}[a.]
		\item Mask out regions of the sky that may be contaminated by point sources or residual foreground emission (primarily from the Galaxy).
		\item Deconvolve the instrumental beam and pixel transfer functions.
		\item Downweight noisy modes to allow later steps of the pipeline to produce optimal
		(in the minimum-variance sense) estimates of the lensing potential and its power spectrum.
		\end{enumerate}
	Implementation details of the filtering procedure are described in Appendix~\ref{sect:pipeline:fl}. The filtering step provides us with filtered CMB temperature and polarization multipoles denoted by $\bar{T}_{\elt m}$, $\bar{E}_{\elt m}$, and $\bar{B}_{\elt m}$.
	\item \textit{Quadratic Estimators}.
	The filtered CMB multipoles are then fed into quadratic estimators designed to extract the statistical anisotropy that is induced by lensing.
	These estimators are formed by summing over pairs of CMB fluctuations with a weight function optimized to detect the off-diagonal contributions to their covariance matrix from (fixed) lenses.
	This step produces maps of the estimated lensing potential $\hat{\phi}$, and is described in detail in Appendix~\ref{sect:pipeline:qe}.
	\item \textit{Power Spectrum Estimation}.
	To estimate the power spectrum of the lensing potential, we exploit the non-Gaussianity induced by lensing.
	At first order in the power spectrum $C_{\elp}^{\phi\phi}$, lensing generates a contribution to the connected 4-point function (or trispectrum) of the observed CMB,
	which we probe using auto- and cross-spectra of the 2-point quadratic estimators $\hat{\phi}$ obtained in step (2) of the pipeline.
	There are two complications here that are worth noting: the auto- and cross-spectra of the lensing estimates have noise biases even in the absence of lensing, which must be carefully estimated and subtracted; and non-Gaussian contamination from unresolved point sources can mimic lensing and bias the reconstructed power spectrum.
	Our approaches to these issues are described in Appendix~\ref{sect:pipeline:clpp}.
	\end{enumerate}
	We now proceed to describe the individual steps of the lensing pipeline in more detail.
	
\subsection{Filtering}
	\label{sect:pipeline:fl}
		At this stage in the pipeline, we have noisy maps of the CMB for both data and simulations.
	To construct minimum-variance estimates of the lensing potential and its power spectrum, we need to filter these maps to downweight noise-dominated modes, as well as to deconvolve the transfer functions due to the finite beam size and pixelization.
	
	To derive the optimal (minimum-variance) filter to apply, we use a simplified model of the data given by
	\be
	\map_i = \sum_{k} \ptg_{ik} s_{k}  + n_i,
	\ee
	where $d_i$ is a single long vector containing the pixelized $T$, $Q$, and $U$ sky maps.
	The `pointing matrix' $\ptg_{i,k}$ takes an input sky signal (consisting of $T$, $E$, and $B$ multipoles collectively indexed by $k$),
	convolves it with a beam and pixel transfer function, and then performs a harmonic transform to map space.
	The harmonic modes of the sky itself are denoted as $s_k$, and $n_i$ is the map noise realization.
	We have used indices $i$ and $k$ for quantities in map-space and harmonic-space respectively.
	Explicitly, the indices $k$ and $i$ are mapped as
	\newcommand{\mind}{M}
	\newcommand{\hind}{X}
	\begin{eqnarray}
	\qquad\qquad k &\mapsto& ( \hind \in \{T, E,B \}, \ell, m ) \\
	\qquad\qquad i &\mapsto& ( \mind \in \{T, Q,U \}, p ),
	\end{eqnarray}
        where $p$ denotes a pixel.
	In this notation, we model the pointing matrix as
	\be
	\ptg_{ (\mind, p), (\hind, \ell m) } = H_{\ell} B_{\ell} R^{\mind}_{\hind, \ell m}( \hat{\boldvec{n}}_p ),
	\ee
	where $H_{\ell}$ is the \healpix\ \citep{gorski2005} \mbox{$N_{\rm side}=2048$} pixel transfer function,
	$B_{\ell}$ is the effective beam transfer function for the map under consideration,
	and $R^{\mind}_{\hind, \ell m}(\hat{\boldvec{n}}_{p})$ represents the real-valued spherical harmonic for field $X$
	evaluated at the center of pixel $p$ in map $M$.

	If both the sky signal and noise are Gaussian random fields
	(with covariance matrices
	$\langle s_k s^*_{k'} \rangle \equiv S_{k k'}$ and
	$\langle n_{i} n_{i'} \rangle \equiv N_{i i'}$, respectively)
	then the inverse-variance filtered estimates of the beam-deconvolved sky signal $\bar{s}$ are given in matrix notation by \citep{Bunn:1994xn}
	\be
	\bar{\vec{s}}
	                     = \tens{S}^{-1} \left[ \tens{S}^{-1} + \ptg^{T} \tens{N}^{-1}  \ptg \right]^{-1} \ptg^{T} \tens{N}^{-1} \vec{d}.
	\label{eqn:cinv}
	\ee
	We use this equation to obtain the inverse-variance filtered $T$, $E$, and $B$ multipoles that are fed into our quadratic estimators, with one additional rescaling discussed below.
	The large bracketed matrix inverse is performed using conjugate descent with a multi-grid preconditioner, following \cite{Smith:2007rg}.
	
	For the purposes of filtering, we use a simplified model for the covariance matrices of $s_k$ and $n_i$
	in which they are diagonal in harmonic space and pixel space, respectively, with
	\begin{eqnarray}
	\qquad \qquad \tens{S} &=& \langle s_{(\hind, \ell m)} s^*_{(\hind', \ell' m')} \rangle = \delta_{\hind \hind'} \delta_{\ell \ell'} \delta_{m m'} C_{\ell}^{\hind\hind} , \nonumber \\
	\tens{N} &=& \langle n_{(\mind, p)} n_{(\mind', p')} \rangle = \delta_{\mind \mind'} \delta_{p p'} N^{\mind}_{p}.
	\label{eqn:cmbcovdefs}
	\end{eqnarray}
	Here,
	$C_{\ell}^{XX}$ are the fiducial theoretical CMB power spectra and
	$N^{\mind}_p$ gives the noise variance for pixel $p$ in map $M$.
	With this diagonal assumption, we have ignored the cross-correlation $C_{\elt}^{TE}$ for the purposes of filtering.
	This is slightly sub-optimal, but has the advantage of allowing the temperature and polarization maps to be filtered independently.
	It also has the benefit of making analytical calculations for the estimator normalization simpler (see Sect.~\ref{sect:pipeline:qe}).
	Also for simplicity, we take $N^{M}_{p}$ to be constant over all unmasked pixels and given by
	\be
	N^{\mind}_p = \left( \frac{\pi N_{\rm lev}^{\mind} }{10\, 800} \right)^2 \frac{N_{\rm pix}}{4\pi} \frac{1}{M_p}.
	\label{eqn:noiselevelmap}
	\ee
	Here $N^{M}_{\rm lev}$ is the map noise level in $\muKarcmin$,
	$N_{\rm pix}$ is the number of pixels in an $N_{\rm side}\!=\!2048$ \healpix\ map,
	and $M_p$ is a mask map (which is zero for masked pixels, and unity otherwise).
	We always take the noise level to be the same for paired $Q$/$U$ maps.
	The mask term in Eq.~\eqref{eqn:noiselevelmap} takes the noise level to infinity for masked pixels.
	In the filtering operation, which only involves the inverse of the noise matrix, this sets masked pixels to zero.
	
	After evaluating Eq.~\eqref{eqn:cinv}, we have a set of inverse-variance filtered sky multipoles $\bar{s}_{X, \elt m}$ for $X\in\{T,E,B\}$.
	The simplified CMB+white-noise model above results in a slightly sub-optimal filtering procedure since it ignores the following:
	\begin{enumerate}[(1)]
	\item variations in the noise level across the sky due to the uneven hit distribution (the Ecliptic poles receive more coverage than the equator for the {\it Planck} scan strategy);
	\item scale-dependence of the noise power; and
	\item foreground power.
	\end{enumerate}
	In \cite{planck2013-p12} we showed that the sub-optimality due to (1) for temperature-based lensing reconstruction was small (accounting for it properly was estimated to improve the uncertainty on the overall lensing amplitude by $4\,\%$).
	The loss will be slightly smaller here for temperature, owing to a more even hit distribution in the full-mission data than in the nominal-mission data used in \cite{planck2013-p12}.
	The loss will be larger for polarization, where noise is a more important part of the total error budget, although we have not evaluated the size of the potential degradation.
	To compensate for (2) and (3) we make a post-correction by rescaling the modes by a quality factor $Q_{\elt}^X$ as a function of scale to obtain final filtered multipole that are closer to optimal.
	We denote these as $\bar{X}_{\elt m}$, and they are given by
	\be
	\bar{X}_{\elt m} = Q_{\elt}^{X} \bar{s}_{X, \elt m}.
	\ee
	These inverse-variance filtered multipoles are the input to the lensing estimators described in Sect.~\ref{sect:pipeline:qe}.
	The choice of $Q_{\elt}^{X}$ is discussed below.
	
	For analytical calculations, it is useful to have an approximation to the filtering procedure that is diagonal in harmonic space
	(i.e. neglecting the small amount of mode-mixing induced by the masking).
	In this approximation, the filtered multipoles are given by
	\be
	\bar{X}_{\elt m} \approx F_{\elt}^{X} X_{\elt m} + \tilde{n}_{\elt m} ,
	\label{eqn:diagapprox}
	\ee
	where $\tilde{n}$ is a noise realization and
	\be
	F_{\elt}^{X} = \frac{Q_{\ell}^{X}}{C_{\ell}^{XX} + N_{\elt}^{XX}}.
	\ee
	Here, $C_{\ell}^{XX}$ are the fiducial CMB auto-spectra used in Eq.~\eqref{eqn:cmbcovdefs}, and the quantity
	\be
	N_{\elt}^{XX} = ( H_{\elt} B_{\elt} )^{-2} \left( \frac{ \pi N_{\rm lev}^{\mind} }{10\, 800} \right)^2
	\ee
	is the pixel- and beam-deconvolved noise power spectrum for the field $X$, where
	$N_{\rm lev}^{M}$ is the appropriate map white-noise level in $\muKarcmin$.
	We choose $Q_{\elt}^{X}$ such that
	\be
	\frac{f_{\rm sky}^{-1}}{(2L+1)} \sum_{m} | \bar{X}_{\elt m} |^2 \approx F_{\elt}^{X},
	\ee
	where $f_{\rm sky} = \sum_p M_p / N_{\rm pix}$ is the unmasked sky fraction and
	the approximation sign indicates a smoothing over multipoles.

\subsection{Quadratic estimators}
	\label{sect:pipeline:qe}
	\newcommand{\isep}{j}

	Ensemble averaging over a fixed realization of the CMB lensing potential, the CMB covariance matrix acquires off-diagonal elements given by
	\be
	\Delta \langle \fielda_{\elt_1 m_1} \fieldb_{\elt_2 m_2} \rangle =
	\sum_{\elp M} (-1)^{M} \threej{\elt_1}{\elt_2}{\elp}{m_1}{m_2}{-M}
	\Wlens^{\fielda \fieldb}_{\elt_1 \elt_2 \elp} \phi_{\elp M},
	\label{eqn:qecov}
	\ee
	with the fields
	\mbox{$\fielda_{\elt m}, \fieldb_{\elt m} \in \{ {T}_{\elt m}$, ${E}_{\elt m}$, ${B}_{\elt m} \}$}.
	The covariance response functions $\Wlens^{XZ}$ for the possible field combinations can be found in \cite{Okamoto:2003zw}. They are linear in the (lensed) CMB power spectra, and are only non-zero if $\ell_1+\ell_2+L$ is even for even-parity combinations, e.g. $TT$ and $TE$, and $\ell_1+\ell_2+L$ is odd for the odd-parity combinations $TB$ and $EB$. Furthermore, the $\Wlens^{XZ}$ are real for the even-parity combinations and imaginary for the odd-parity.
	We use the off-diagonal covariance to estimate the lensing potential;
	after the inverse-variance-filtered sky maps are generated, pairs of fields
	\mbox{$\bar{X}_{\elt m}$, $\bar{Z}_{\elt m} \in \{ \bar{T}_{\elt m}$, $\bar{E}_{\elt m}$, $\bar{B}_{\elt m} \}$}
	are fed into quadratic estimators to estimate the lensing potential.
	A general quadratic estimator is given as a function of input fields $\bar{X}$ and $\bar{Z}$ as
	\begin{multline}
	\bar{x}_{\elp M}[\bar{X}, \bar{Z}] = \frac{(-1)^{M}}{2}
	\sum_{\elt_1 m_1, \elt_2 m_2}
	\threej{\elt_1}{\elt_2}{\elp}{m_1}{m_2}{-M}
	\\ \times
	W^{x}_{\elt_1 \elt_2 \elp}
	\bar{X}_{\elt_1 m_1} \bar{Z}_{\elt_2 m_2},
	\label{eqn:xbar}
	\end{multline}
	where $W^{x}_{\elt_1 \elt_2 \elp}$ is a set of weight functions that define the estimator $x$, along
	with the input inverse-variance filtered multipoles $\bar{X}$, $\bar{Z}$ for a particular map or pair of maps.
	The choice of $\bar{T}$, $\bar{E}$, or $\bar{B}$ for $\bar{X}$ and $\bar{Z}$ depends on the specific estimator $x$.
	In most of what follows we shall drop the $[\bar{X}, \bar{Z}]$ unless it is necessary to avoid ambiguity.
		
	Optimal lensing estimators use a matched filter for the lensing-induced covariance of Eq.~\eqref{eqn:qecov}.
	We denote these as $W^{XZ}$ for the fields $X$ and $Z$.
	The temperature-only optimal estimator, for example, uses
	\mbox{$W^{TT} = \left. \Wlens^{TT} \right|_{\rm fid}$}, where the `fid' subscript indicates that the lensing weight function is evaluated for the fiducial cosmological model given at the end of Sect.~\ref{sec:data_and_methodology}. For the odd-parity combinations, the matched filter is minus the covariance response function, e.g. $W^{TB} = - \left. \Wlens^{TB} \right|_{\rm fid}$. Neglecting the lensing of any primordial $B$-mode signal
	(certainly a valid approximation for scales $\elt_B > 100$ with {\it Planck} polarization sensitivity), generally
	there are eight possible lensing estimators based on the possible combinations of $T$, $E$, and $B$. (This reduces to five estimators for our baseline analysis that uses a foreground-cleaned map of the CMB in temperature and polarization, and for which there is no distinction between $TE$ and $ET$, for example.)
	
	The quadratic estimators defined by Eq.~\eqref{eqn:xbar} will receive contributions from non-lensing sources such as
	masking,
	beam asymmetry, and
	inhomogeneity of the instrumental noise.
	We determine and correct for this bias by averaging the reconstruction
	$\bar{x}_{LM}$ over Monte Carlo simulations that include these effects,
	thereby estimating a `mean-field' $\bar{x}_{LM}^{MF}$ that we then subtract.
	
	The response of a quadratic estimator to the covariance given in Eq.~\eqref{eqn:qecov} (averaged over CMB realizations with a single mode $\phi_{LM}$ of the lensing potential held fixed) is, in the diagonal approximation of Eq.~\eqref{eqn:diagapprox},
	\be
	\langle \bar{x}_{LM} \rangle = \resp_{\elp}^{x \phi} \phi_{\elp M},
	\ee
	where
	\be
	{\cal R}_{\elp}^{x \phi} =
	\frac{1}{2(2\elp+1)} \sum_{\elt_1 \elt_2}
	W^{x}_{\elt_1 \elt_2 \elp} \Wlens^{X Z}_{\elt_1 \elt_2 \elp} F_{\elt_1}^{X} F_{\elt_2}^{Z}.
	\label{eqn:rlpp}
	\ee
	As a baseline, all of the results in this paper are obtained using the fiducial model at the end of Sect.~\ref{sec:data_and_methodology} to calculate the $\Wlens^{XZ}$ that appear in the estimator response.
	However, for cosmological parameter sampling with the lensing likelihood the normalization is recalculated using the appropriate CMB power spectra as the parameter space is explored. This is discussed further in Appendix~\ref{sect:pipeline:likelihood}.
	
	Putting the above together, we form estimates of the lensing potential as
	\be
	\hat{\phi}^{x}_{LM} = \frac{1}{\resp_L^{x\phi}} ( \bar{x}_{LM}^{\phantom{MF}} - \bar{x}_{LM}^{MF} ).
	\label{eqn:phihat}
	\ee
	We can also sum the individual estimators into a combined minimum-variance estimator (MV) as
	\be
	\hat{\phi}^{\rm MV}_{LM} =
	\frac{
		\sum_{x}  {
			\hat{\phi}^{x}_{LM} \resp_L^{x\phi}
		}
	}{
		\sum_{x} {
			\resp_L^{x\phi}
		}
	},
	\ee
	where the sum is taken over the eight lensing estimators ($TT$, $EE$, $TE$, $TB$, $EB$, $ET$, $BT$, and $BE$).
	
	Finally, we note an important implementation detail.
	Naively, evaluating an estimator of the form in Eq.~\eqref{eqn:xbar} requires
	\mbox{${\cal O}(\elt_{\rm max}^4 \elp_{\rm max}^2)$} operations, which would be very computationally expensive at {\it Planck} resolution.
	Fortunately, all of the weight functions used in this paper can be re-written as a sum of separable terms with the following form
	\begin{multline}
	W^{x}_{\elt_1 \elt_2 \elp} = \sum_{\isep}
	(-1)^{s^{x, \isep}_L}
	\sqrt{\frac{(2\elt_1 + 1)(2\elt_2 + 1)(2\elp+1)}{4\pi}}
	\\ \times
	\threej{\elt_1}{\elt_2}{L}{-s_1^{x, \isep}}{-s_2^{x, \isep}}{s_{L}^{x, \isep}}
	w_{\elt_1}^{x, \isep} w_{\elt_2}^{x, \isep} w_{\elp}^{x, \isep}.
	\label{eqn:qeseparable}
	\end{multline}
	This separability leads to estimators that can be evaluated with ${\cal O}( \isep_{\rm max} \elt_{\rm max}^2 \elp_{\rm max} )$ operations in position space as
	\begin{multline}
	\bar{x}_{\elp M} = \frac{1}{2}
	\int d\hatn  \sum_{\isep} \yslm{s_{L}^{x, \isep}}{L}{M}^*(\hatn) w_{\elp}^{x, \isep}
	\\ \times
	\left[ \sum_{\elt_1 m_1} \yslm{s_1^{x, \isep}}{\elt_1}{m_1}(\hatn) w_{\elt_1}^{x, \isep} \bar{X}^{x,\isep}_{\elt_1 m_1} \right]
	\\ \times
	\left[ \sum_{\elt_2 m_2} \yslm{s_2^{x, \isep}}{\elt_2}{m_2}(\hatn) w_{\elt_2}^{x, \isep} \bar{Z}^{x,\isep}_{\elt_2 m_2} \right].
	\end{multline}
	The separability of the weight functions also allows the response functions to be calculated quickly analytically;
	using the position-space approach given in \cite{Dvorkin:2008tf} they can be evaluated in
	\mbox{${\cal O}( l_{\rm max}^2 )$} operations.
	A sample implementation of these weights can be found at \href{http://github.com/dhanson/quicklens}{http://github.com/dhanson/quicklens} in the file \href{https://github.com/dhanson/quicklens/blob/master/quicklens/qest/lens.py}{quicklens/qest/lens.py}.
	\newcolumntype{H}{>{\setbox0=\hbox\bgroup}c<{\egroup}@{}}

\subsection{Power spectrum estimation}
	\label{sect:pipeline:clpp}
\begin{figure}[t]
	\centering
	\includegraphics[width=\columnwidth]{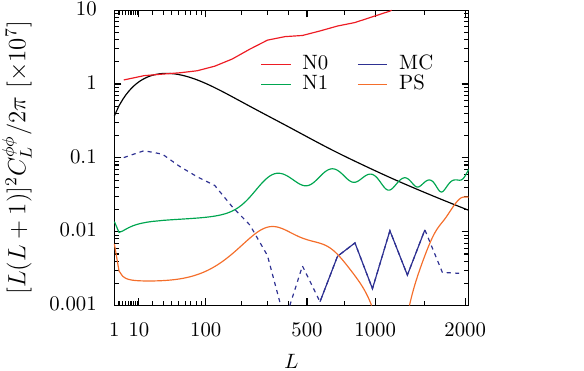}
	\vspace{-0.2in}
	\caption{
	Lens reconstruction bias terms for the MV lensing estimator applied to the \planck\ 2015 \smica\ maps.
	Dashed lines indicate regions where the bias term is negative.
	The fiducial $\Lambda$CDM theory power spectrum $C_{\elp}^{\phi\phi,\ {\rm fid}}$ is plotted as the black solid line.
	\label{fig:clpp_bias_terms}
	}
	\end{figure}

		Ensemble averaging over realizations of both the CMB lensing potential
	and the primary temperature and polarization fluctuations, the CMB becomes non-Gaussian.
	At first order in the lensing potential power spectrum $C_L^{\phi\phi}$ this non-Gaussianity
	is manifest as a connected 4-point function \citep{Hu:2001fa}
	\begin{multline}
	\langle \fielda_{\elt_1 m_1} \fieldb_{\elt_2 m_2} \fieldc_{\elt_3 m_4} \fieldd_{\elt_4 m_4} \rangle_c =
	\sum_{LM}
	\threej{\elt_1}{\elt_2}{\elp}{m_1}{m_2}{M}
	\threej{\elt_3}{\elt_4}{\elp}{m_3}{m_4}{-M}
	\\ \times (-1)^M
	C_{\elp}^{\phi\phi} \Wlens^{{\fielda \fieldb}}_{\elt_1 \elt_2 \elp} \Wlens^{{\fieldc \fieldd}}_{\elt_3 \elt_4 \elp}
	+ \text{2 perms}.
	\label{eqn:trispec_clpp}
	\end{multline}
	As might be expected given the presence of the covariance response functions $\Wlens$ in this expression,
	estimators to extract $C_L^{\phi\phi}$ from the trispectrum can be written using the quadratic estimators above as building blocks.
	We write the cross-spectrum of two quadratic estimators explicitly as
	\be
	C^{\hat{\phi} \hat{\phi}}_{\elp, xy}[ \bar{X}, \bar{Z}, \bar{C}, \bar{D} ] \equiv
	\frac{f_{\rm sky}^{-1}}{2\elp+1}
	\sum_{M} \hat{\phi}_{\elp M}^{x}[\bar{X}, \bar{Z}] \hat{\phi}_{\elp M}^{y*}[\bar{C}, \bar{D}],
	\label{eqn:clpphatest}
	\ee
	where
	$f_{\rm sky} = \sum_p M_p / N_{\rm pix}$ is the unmasked sky fraction, with $M_p$ the mask map used in Eq.~\eqref{eqn:noiselevelmap}.
	We form estimates of the lensing potential power spectrum (based on estimators $x$ and $y$) as
	\begin{multline}
	\hat{C}_{L, xy}^{\phi\phi} =
	C^{\hat{\phi} \hat{\phi}}_{\elp, xy}
	- \left. \Delta C_{\elp, xy}^{\hat{\phi} \hat{\phi}} \right|_{\textsc{N0}}
	\\
	- \left. \Delta C_{\elp, xy}^{\hat{\phi} \hat{\phi}} \right|_{\textsc{N1}}
	- \left. \Delta C_{\elp, xy}^{\hat{\phi} \hat{\phi}} \right|_{\textsc{MC}}
	- \left. \Delta C_{\elp, xy}^{\hat{\phi} \hat{\phi}} \right|_{\textsc{PS}}.
	\label{eqn:clppest}
	\end{multline}
	There are several correction terms here, which are discussed in more detail below.
	We also plot them for the MV reconstruction based on the \planck\ 2015 \smica\ maps in Fig.~\ref{fig:clpp_bias_terms}.
	
	The $\Delta C_{\elp, xy}^{\hat{\phi} \hat{\phi}} |_{\textsc{N0}}$ term represents the disconnected contribution to the 4-point function, which would be non-zero even in the absence of lensing.
	We estimate this term by replacing some of the data fields with those from two sets of independent simulations
	(labelled $\textsc{MC1}$ and $\textsc{MC2}$), and then average over realizations as
	\begin{multline}
	\left. \Delta C_{L,xy}^{\hat{\phi}\hat{\phi}} \right|_{\textsc{N0}} =
	\Bigg<
	- C_{\elp, xy}^{\hat{\phi}\hat{\phi}}[ \bar{X}_{\textsc{MC1}}, \bar{Z}_{\textsc{MC2}}, \bar{C}_{\textsc{MC2}}, \bar{D}_{\textsc{MC1}} ]
	\\
	+ C_{\elp, xy}^{\hat{\phi}\hat{\phi}}[ \bar{X}_{\textsc{MC1}}, \bar{Z}, \bar{C}_{\textsc{MC1}}, \bar{D} ]
	+ C_{\elp, xy}^{\hat{\phi}\hat{\phi}}[ \bar{X}_{\textsc{MC1}}, \bar{Z}, \bar{C}_{}, \bar{D}_{\textsc{MC1}} ]
	\\
	+ C_{\elp, xy}^{\hat{\phi}\hat{\phi}}[ \bar{X}, \bar{Z}_{\textsc{MC1}}, \bar{C}_{\textsc{MC1}}, \bar{D} ]
	+ C_{\elp, xy}^{\hat{\phi}\hat{\phi}}[ \bar{X}, \bar{Z}_{\textsc{MC1}}, \bar{C}_{}, \bar{D}_{\textsc{MC1}} ]
	\\
	- C_{\elp, xy}^{\hat{\phi}\hat{\phi}}[ \bar{X}_{\textsc{MC1}}, \bar{Z}_{\textsc{MC2}}, \bar{C}_{\textsc{MC1}}, \bar{D}_{\textsc{MC2}} ]
	\Bigg>_{\textsc{MC}1,\ \textsc{MC}2}.
	\end{multline}
	This method for determining the disconnected bias
	can be derived from the Edgeworth expansion of the lensed CMB
	(\citealt{Regan:2010cn};
	neglecting the $T\phi$ and $E\phi$ correlations).
\refchange{
If the simulations have a covariance differing from reality by a linear (in general, anisotropic) correction, this form of the $\textsc{N0}$ estimator using a mixture of data and simulations is insensitive to the correction at linear order \citep{Namikawa:2012pe}.
The largest difference we see between simulations and data is about
$5\%$ in power before correction, so the quadratic error should be at
most about $(0.05)^2 =  0.0025$, corresponding to an error of less
than $0.4\,\sigma$ for aggressive bins at $L\le 400$ rising to around $1.3\,\sigma$ at $L=2000$.
Since we also match the data calibration, and add isotropic foreground
and noise power to the simulations to match the observed amplitude and
shape of the power spectrum of the data, the isotropic part of the
error is actually close to zero. A residual error only arises at
second order due to our inability to capture any statistical anisotropy in the missing
foreground and noise power, and hence should be much smaller than the
above estimates. This makes the $\textsc{N0}$ subtraction rather robust, despite it being substantially larger than the signal on small scales.
}	
%The mixture of data and simulations to estimate this term makes it stable with respect to small infidelities in the Monte Carlo simulations used to determine the bias \citep{Namikawa:2012pe}.
Furthermore, the mixed form reduces correlations of the lensing power spectrum estimates between different multipoles~\citep{2011PhRvD..83d3005H} and with the measured CMB power spectra~\citep{Schmittfull:2013uea}.

	The $\Delta C_L^{\hat{\phi}\hat{\phi}} |_{\textsc{N1}}$ term corrects for the non-Gaussian secondary contractions
	(the other permutations in Eq.~\ref{eqn:trispec_clpp}) due to lensing, as discussed in \cite{Kesden:2003cc}.
	We evaluate it using the flat-sky approximation in 2D Fourier space as
	\begin{multline}
	\left. \Delta C_{L,xy}^{\hat{\phi}\hat{\phi}} \right|_{\textsc{N1}} =
	\frac{1}{\resp_L^{x\phi} \resp_L^{y\phi}}
	\int \frac{d^2 \vec{\elt}_1}{(2\pi)^2}
	\int \frac{d^2 \vec{\elt}'_1}{(2\pi)^2}
	\\ \hspace{-0.5in} \times
	\Big[ C^{\phi\phi}_{ | \vec{\elt}_1 - \vec{\elt}'_1 |}
	\Wlens^{XC}(-\vec{\elt}_1, \vec{\elt}'_1) \Wlens^{ZD}(-\vec{\elt}_2, \vec{\elt}'_2)
	\\ \qquad +
	C^{\phi\phi}_{ | \vec{\elt}_1 - \vec{\elt}'_2 | }
	\Wlens^{XD}(-\vec{\elt}_1, \vec{\elt}'_2) \Wlens^{ZC}(-\vec{\elt}_2, \vec{\elt}'_1) \Big]
	\\ \times F^{X}_{|\vec{\elt}_1|} F^{Z}_{|\vec{\elt}_2|} F^{C}_{| \vec{\elt}'_1 |} F^{D}_{| \vec{\elt}'_2 |}
	\\ \times
	W^{x} (\vec{\elt}_1, \vec{\elt}_2) W^{y} (\vec{\elt}'_1, \vec{\elt}'_2),
\label{eq:None}
	\end{multline}
	where $\vec{\elt}_1 + \vec{\elt}_2 = \vec{\elt}'_1+ \vec{\elt}'_2 = \vec{L}$.
	Here the weight functions are flat-sky lensing weight functions; they can be found in \cite{Hu:2001kj}.

	We characterize any differences between the average lensing power obtained on simulations with the input power.
	This is determined by
	\begin{multline}
	\left. \Delta C_L^{\hat{\phi}\hat{\phi}} \right|_{\textsc{MC}} \equiv
	\Bigg<
	C_{\elp, xy}^{\hat{\phi}\hat{\phi}}[ \bar{X}_{\textsc{MC1}}, \bar{Z}_{\textsc{MC1}}, \bar{C}_{\textsc{MC1}}, \bar{D}_{\textsc{MC1}} ]  \\
	- C_{\elp, xy}^{\hat{\phi}\hat{\phi}}[ \bar{X}_{\textsc{MC1}}, \bar{Z}_{\textsc{MC2}}, \bar{C}_{\textsc{MC1}}, \bar{D}_{\textsc{MC2}} ]
	\\
	- C_{\elp, xy}^{\hat{\phi}\hat{\phi}}[ \bar{X}_{\textsc{MC1}}, \bar{Z}_{\textsc{MC2}}, \bar{C}_{\textsc{MC2}}, \bar{D}_{\textsc{MC1}} ]
	\\
	- \left. \Delta C_{L,xy}^{\hat{\phi}\hat{\phi}} \right|_{\textsc{N1}} - \left. C_L^{\phi\phi} \right|_{\rm fid}
	\Bigg>_{\textsc{MC1},\ \textsc{MC2}}.
	\end{multline}
	The MC correction is included to account for possible issues in the estimation of the normalization,
	mixing of power between power spectrum bins, and errors in the calculation of the $\textsc{N1}$ bias.
	Although in principle several of these effects should be included as multiplicative rather than additive corrections,
	we find that the MC correction is generally small enough (less than $10\,\%$ of the fiducial lensing power spectrum) that
	it does not matter at our experimental sensitivity whether we apply the correction additively or multiplicatively.

	Finally, we make a correction $\Delta C_{\elp, xy}^{\hat{\phi}\hat{\phi}} |_{\textsc{PS}}$
	for the non-Gaussianity due to the shot-noise of unresolved point sources in temperature.
	This is given by
	\be
	\left. \Delta C_{\elp, xy}^{\hat{\phi}\hat{\phi}} \right|_{\textsc{PS}} =
	\widehat{S^4}
	\frac{ {\cal R}_L^{x S^2} {\cal R}_L^{y S^2} }{ {\cal R}_L^{x \phi} {\cal R}_L^{y \phi} },
	\ee
	where $\widehat{S^4}$ is an estimate of the shot-noise amplitude and
	$S^2$ denotes the weight function (non-zero only for $TT$) defined by
	\be
	W^{S^2}_{\elt_1 \elt_2 L} =
	\sqrt{ \frac{ (2\elt_1 + 1)(2 \elt_2 +1)(2L +1) }{ 4\pi } }
	\threej{\elt_1}{\elt_2}{L}{0}{0}{0}.
	\ee
        The $\resp^{xS^2}$ is given by Eq.~\eqref{eqn:rlpp} but with $\Wlens_{\ell_1 \ell_2 L}$ replaced by $W^{S^2}_{\ell_1 \ell_2 L}$. Following \cite{Osborne:2013nna}, the estimate for the shot-noise trispectrum amplitude is obtained in an analogous way to Eq.~\eqref{eqn:clppest}, but using the quadratic estimator formed with the point-source weight function above rather than the lensing weight function.
	The overall source amplitude is then estimated as
	\be
	\widehat{S^4} = \frac{
		\sum_{\elp} (2\elp+1) \left( \resp_{\elp}^{S^2 S^2} \right)^2 \hat{C}_{\elp}^{S^2 S^2}
		}{
		\sum_{\elp} (2\elp+1) \left( \resp_{\elp}^{S^2 S^2} \right)^2
		},
	\ee
        where the sums are taken over $100 \le \elp \le 2048$.
        Note that $\resp^{S^2 S^2}$ is given by Eq.~\eqref{eqn:rlpp} but with both $\Wlens_{\ell_1 \ell_2 L}$ and $W^x_{\ell_1 \ell_2 L}$ replaced by $W^{S^2}_{\ell_1 \ell_2 L}$.
	
	We use Eq.~\eqref{eqn:clppest} to estimate the lensing potential power spectrum,
	which we then bin into bandpowers for plotting as well as constructing a lensing likelihood.
	We use Monte-Carlo simulations of the power spectrum estimation to construct error bars for these bandpowers.
	There is one technical point here, which is that the realization-dependent
	$\Delta C_{\elp, xy}^{\phi\phi} |_{\textsc{N0}}$
	correction term can be cumbersome to compute.
	Instead, for the purpose of determining error bars, we use a semi-analytical approximation to this term given by
	\begin{multline}
	\left. \Delta C_{\elp, xy}^{\phi\phi} \right|_{\textsc{N0},\ {\rm analytic}}[\bar{X}, \bar{Z}, \bar{C}, \bar{D}] =
	\frac{1}{\resp_{\elp}^{x\phi} \resp_{\elp}^{y\phi}}
	\frac{1}{4(2\elp+1)}
	\\ \times
	\sum_{\elt_1 \elt_2} W^{x}_{\elt_1 \elt_2 \elp}
	\Big(
		W^{y\ast}_{\elt_1 \elt_2 \elp} \bar{C}_{\elt_1}^{XC} 	\bar{C}_{\elt_2}^{ZD} +
		(-1)^{\ell_1 + \ell_2 + L} W^{y\ast}_{\elt_2 \elt_1 \elp} \bar{C}_{\elt_1}^{XD} 	\bar{C}_{\elt_2}^{ZC}
	\Big),
	\end{multline}
	where the empirical power spectra of the filtered fields are
	\be
	\bar{C}_{\elt}^{XZ} = \frac{f_{\rm sky}^{-1}}{2L+1} \sum_{m}  \bar{X}_{\elt m} \bar{Z}_{\elt m}^* .
	\ee

\section{Bias-hardened estimators}
	\label{sect:pipeline:bias_hardened}
	The quadratic lensing estimators of Appendix~\ref{sect:pipeline:qe} can be biased by non-lensing sources of statistical anisotropy such as beam asymmetry, noise inhomogeneity, and masking.
	In our fiducial analyses, we estimate these mean-field biases using Monte-Carlo simulations. However, this approach can be sensitive to the fidelity of the simulations, particularly for the reconstruction on large angular scales where the mean-field biases can be orders of magnitude above the lensing signals of interest.
	As a cross-check, we also take a more data-dependent approach to correcting these biases.
	As with lensing, most sources of statistical anisotropy in a map can be associated with a covariance response function $\Wlens^{z}_{\elt_1 \elt_2 \elp}$, and a field $z_{\elp M}$ that describes the spatial dependence of the anisotropy.
	It is possible to construct quadratic estimators for $z_{\elp M}$, and then calculate and subtract the resulting lensing bias.
	This procedure amounts to constructing special `bias-hardened' estimators \citep{Namikawa:2012pe}.
	Given a source of bias $z$, we construct a bias-hardened estimator with weight function
	\be
	W^{\phi - z}_{\elt_1 \elt_2 \elp} = W^{\phi}_{\elt_1 \elt_2 L} - \frac{ \resp_{\elp}^{\phi z} }{ \resp_{\elp}^{zz} } W^{z}_{\elt_1 \elt_2 L},
	\label{eqn:wbiashardened}
	\ee
where $\resp_\elp^{\phi z}$ and $\resp_\elp^{zz}$ are generalizations of Eq.~\eqref{eqn:rlpp}, and $W^{z}_{\elt_1 \elt_2 L}$ is the matched filter for
$\Wlens^z_{\elt_1 \elt_2 \elp}$.
	This method is potentially less susceptible to errors in the mean-field estimate (due to inaccuracies in the simulations used to obtain it), although it has several caveats that are discussed in more detail in \cite{Namikawa:2012pe}.

\section{Power spectrum likelihood}
	\label{sect:pipeline:likelihood}
		Here, we provide more detail on the components of our lensing likelihood (given in Eq.~\ref{eqn:likelihood}, and copied below for convenience)
	\be
		-2 \log {\cal L}_{\phi} = \bin_i^{\elp}
		( \hat{C}_{\elp}^{\phi\phi} - C_{\elp}^{\phi\phi, {\rm th}} )
		\left[ \Sigma^{-1} \right]^{ij} \bin_j^{\elp'}
		( \hat{C}_{\elp'}^{\phi\phi} - C_{\elp'}^{\phi\phi, {\rm th}} ).
\label{eq:binlike}
	\ee
	The bandpower binning function $\bin_{i}^{L}$ is given by
	\be
	\bin_{i}^{L} =
	\frac{
		\clppfid V_L^{-1}
		}{
		\sum_{\elp' = \elp_{\rm min}^{i}}^{\elp_{\rm max}^{i}} \left(C_{L'}^{\phi\phi,\, {\rm fid}}\right)^2 V_{\elp'}^{-1}
		},
	\ee
	where $V_L$ is an approximation to the variance of the power spectrum estimate.
	For a given cross-spectrum between estimators $x$ and $y$ we use
	\be
	V_L^{-1} = \frac{2L+1}{2 f^{-1}_{\rm sky}} \resp_{L}^{x \phi} \resp_L^{y \phi},
	\ee
	which produces minimum-variance estimates of the lensing amplitude for each bin
	in the limit that the sample variance of the lenses is negligible
	(a reasonable approximation for {\it Planck}, which is noise dominated on any individual mode).
	
	The dependence of the likelihood on cosmological parameters enters through the `theory' spectrum
$C_L^{\phi\phi, {\rm th}}$. This is the expected value of the estimated spectrum at each point in parameter space. It depends on the cosmological parameters $\vtheta$ in several ways:
\begin{itemize}
\item directly, through the theory spectrum $\left. C_L^{\phi\phi}\right|_\vtheta$;
\item indirectly (but linearly) on $\left. C_L^{\phi\phi}\right|_\vtheta$, via the theory-dependence of $\None$ (Eq.~\ref{eq:None}); and
\item indirectly, and non-linearly, on the CMB power spectra $C_\elt^{TT}$, $C_\elt^{TE}$, and $C_\elt^{EE}$ through the  estimator normalization and $\None$ (from the theory-dependent covariance response $\Wlens$ of Eq.~\ref{eqn:qecov}).
\end{itemize}
 We neglect any other theory dependence, for example in the MC correction that is hard to quantify.
	In detail, for a given set of cosmological parameters $\vtheta$, we should calculate
	\be
	C_{\elp}^{\phi\phi, {\rm th}} =
	\frac{ \left. (\resp_{\elp}^{x \phi} \resp_{\elp}^{y \phi}) \right|_{\vtheta }}{ \left. (\resp_{\elp}^{x \phi} \resp_{\elp}^{y \phi})\right|_\fid }
 \left. C_{\elp}^{\phi\phi} \right|_{\vtheta} - \left. \Delta C_{\elp, xy}^{\phi\phi} \right|_{\textsc N1,\ {\rm fid}}  + \left. \Delta C_{\elp, xy}^{\phi\phi} \right|_{\textsc N1,\ \vtheta} ,
	\label{eqn:clppth}
	\ee
	where the $\Wlens$ component of the response functions $\resp$ in the numerator, as well as the $\None$ term, are calculated using the CMB power spectra for parameters $\vtheta$.

Fully recalculating everything for each point in parameter space is prohibitively slow, but for small deviations from the fiducial model we can use a linearized approximation. The dependence on $\left. C_L^{\phi\phi}\right|_\vtheta$ is already linear, and expanding the CMB power spectrum dependence to linear order about the fiducial model
we can write
\begin{multline}
	C_{\elp}^{\phi\phi, {\rm th}} \approx
 \left. C_{\elp}^{\phi\phi} \right|_{\vtheta}
+
\frac{d \ln (\resp_{\elp}^{x \phi} \resp_{\elp}^{y \phi}) }{d C_{\ell'}^j}
\left(\left.C_{\ell'}^j\right|_\vtheta -\left.C_{\ell'}^j\right|_{\rm fid} \right) \left.C_L^{\phi\phi}\right|_{\rm fid} \\
+
M^{(1)\phi}_{LL',xy} \left( \left.C_{L'}^{\phi\phi}\right|_\vtheta - \left.C_{L'}^{\phi\phi}\right|_{\rm fid} \right)
+
\frac{\left. d \Delta C_{\elp, xy}^{\phi\phi} \right|_{\textsc N1}}{d C_{\ell'}^j}
\left(\left.C_{\ell'}^j\right|_\vtheta -\left.C_{\ell'}^j\right|_{\rm fid} \right),
\end{multline}
where $j$ sums over the various CMB power spectra.
The matrix $M^{(1)\phi}_{LL'}$ (from Eq.~\ref{eq:None}) gives the exact linear dependence of $\None$ on the lensing potential for fixed CMB power spectra, and can be pre-computed along with the other derivative matrices for the fiducial model. The binned likelihood of Eq.~\eqref{eq:binlike} then depends on a binned `theory' power spectrum given in the linear approximation by
\be
 \bin_i^{\elp} C_{\elp}^{\phi\phi, {\rm th}}
\approx  \bin_i^{\elp} \left.C_{\elp}^{\phi\phi}\right|_\vtheta + M_i^{a,\ell'}\left(
 \left.C^a_{\ell'}\right|_\vtheta -  \left.C^a_{\ell'}\right|_{\rm fid}\right),
 \label{eq:fastlike}
\ee
where now $a$ sums over both the $C_{L}^{\phi\phi}$ and CMB power spectra terms, and $M_i^{a,\ell'}$ can be pre-computed in the fiducial model.

The \planck\ lensing likelihood consists of bandpower estimates
$\bin_i^{\elp}\hat{C}_{\elp}^{\phi\phi}$ and the covariance $\Sigma_{ij}$, a set of binning functions $\bin_i^{\elp}$,  a set of linear-correction kernels $M_i^{a,\ell'}$, and the band values of the linear-correction kernel applied to the fiducial spectrum $M_i^{a,\ell'} C^a_{\ell'} |_{\rm fid}$. Note that since the lensing estimator is independent of the CMB calibration, for a theory model $C^j_\ell$ and map calibration parameter $y_{\rm cal}$, the theory CMB power spectra that are used in Eq.~\eqref{eq:fastlike} are those matched to the data, i.e.
$C^j_\ell |_\vtheta  = C^j_\ell/y_{\rm cal}^2$.

When evaluating `lensing-only' likelihoods, where the CMB power spectra are only very weakly indirectly constrained, we fix the $C^j_\ell |_\vtheta$ used in the linear correction to a best-fit to the full data. When calculating joint lensing and CMB results, the correction depends on the CMB power spectra at each point in model space.

\section{CIB model}
\label{app:cib_model}

\begin{figure}[!ht]
	\centering
	\includegraphics[width=\columnwidth]{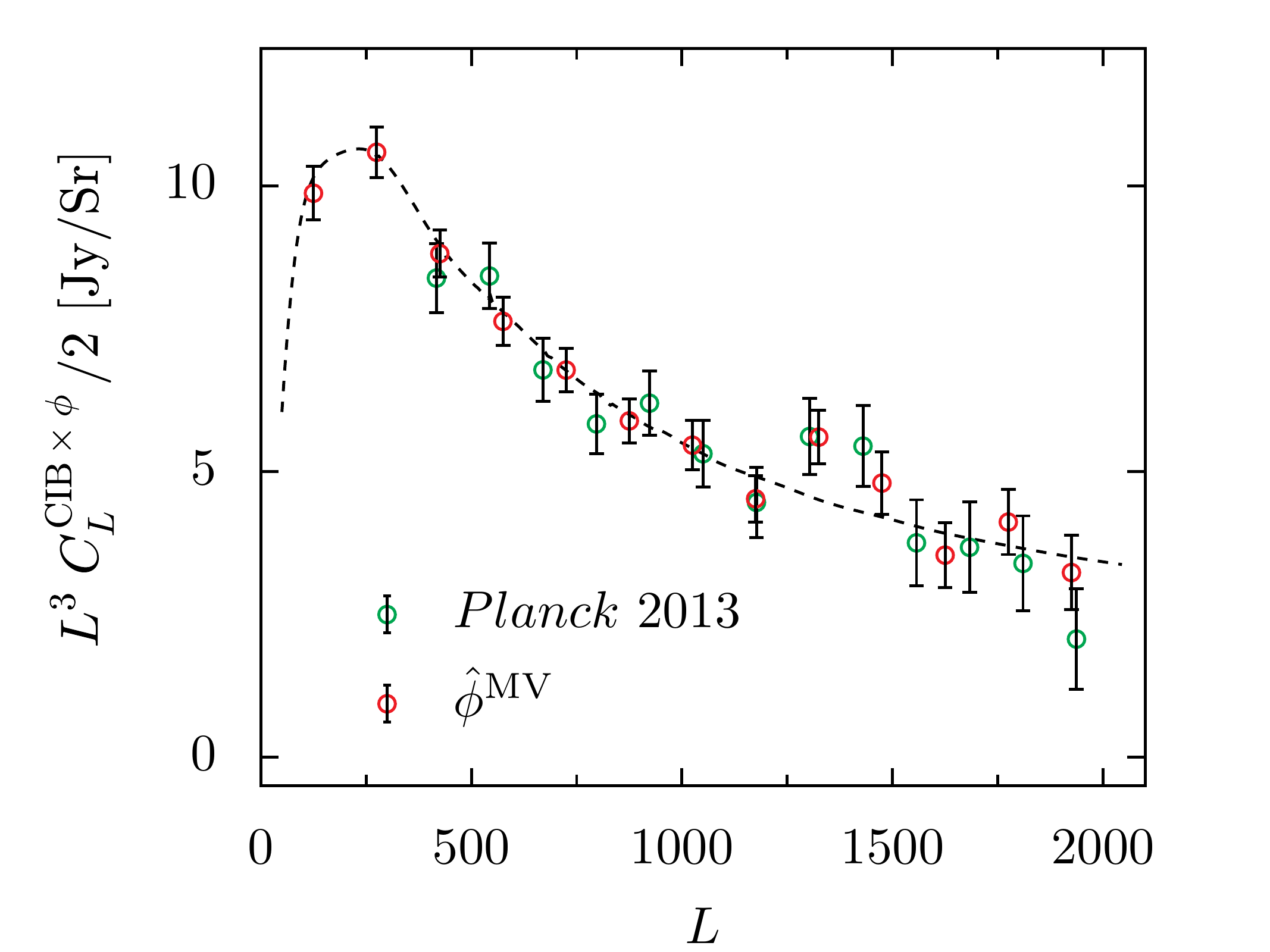}
	\vspace{-0.2in}
	\caption{
	Cross-correlation between the MV lensing potential estimate and the {\it Planck} 545\,GHz channel (red points).
	The cross-correlation uses a Galaxy mask leaving $40\,\%$ of the sky.
	The earlier result from \cite{planck2013-p13} using a $TT$ lensing estimator is shown with green points.
	The simple CIB model discussed in Appendix~\ref{app:cib_model}
        is plotted as the black dashed line.
	\label{fig:qexp}
	}
\end{figure}

The cosmic infrared background (CIB) is a diffuse sky signal which
begins to dominate over the CMB at frequencies
\mbox{$\nu \gtrsim 300\,{\rm GHz}$}.
It is generated by dust that is heated by UV light from young stars and then re-radiates thermally in the infrared.
The CIB contains approximately half of the total extragalactic stellar flux, and has excellent redshift overlap with the CMB lensing potential \citep{Song:2002sg}.
In Sect.~\ref{sect:lensing_bmode_power_spectrum} we use the CIB as a tracer of the lensing potential to estimate the lensing-induced B-mode signal.
This requires a model for the cross-correlation between the CIB and the lensing potential.
We calculate this in the Limber approximation as \citep{Limber1954}
\be
C_L^{\text{\textsc{cib}-}\phi} = \int d\chi \frac{1}{\chi^2} K_L^{\textsc{cib}}(\chi) K_L^{\phi}(\chi) P(k=L/\chi; \eta_0-\chi),
\label{eqn:limber}
\ee
where $P(k; \eta)$ is the (non-linear) matter power spectrum at comoving wavenumber $k$ and conformal time $\eta$.
The lensing kernel in flat models is given by
\be
K_\ellp^{\phi}(\chi) = -\frac{ 3 \Omega_{\rm m} H_0^2 }{a(\chi)} \left(\frac{\chi}{\ellp}\right)^2 \left( \frac{\chi_* - \chi}{\chi_* \chi} \right) ,
\ee
where $a(\chi)$ is the scale factor at conformal time
$\eta_0-\chi$
and
$\chi_*$ denotes the last scattering surface at $z\approx 1100$.
To model the CIB, we use the simple SSED model of \cite{Hall:2009rv}, which has a kernel given by
\be
K_L^{\textsc{cib}}(\chi) = b_c
\frac{\chi^2}{(1+z)^2} \exp\left( -\frac{ (z-z_c)^2 }{2 \sigma_z^2 } \right) f_{\nu (1+z)},
\label{eqn:kcib}
\ee
where $b_c$ is an overall normalization and $z_c = \sigma_z = 2$ describe the redshift distribution of the CIB intensity. The redshift $z$ is evaluated at conformal time $\eta_0-\chi$.
The spectral energy distribution (SED) of a typical CIB source is described by $f_{\nu}$. It is modelled as a modified blackbody with temperature $T=34\,\rm{K}$, and spectral index $\beta=2$ up to a threshold $\nu' \approx 4955\, \rm{GHz}$,
where it transitions to a power-law decay with index $\alpha=2$:
\be
f_{\nu} =
\begin{cases}
\left[ \exp\left( \frac{h \nu}{k T} \right) - 1 \right]^{-1} \nu^{\beta + 3} & (\nu \le \nu') , \\
\left[ \exp\left( \frac{h \nu'}{k T} \right) - 1 \right]^{-1} {\nu'}^{\beta + 3} \left( \frac{\nu}{ \nu'} \right)^{-\alpha}  & (\nu > \nu').
\end{cases}
\ee
The precise value of $\nu'$ is chosen to match smoothly the slope of $f_{\nu}$ on both sides of the transition frequency.
In this model, the modified blackbody component of the SED is generally what we observe;
even at an observation frequency of $1200\,{\rm GHz}$ (i.e. $250\,\mu{\rm m}$), the power-law transition does not occur until $z \approx 4$.
Our theory curve is plotted in Fig.~\ref{fig:qexp}, as well as
the measured cross-correlation of the {\it Planck} 2015 545\,GHz map with the MV lensing potential estimate (over $40\,\%$ of the sky).
In addition we have plotted the bandpowers from the rigorous analysis presented in the {\it Planck} 2013 data release \citep{planck2013-p13}.
We have set the normalization $b_c$ for the theory curve such that the overall amplitude is consistent with both measurements;
calculating $\chi^2$ values against the theory curve we find
PTEs of
$73\,\%$ for the MV lensing potential estimator and
$49\,\%$ for the
for the 13 plotted bins we find a
$\chi^2$ PTE with respect to the theory curve of $49\,\%$.

\section{Parameter dependence of $\clpp$ in \lcdm\ models}
\label{app:clpp_param_depend}
In this appendix we discuss the dependence of the lensing potential power spectrum at multipoles $\ellp \gtrsim 100$ on the parameters of \lcdm\ models, providing results that are used in the discussion of the `lensing-only' constraints in Sect.~\ref{sec:parameters}. Since these constraints adopt a prior on $\ns$ that is rather tighter than the resulting posterior widths for $\Omega_{\rm m}h^2$ and $h$, we ignore dependencies on $\ns$ throughout this appendix. We caution the reader that for joint constraints, or fits to the CMB power spectra alone, fractional variations in $\ns$ can be comparable to those in $\Omega_{\rm m}h^2$ and $h$, so the dependence of $\clpp$ on $\ns$ cannot be ignored.
The dependence of $\clpp$ on the matter density was recently discussed in~\citet{Pan:2014xua}; we follow their discussion closely.

For multipoles $L \gtrsim 20$, the angular power spectrum of the lensing potential can be Limber-approximated as
\begin{equation}
\ellp^4 \clpp = 4 \int_0^{\chi_*} d\chi \, (k^4 P_{\Psi})(k=\ellp/\chi;\eta_0-\chi)
(1-\chi/\chi_*)^2
\end{equation}
in spatially-flat models. Here, $P_\Psi(k;\eta)$ is the equal-time dimensional power spectrum of the Weyl potential $\Psi$. Well after baryon decoupling, we have\footnote{We ignore the small effects of non-zero neutrino mass in the base \lcdm\ model, so that the linear growth of structure is scale-independent well after baryon decoupling.}
\begin{equation}
k^4 P_\Psi(k;\eta) \propto k \mathcal{P}_{\mathcal{R}}(k) T^2(k) g^2(a) \, ,
\end{equation}
where $\mathcal{P}_{\mathcal{R}}(k)$ is the dimensionless power spectrum of the primordial curvature perturbation $\mathcal{R}$, $T(k)$ is the transfer function,
and $g(a)$ is the growth function for the Weyl potential at scale factor $a$ normalized to unity at high redshift. In $\Lambda$CDM,
\begin{equation}
g(a) = \frac{5 \Omega_{\rm m} E(a)}{2a} \int_0^a \frac{da'}{\left[a' E(a')\right]^3} \, ,
\end{equation}
where $\Omega_{\rm m}$ is the matter fraction at the present and $E(a)\equiv H(a)/H_0$. For $\Omega_{\rm m}=0.3$, the current value of the growth function is $0.78$ and $d \ln g / d \ln \Omega_{\rm m} = 0.23$. Close to $\Omega_{\rm m}=0.3$, we can approximate the cosmology dependence of the growth function at redshift zero as $g(a=1) \approx 0.78 (\Omega_{\rm m}/0.3)^{0.23}$. For $\ellp > 100$, the lensing potential mostly probes early enough times that $g(a) \approx 1$ is a good approximation~\citep{Pan:2014xua}. However, if we wish to express the amplitude of $P_\Psi(k)$ at early times in terms of $\sigma_8$ (at the present), we require $g(a=1)$.

The transfer function is normalized to unity for $k \ll k_{\rm eq}$, where
$k_{\rm eq} \sim 10^{-2} \, \mathrm{Mpc}^{-1}$ the inverse of the comoving Hubble parameter at matter-radiation equality. For $k \gg k_{\rm eq}$ we have $T(k) \sim (k/k_{\rm eq})^{-2} \ln(k/k_{\rm eq})$. Keeping the radiation energy density fixed today, $k_{\rm eq} \propto \omega_{\rm m}$, where $\omega_{\rm m} \equiv \Omega_{\rm m}h^2$. As well as $k_{\rm eq}$, the other scales that enter the transfer function are the sound horizon at the drag epoch (when baryons dynamically decouple) $r_{\rm drag}$, and the diffusion scale $k_{\rm D}$. The broad-band shape is controlled by $k_{\rm eq}$, while $r_{\rm drag}$ determines the scale below which baryon suppression effects are important (see e.g.~\citealt{Eisenstein:1997ik}). The non-zero baryon fraction $\omega_{\rm b}/\omega_{\rm m}$ tends to suppress the transfer function on small scales since (essentially) unclustered baryons impede the growth of the CDM density perturbation up to the time after decoupling when baryon infall into the CDM potential wells is complete. The further effect of baryons is to imprint baryon acoustic oscillations in the transfer function for $k r_{\rm drag} > \pi$ and $k \lea k_{\rm D}$. Baryon acoustic oscillations have little impact on $\clpp$ since they are washed out by the line-of-sight integration. We therefore ignore them here and, where an explicit form for the transfer function is required, for illustrative purposes we use the `no-wiggle' fit of~\citet{Eisenstein:1997ik}. For the range of parameters of interest, we can assume that the physical baryon density $\omega_{\rm b}$  is fixed, so the transfer function can be thought of as a function of $k/k_{\rm eq}$, $k r_{\rm drag}$ and $\omega_{\rm m}$.

Ignoring departures of the growth function from unity and approximating $\mathcal{P}_{\mathcal{R}}(k)$ as scale-invariant with amplitude $A_{\rm s}$, we can write
\begin{equation}
\ellp^4 \clpp \propto A_{\rm s} \elleq \int_0^1 dx \, \left(\frac{\ellp}{\elleq x}\right) T^2 \left(\frac{\ellp}{x \chi_*}\right) (1-x)^2 \, ,
\label{eq:lenspower}
\end{equation}
where $x=\chi/\chi_*$ and $\elleq \equiv k_{\rm eq} \chi_* \approx 140$. The $A_{\rm s} \elleq$ prefactor has a simple physical interpretation: CMB photons are typically deflected by $\elleq$ lenses of size $k_{\rm eq}^{-1}$, and each mean-squared deflection is proportional to $A_{\rm s}$, giving a mean-squared deflection proportional to $A_{\rm s} \elleq$. Given the $k$-dependencies of the transfer function through $k/k_{\rm eq}$ and $k r_{\rm drag}$, and its direct parameter-dependence on $\omega_{\rm m}$, we see that in \lcdm, the lensing power spectrum at $\ellp \gtrsim 100$ is (mostly) determined by only $A_{\rm s}$, $\ellp/\elleq$, $\ellp/\ell_{\rm drag}$, and $\omega_{\rm m}$. Here, the multipole $\ell_{\rm drag} \equiv \chi_*/r_{\rm drag}$ characterizes the BAO scale at last-scattering. The integral in Eq.~\eqref{eq:lenspower} determines the shape of the lensing power spectrum. Defining
\begin{equation}
f\left(\frac{L}{\elleq},\frac{L}{\ell_{\rm drag}},\omega_{\rm m}\right) \equiv \int_0^1 dx \, \left(\frac{\ellp}{\elleq x}\right) T^2 \left(\frac{\ellp}{\chi_* x}\right) (1-x)^2 \, ,
\end{equation}
the dependence on the cosmological parameters (e.g. $\elleq$) in the vicinity of the fiducial model is, at any $\ellp$, approximately a product of power laws with exponents
\begin{eqnarray}
n_\ellp^{\rm eq} &\equiv& - \left. \partial \ln f_\ellp / \partial \ln (L/\elleq) \right|_{\rm fid} \, , \\
n_\ellp^{\rm drag} &\equiv& - \left. \partial \ln f_\ellp / \partial \ln (L/\ell_{\rm drag})\right|_{\rm fid} \, , \\
n_\ellp^{\rm m} &\equiv&  \left. \partial \ln f_\ellp / \partial \ln \omega_{\rm m} \right|_{\rm fid} \, ,
\end{eqnarray}
each evaluated for the fiducial cosmology. Note also that
$n_\ellp^{\rm eq}$ and $n_\ellp^{\rm drag}$ determine the local power-law slope of $\clpp$, with
\begin{equation}
 n_\ellp^{\rm eq}+n_\ellp^{\rm drag} = - d\ln (\ellp^4 C_\ellp^{\phi\phi}) / d\ln \ellp \, .
\end{equation}
On the angular scales where Eq.~\eqref{eq:lenspower} is valid, only modes with $k > k_{\rm eq}$ contribute. For these scales, $(k/k_{\rm eq})T^2(k)$ increases with $k_{\rm eq}$ giving $n_\ellp^{\rm eq} >  0$ (and monotonic increase of $n_\ellp^{\rm eq} $ with $\ellp$).
Around $\ellp = 200$, the scales that dominate the $S/N$ of a lensing amplitude measurement from \planck, $n_\ellp^{\rm eq} \approx 0.9$. This, and $A_{\rm s}$, dominate the parameter dependence of $\clpp$. The effect of baryon suppression is reduced as $\omega_{\rm m}$ increases at fixed $\omega_{\rm b}$, so $n_\ellp^{\rm m} > 0$ and also rises monotonically with $\ellp$. Around $\ellp =200$, we find $n_\ellp^{\rm m} \approx 0.3$, which gives a small, but non-negligible, further parameter dependence to $\clpp$. Finally, the dependence on $\ell_{\rm drag}$ is very small, since $n_\ellp^{\rm drag} \leq 0.15$ (with peak value around $\ellp \approx 100$, roughly equal to $\ell_{\rm drag}$). Putting these pieces together, we find a dependence of $\ellp^4 \clpp$ on cosmological parameters scaling approximately as
\begin{equation}
\ellp^4 \clpp \propto A_{\rm s} \elleq^{n_\ellp^{\rm eq} +1} \ell_{\rm drag}^{n_\ellp^{\rm drag}} \omega_{\rm m}^{n_\ellp^{\rm m}} \, ,
\label{eq:clppdependscales}
\end{equation}
for $\ellp \gtrsim 100$, where the dependence on $\ell_{\rm drag}$ is very weak.

We can express the parameter dependence of $\clpp$ in terms of the usual \lcdm\ parameters. The comoving distance to last-scattering, $\chi_*$, is
\begin{equation}
\chi_* = \int_{a_*}^1 \frac{da}{a^2 H(a)} \, ,
\end{equation}
where $a_*$ is the scale factor at last-scattering (whose dependence on cosmology can be ignored here).
At fixed radiation energy density today, the dependence on $\Omega_{\rm m}$ and $h$ is approximately
\begin{equation}
\chi_* \propto \Omega_{\rm m}^{-0.4} h^{-1} \, ,
\end{equation}
around $\Omega_{\rm m} = 0.3$ and $h=0.7$.\footnote{We ignore the small dynamical effects of radiation at high redshift.} This differs from the $(\Omega_{\rm m} h^2)^{-1/2}$ of an Einstein-de Sitter universe due to the late-time effects of $\Lambda$. Using $k_{\rm eq} \propto \Omega_{\rm m} h^2$, we have
\begin{equation}
\elleq \propto \Omega_{\rm m}^{0.6} h \, .
\end{equation}
At fixed baryon density, the sound horizon at the drag epoch scales like $r_{\rm drag} \propto \omega_{\rm m}^{-0.25}$ in the vicinity of the fiducial model, so that
\begin{equation}
\ell_{\rm drag} \propto \Omega_{\rm m}^{-0.15} h^{-0.5} \, .
\end{equation}
Using these dependencies in Eq.~(\ref{eq:clppdependscales}), we find
\begin{equation}
\ellp^4 \clpp\propto A_{\rm s} \left(\Omega_{\rm m}^{0.6} h\right)^{n_\ellp + 1} \Omega_{\rm m}^{0.15\times n_\ellp^{\rm drag}-0.2\times n_\ellp^{\rm m}}
\, ,
\label{eq:clppdepend}
\end{equation}
where
\begin{equation}
n_\ellp \equiv n_\ellp^{\rm eq}-0.5\times n_\ellp^{\rm drag} +2n_\ellp^{\rm m}
\end{equation}
is approximately $1.5$ at $\ellp=200$. The remaining dependence on $\Omega_{\rm m}$ is very weak, with $0.15\times n_\ellp^{\rm drag}-0.2\times n_\ellp^{\rm m} \approx -0.05$ at $\ellp = 200$. Equation~(\ref{eq:clppdepend}) gives the main dependence of the lensing power spectrum on the cosmological parameters of \lcdm\ models at multipoles $\ellp \gtrsim 100$.

\subsection{Normalization by $\sigma_8$}

The dependence of the lensing power spectrum on $A_{\rm s}$ can be eliminated in favour of $\sigma_8$. We use the Poisson equation to relate the dimensional matter power spectrum at $a=1$ to that of the Weyl potential, giving
\begin{equation}
P_\delta (k;a=1) \propto \omega_{\rm m}^{-2} k^4 P_\Psi(k;a=1) \, .
\end{equation}
It follows that the variance of the density contrast in spheres of radius $R$ is
\begin{align}
\sigma_R^2 &\propto \omega_{\rm m}^{-2} g^2(a=1) \int dk\, k^3 \mathcal{P}_{\mathcal{R}}(k) T^2(k) W^2(kR) \nonumber \\
&\propto A_{\rm s} k_{\rm eq}^4 \omega_{\rm m}^{-2} g^2(a=1) \int_0^\infty
dx \, x^3 T^2(x k_{\rm eq}) W^2(x k_{\rm eq} R) \, ,
\label{eq:sigma8}
\end{align}
where we have assumed that $\mathcal{P}_\mathcal{R}(k)$ is scale-invariant in passing to the second line. The window function
\begin{equation}
W(kR) \equiv \frac{3 j_1(kR)}{kR} = \frac{3}{(kR)^3}(kR \cos kR-\sin kR) \, ,
\end{equation}
is the Fourier transform of a normalized spherical top-hat function of radius $R$; following convention, we take $R = 8 h^{-1} \, \mathrm{Mpc}$.

The integral in Eq.~\eqref{eq:sigma8} is a function of $k_{\rm eq}R$, $k_{\rm eq} r_{\rm drag}$, and $\omega_{\rm m}$ at fixed baryon density (with the latter two from the transfer function). Evaluating the logarithmic derivative with respect to these parameters at the fiducial values, using the `no-wiggle' transfer function from~\citet{Eisenstein:1997ik}, we find that the integral scales as $(k_{\rm eq}R)^{-1.4}
\omega_{\rm m}^{0.45}$, with only a very weak further dependence on $k_{\rm eq} r_{\rm drag}$, in the vicinity of the fiducial model. Finally, the parameter dependence of $\sigma_8^2$ is
\begin{align}
\sigma_8^2 &\propto A_{\rm s} k_{\rm eq}^4 \omega_{\rm m}^{-2} g^2(a=1) (k_{\rm eq} h^{-1})^{-1.4} \omega_{\rm m}^{0.45} \nonumber \\
&\propto  A_{\rm s} \Omega_{\rm m}^{1.5} h^{3.5} \, .
\label{sigma8Approx}
\end{align}
This parameter dependence agrees well with direct finite-differencing of $\sigma_8$ calculations with \camb.

\raggedright
\end{document}